\documentclass[prd,superscriptaddress,nofootinbib,showkeys]{revtex4-1}
\usepackage{head}
\usepackage{setspace}
\usepackage{booktabs}
\usepackage{lineno,hyperref}
\usepackage{float}
\usepackage{xcolor,ulem}

\begin{document}
\title{The impact of  ATLAS and CMS  single differential top-quark pair measurements at $\sqrt{s}=8$ TeV on CTEQ-TEA PDFs}

\author{Musajan Kadir}
\affiliation{
	School of Physics Science and Technology, Xinjiang University,
	Urumqi, Xinjiang 830046 China }

\author{Alim Ablat}
\affiliation{
	School of Physics Science and Technology, Xinjiang University,
	Urumqi, Xinjiang 830046 China }

\author{Sayipjamal Dulat}
\email{sdulat@hotmail.com}
\affiliation{
	School of Physics Science and Technology, Xinjiang University,
	Urumqi, Xinjiang 830046 China }

\author{Tie-Jiun Hou}
\email{houtiejiun@mail.neu.edu.cn}
\affiliation{
  Department of Physics, Northeastern University, Shenyang, Liaoning, China}

\author{Ibrahim Sitiwaldi}
\affiliation{
	School of Physics Science and Technology, Xinjiang University,
	Urumqi, Xinjiang 830046 China }

\date{\today}

\begin{abstract}

	By applying the Error PDF Updating Method, we analyze the impact of the absolute and normalized single differential cross-sections for top-quark pair production data from the ATLAS and CMS experiments at the Large Hadron Collider, at a centre-of-mass energy of $\sqrt{s}= 8$ TeV, on the CT14HERA2 PDFs.
	We find that  the top quark pair single differential distributions  provide minor constraints on the CT14HERA2 gluon PDF when the nominal CT14HERA2 inclusive jet production data are included in the fit. Larger constraints on the gluon distribution are present when the jet data are removed (CT14HERA2mJ) and/or when increased weights are given to the top data in the CT14HERA2 fits.  The weighted $t\bar t$ data provide significant constraints on the CT14HERA2mJ gluon PDF, that are comparable to those obtained from inclusive jet production data. Furthermore, we examine the top quark mass sensitivity of the top-quark pair single differential distributions.

\end{abstract}

\keywords{top-quark pair production; gluon PDF; jet data}

\maketitle
\newpage
\tableofcontents
\newpage
\setcounter{page}{0}

\section{Introduction} \label{sec:Introduction}

Precise measurements of and  predictions for  $t \bar t$ 
pair production are crucial for tests of the standard model and for searches for new physics beyond the standard model~\cite{Frederix:2007gi}.
Thus, an understanding of the uncertainties due to an imperfect knowledge of parton distribution functions (PDFs) is crucial. The large integrated luminosity and the high centre-of-mass energy of the Large Hadron Collider (LHC), provide a large sample of  $t \bar t$ events. 
The dominant production mechanism for $t\bar t$ pair production at the LHC is through gluon-gluon fusion, and thus $t\bar t$ data have the potential for constraining the gluon PDF, especially at high $x$.
In the analysis carried out in Ref.~\cite{Guzzi:2014wia} theory predictions for the final-state top quark $p_T^t$ and $y_t$ distributions in $t \bar t$ pair production at the LHC 
at approximate next-to-next-to-leading order (aNNLO) in QCD are used to study the impact of ATLAS and CMS 7 TeV differential cross sections measurements on proton PDFs.
The aNNLO theory prediction in~\cite{Guzzi:2014wia} uses methods of QCD threshold resummation beyond the leading logarithmic accuracy and is implemented in the xFitter platform~\cite{Bertone:2016ywq}  by using fastNLO tables to facilitate the global PDF analysis.  
A moderate improvement on the uncertainty on the gluon distribution at high $x$ was observed.
The more recent analyses of Refs.~\cite{Czakon:2016dgf,Czakon:2017dip}
have also provided fastNLO tables for the exact NNLO predictions
 for the  invariant mass
of the top-quark pair, the average transverse momentum of the $t,\bar{t}$ quark, the average rapidity of the  $t,\bar{t}$ quark, and the rapidity of the top-quark pair, i.e. the distributions measured by the ATLAS and CMS experiments. The fastNLO tables are at NNLO in QCD with $m_t =173.3$ GeV,  renormalization scale  and factorization scales  
$\mu_R=\mu_F= H_T/4$, $H_T=\sqrt {m^2_t + p^2_{T,t}} + \sqrt {m^2_t + p^2_{T,\bar t}}$ for $m_{t\bar t}$, $y_{t\bar t}$ and $y_t$ distributions, and 
$\mu_R=\mu_F= m_T/4 =\frac{1}{4}\sqrt {m^2_t + p^2_{T}}$ for $p_{T}$ distribution of the average top/antitop quark. And in their calculation they use the same binning (see  Table~\ref{tab:bin} ) as
the  ATLAS~\cite{Aad:2015mbv} and CMS \cite{Khachatryan:2015oqa}  8 TeV measurements of top-quark pair differential cross-sections.
\begin{table}[h]\caption{Summary of the  fastNLO tables provided in the work~\cite{Czakon:2017dip}.}\label{tab:bin} 
	\begin{center}
		\renewcommand{\arraystretch}{1.5}
		\begin{tabular}{llc}
			\hline
			Observable & Binning & $\mu_{\mathrm{F}}=\mu_{\mathrm{R}}$
			\\ \hline
			$d\sigma/dm_{t\bar{t}}\;\;[\mathrm{GeV}]$ & $\{345,\,400,\,470,\,550,\,650,\,800,\,1100,\,1600\} $ &  $H_T/4$ 
			\\
			$d\sigma/dy_t$ & $\{-2.5,\,-1.6,\,-1.2,\,-0.8,\,-0.4,\,0.0,\,0.4,\,0.8,\,1.2,\,1.6,\,2.5
			\} $ &  $H_T/4$ \\
			$d\sigma/dy_{t\bar{t}}$ & $\{-2.5,\,-1.3,\,-0.9,\,-0.6,\,-0.3,\,0.0,\,0.3,\,0.6,\,0.9,\,1.3,\,2.5
			\} $ &  $H_T/4$ \\
			$d\sigma/dp^t_T \;\;[\mathrm{GeV}]$ & $\{0,\, 60,\, 100,\, 150,\, 200,\, 260,\, 320,\, 400,\, 500\} $ &  $m_T/2$ \\
			\hline
		\end{tabular}
	\end{center}
\end{table}

In this paper we study the impact of the ATLAS~\cite{Aad:2015mbv} and CMS \cite{Khachatryan:2015oqa} measurements of top-quark pair differential cross-sections 
data on the CT14HERA2~\cite{Hou:2016nqm} and CT14HERA2mJ PDFs, and thus in Table~\ref{Tab:chi2} we  provide relevant basic information. For the measurements in Table~\ref{Tab:chi2}, the ATLAS  experiment has  provided statistical, fifty six correlated systematic errors including luminosity errors.
CMS collaboration provided the statistical errors, along with eleven correlated systematic errors, including the luminosity errors. 

\begin{table}[htbp]
	\begin{center}\caption{Number of data points and $\chi^2/N_{pts}$ for inclusive jet and top-quark pair data, after {\tt \texttt{ePump}} updating from the CT14HERA2 and CT14HERA2mJ PDFs.}\vspace{0.4cm}\label{Tab:chi2}
	\begin{tabular}{|c|c c|c|c|c|c|}
		\hline
		\multirow{2}{*}{Detector} & \multicolumn{2}{c|}{\multirow{2}{*}{Observable}} & \multirow{2}{*}{$N_{pts}$} & \multicolumn{2}{c|}{$\chi^2/N_{pts}$(CT14HERA2)} & \multirow{2}{*}{$\chi^2/N_{pts}$(CT14HERA2mJ)} \\ \cline{5-6}
		& \multicolumn{2}{c|}{}  &  &weight=1.0 & weight=9.0 & \\ \hline
		CDF    & inclusive jet &  ~\cite{Aaltonen:2008eq} &  72  & 1.46 & - & 1.50  \\ \hline
	    D0     & inclusive jet &  ~\cite{Abazov:2008ae}   &  110    & 1.03 & - & 1.03  \\ \hline
		ATLAS  & inclusive jet &  ~\cite{Aad:2011fc}      &  90   & 0.57 & - & 0.57  \\ \hline
		CMS    & inclusive jet &  ~\cite{Chatrchyan:2012bja} &  133  & 0.89 & - & 0.93  \\ \hline
		\multirow{4}{*}{ATLAS} & $\frac{1}{\sigma}  \frac{d\sigma}{d|y_{t\bar t}|}, \frac{d\sigma}{d|y_{t\bar t|}}$ & ~\cite{Aad:2015mbv} & 5 &  2.21, 3.83         & 1.18, 1.48 & 5.21, 7.29 \\ \cline{2-7} 
		& $\frac{1}{\sigma}  \frac{d\sigma}{dm_{t\bar t}}, \frac{d\sigma}{dm_{t\bar t}}$ & ~\cite{Aad:2015mbv}& 7 & 0.25, 0.45 & 0.25, 0.42 & 0.35, 0.40 \\ \cline{2-7} 
		&$\frac{1}{\sigma}  \frac{d\sigma}{d|y_t|}, \frac{d\sigma}{d|y_t|}$& ~\cite{Aad:2015mbv}& 5 & 2.40, 2.83 & 1.45, 1.62 & 5.34, 5.79 \\ \cline{2-7} 
		&$\frac{1}{\sigma}  \frac{d\sigma}{dp^t_T}, \frac{d\sigma}{dp^t_T}$&  ~\cite{Aad:2015mbv} & 8 & 0.39, 0.34 & 0.38, 0.33 & 0.38, 0.32 \\ \hline
		\multirow{4}{*}{CMS} & $\frac{1}{\sigma}  \frac{d\sigma}{dy_{t\bar t}}$ & ~\cite{Khachatryan:2015oqa} & 10 & 2.31 & 1.07 & 3.34 \\ \cline{2-7} 
		&$ \frac{1}{\sigma}  \frac{d\sigma}{dm_{t\bar t}}$ & ~\cite{Khachatryan:2015oqa} & 7 & 7.69 & 3.96          & 9.30 \\ \cline{2-7} 
		&$\frac{1}{\sigma}  \frac{d\sigma}{dy_t}$ &          ~\cite{Khachatryan:2015oqa} & 10 & 2.52 & 2.50     & 3.32 \\ \cline{2-7} 
		&$\frac{1}{\sigma}  \frac{d\sigma}{dp^t_T}$&                   ~\cite{Khachatryan:2015oqa}& 8 & 3.55 & 2.20    & 4.82 \\ \hline
	\end{tabular}
    \end{center}
\end{table}

In Tables~\ref{Tab:chi2} we provide the number of data points and $\chi^2/N_{pts}$ for inclusive jet and top-quark pair data, after {\tt \texttt{ePump}} updating from the CT14HERA2 and CT14HERA2mJ PDFs. For the top-quark pair data the $\chi^2/N_{pts}$ decreases for CT14HERA2 than for CT14HERA2mJ PDFs. This shows that original CT14HERA2mJ PDFs is enhanced in quality after including inclusive jet data. But it is not the case for the ATLAS absolute and normalized $p^t_T$ , as well as absolute $m_{t\bar t}$ distributions, $\chi^2$ also did not decrease visibly in these distributions even when we set the weight=9. The  $\chi^2/N_{pts}$ of the four inclusive jet distributions are very different from the top-quark distributions. 
$\chi^2/N_{pts}$ is far larger in CMS normalized $t\bar t$ distributions than in ATLAS normalized $t\bar t$ distributions. It means, at least, there are some comparable difference between CMS and ATLAS data for the same measurements, while we are using the same theoretical predictions.

Ref.~\cite{Czakon:2016olj,Ball:2017nwa} has previously
studied the impact of top-quark pair differential distributions measured by ATLAS~\cite{Aad:2015mbv} and CMS \cite{Khachatryan:2015oqa} at 8 TeV on the gluon PDF within the NNPDF framework. They found that the differential distributions from top-quark pair production provide relatively strong constraints on the large-x gluon.
Within the  MMHT framework, Ref.~\cite{Bailey:2019yze} found that  the impact of the ATLAS~\cite{Aad:2015mbv} data on the  gluon PDF is relatively weak. 
With the  CMS data~\cite{Khachatryan:2015oqa},  
they found  that both $y_t$ and  $y_{t\bar t}$ distributions have a noticeable
impact on the gluon PDF at high x, with the impact of the  $y_{t\bar t}$  larger than that of $y_t$.
This paper examines in detail the impact of the LHC $t \bar{t}$ data in the CTEQ-TEA framework. 

Despite improvements, such as the use of fastNLO tables, global PDF fitting is still very CPU-intensive. In Ref.~\cite{Schmidt:2018hvu},  a software package, {\tt \texttt{ePump}} (error PDF Updating Method Package) ~\cite{Schmidt:2018hvu}, has been developed which can provide both the updated best-fit PDF and the updated eigenvector PDFs from a PDF set previously obtained by a global PDF analysis. {\tt \texttt{ePump}} has been previously used~\cite{Willis:2018yln}, \cite{Hou:2019gfw}  \cite{Yalkun:2019gah}~\cite{Czakon:2019yrx} to perform analyses that have the potential to reduce PDF uncertainties at the LHC. 

In this paper, we use {\tt \texttt{ePump}} to study the impact of the LHC 8 TeV single differential top-quark pair distribution
data from ATLAS~\cite{Aad:2015mbv} and CMS~\cite{Khachatryan:2015oqa} on  the gluon PDFs starting from the global PDF sets CT14HERA2~\cite{Hou:2016nqm} and CT14HERA2mJ. 
CT14HERA2 is an updeted version of the CT14NNLO PDFs~\cite{Dulat:2015mca} with 
the HERA Run I data  replaced by the combined HERA I+II data~\cite{Abramowicz:2015mha}. The  CT14HERA2 PDF fit contains inclusive jet data from the Tevatron and from the LHC. Since inclusive jet data also provide constraints on the gluon distribution, additional PDFs, titled as CT14HERA2mJ, were constructed without the jet data by a full PDF global
analysis, in order to examine more closely the impact of the $t \bar{t}$ data alone, and in combination with the jet data.  

The absolute and normalized (to the total $t \bar{t}$ cross section) single differential $t\bar t$ measurements from ATLAS
in the variables $|y_{t\bar t}|$, $dm_{t\bar t}$, $p^t_T$ and $|y_t|$, 
and the  normalized single differential $t\bar t$ measurements from CMS in the variables
$y_{t\bar t}$, $m_{t\bar t}$, $p^t_T$ and $y_t$ are listed in Table~\ref{Tab:chi2}.
We also show the number of data points for jet data that are included in the CT14HERA2 fit. 
The values of $\chi^2/N_{pts}$ in the
Table~\ref{Tab:chi2} are calculated by using {\tt \texttt{ePump}} to update the CT14HERA2 and CT14HERA2mJ PDFs with the inclusion of each individual $t\bar t$ data sets. These will be discussed in detail later in this paper.

This paper is organized as follows.
In section~\ref{sec:ATLAS8updateHERA2}, 
we first calculate the degree of correlation between the CT14HERA2 gluon PDF and the ATLAS and CMS 8 TeV $t \bar{t}$ data.  Then we compare the {\tt \texttt{ePump}} updated gluon PDFs obtained by adding that data one by one, to the original CT14HERA2 PDFs. 
The corresponding NNLO theory for $t \bar{t}$ predictions using the updated PDFs are then compared to the corresponding ATLAS and CMS measurements. The impact of the updated PDFs on the  Higgs boson  gluon-gluon fusion cross section $\sigma_H (gg \rightarrow H)$ is then discussed.
In section \ref{sec:tension},
tensions between the ATLAS and CMS 8 TeV  absolute and normalized single differential $t\bar t$ data  with the  other data sets in the CT14HERA2 PDFs are described.
In section \ref{sec:ATLAS8updateHERA2mJ}, 
using the same method utilized with the CT14HERA2 PDFs,  we analyze the impact of the ATLAS and CMS  8 TeV  single differential $t\bar t$ measurements on the CT14HERA2mJ PDFs. 
In section \ref{sec:jetupdateHERA2mJ}, 
we compare the impact from the CMS 7 TeV inclusive jet data
and from the $t \bar{t}$ data. 
In section \ref{mtop}, the impact of the value of the top quark mass on the single differential $t \bar{t}$ cross section predictions is analyzed. 
Our conclusions are presented in section \ref{sec:conclusion}.

Before beginning the full discussion of the analysis,   we summarize the notations used in this paper: 
\begin{itemize}
	\item The suffix ``.54" in CT14HERA2.54 indicates that the error band is obtained with 54 eigen-vector PDF sets rather than with the entirety of the  56 PDF sets. The last two sets are omitted, which expand the uncertainty for the small $x$ gluon, a region not relevant for this study.
	\item CT14HERA2mJ PDFs are obtained  after excluding the four jet data sets present in the CT14HERA2 PDFs.
	\item The {\tt \texttt{ePump}} updated CT14HERA2.54 (CT14HERA2mJ) PDFs using the ATLAS  8 TeV  absolute and normalized  data in the $|y_{t\bar t}|$, $m_{t\bar t}$, $|y_t|$ and $p^t_T$ distributions are denoted by attaching the suffix XXX and NXXX to the absolute and normalized distributions, respectively.
\end{itemize}

\section{The impact of  ATLAS and CMS 8 TeV $t\bar{t}$ data  on CT14HERA2 PDFs}\label{sec:ATLAS8updateHERA2}

In this section we examine the impact of the ATLAS~\cite{Aad:2015mbv} and  CMS~\cite{Khachatryan:2015oqa} 8 TeV $t\bar{t}$ data in the CT14HERA2.54 global PDF fit 
and  fastNLO theory at NNLO in QCD~\cite{Czakon:2016dgf,Czakon:2017dip}. 
The integrated luminosities of the ATLAS and CMS 8 TeV measurements are 20.3 fb$^{-1}$~\cite{Aad:2015mbv} and 19.7 fb$^{-1}$, respectively.

\subsection{Correlation between the CT14HERA2 gluon PDF and  $t\bar t$ data}

\begin{figure}[H]
	\centering
	\includegraphics[width=0.49\textwidth]{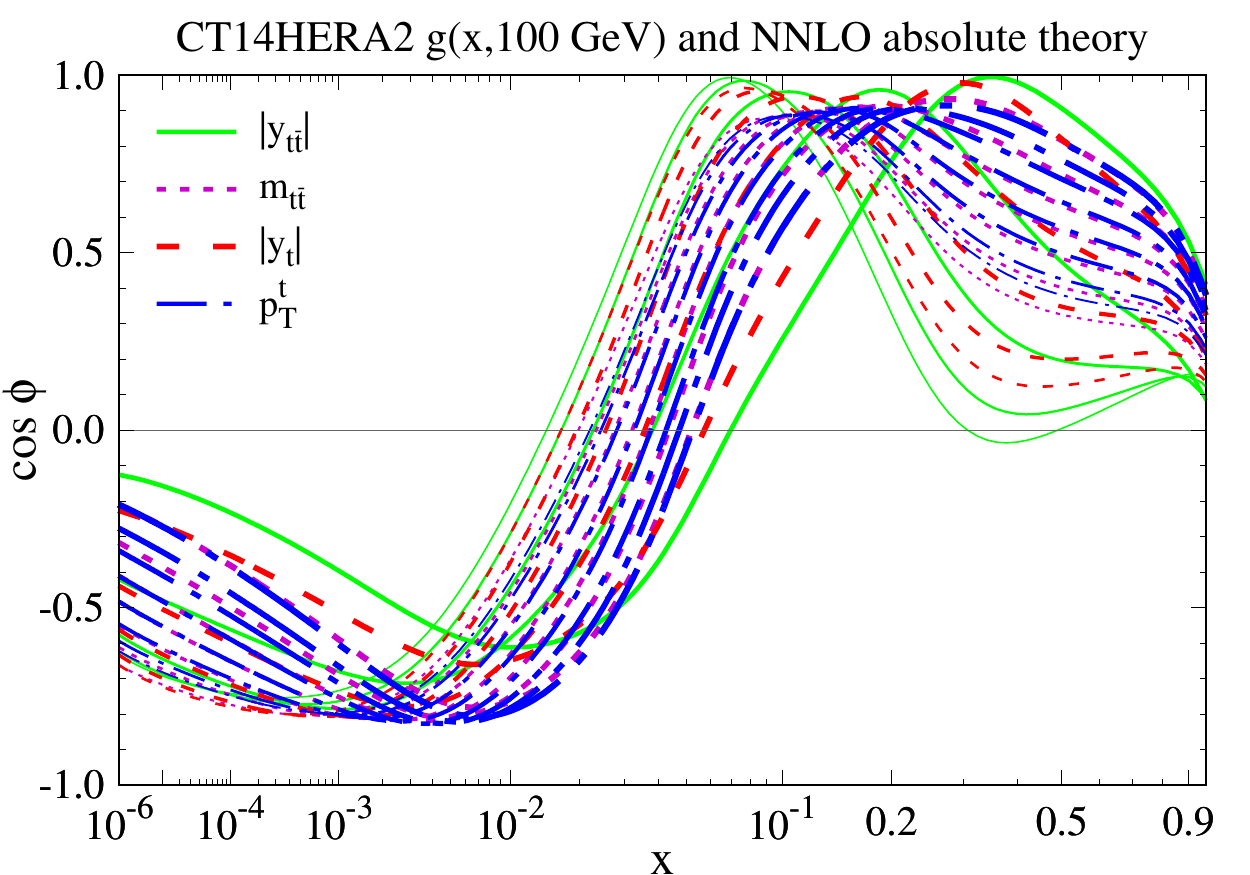}\vspace {0.4cm}
	\includegraphics[width=0.49\textwidth]{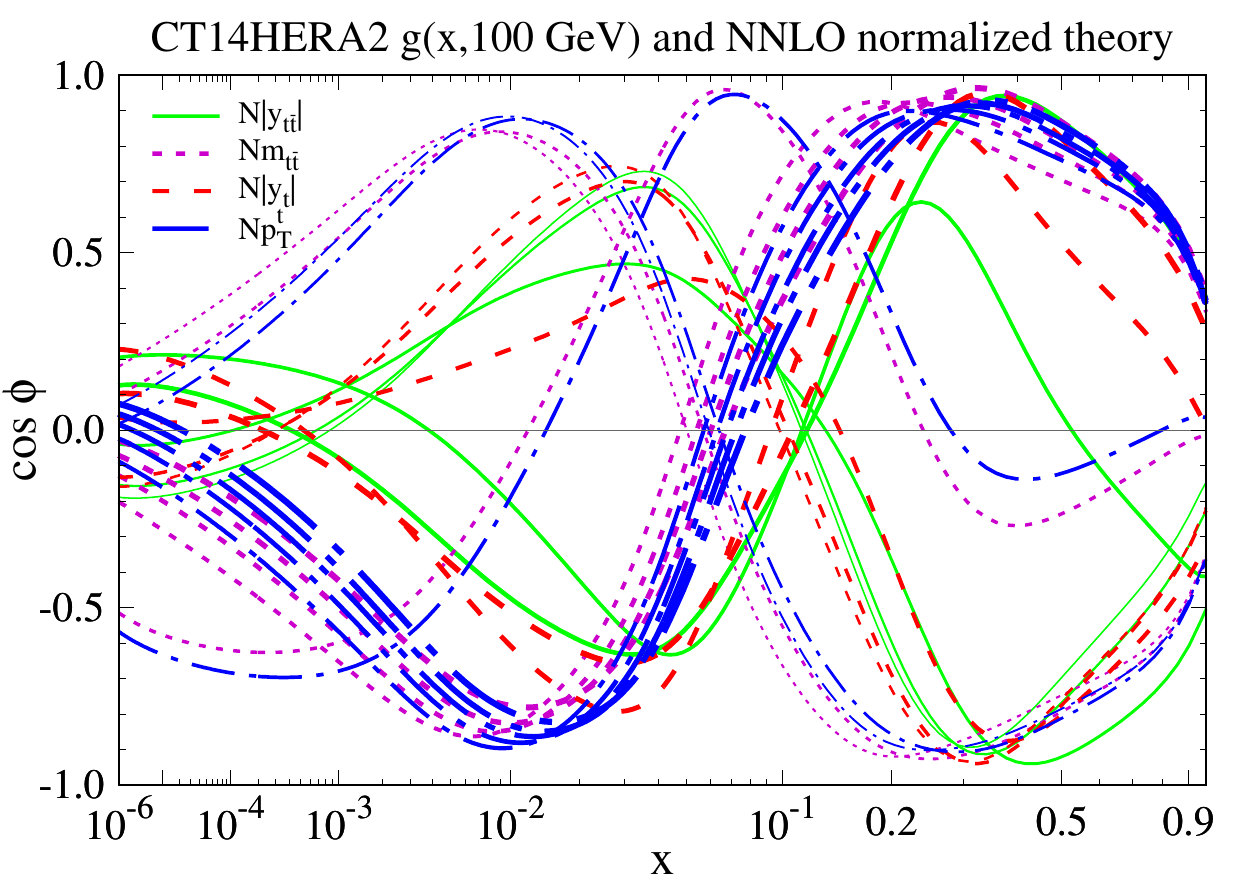} \\
	\includegraphics[width=0.49\textwidth]{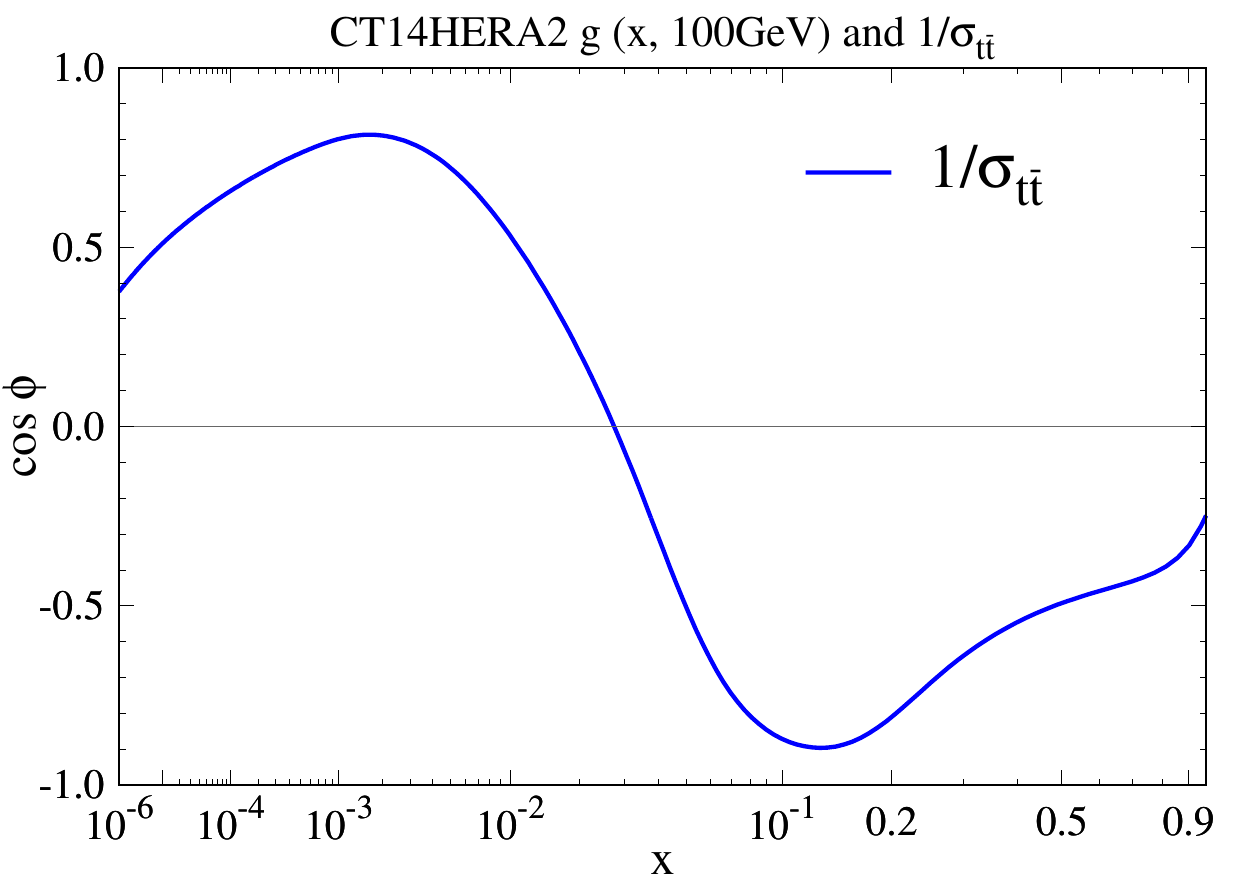}\vspace{0.5cm}
	\caption{Correlation cosine $\cos \phi$ between CT14HERA2.54 $g(x,Q= 100~GeV)$ PDF and fastNLO predictions for each bins of the $t\bar t$ differential distribution  absolute (left) and normalized (right), as well as inverse of the total cross sections (bottom).
		Note that the thickness of the line for each distribution changes from thin to thick, which corresponds to the from first bin to the last bin.  }\label{cosphi-CT14HERA2-gluon-ttbarTheory}
\end{figure}

The correlation between 
a specific absolute or normalized $t\bar t$ data point
and $g(x, Q )$ gluon PDF at a given $x$ and $Q$ value is represented by the correlation cosine
$\cos\phi$ \cite{Nadolsky:2008zw,Campbell:2017hsr}. 
Here, the quantity of $\cos\phi$  characterizes whether the data point and the PDF
is correlated ($\cos\phi \sim 1$), anti-correlated ($\cos\phi \sim -1$) or uncorrelated ($\cos\phi \sim 0$).
Large positive and negative values of $\cos\phi$ indicate 
direct sensitivity of the $t\bar t$ data point to the gluon PDF in a particular region in $x$.
In Fig.~\ref{cosphi-CT14HERA2-gluon-ttbarTheory}, the correlation coefficient between CT14HERA2.54 g($x, Q = 100$ GeV)  PDF and the absolute (left) and normalized (right) differential $t\bar t$ data   is distinguished by varying the type of line used. Each data point is represented by its own correlation curve.
Solid green lines, magenta dotted lines, red dashed lines, dark blue long-dashed-dotted lines correspond to the LHC 8 TeV absolute (left) and normalized (right) differential $t \bar t$  cross-section data 
as a function of the $|y_{t\bar t}|$, $m_{t\bar t}$, $|y_t|$, and $p^t_{T}$, respectively.
We observe that, due to the kinematic range, the absolute $t \bar{t}$ distributions are highly correlated to the gluon PDF for $ 0.08 \lesssim x \lesssim 0.4$ and highly anti-correlated for $10^{-4} \lesssim x \lesssim 10^{-2}$. We also observe that, due to the total $t \bar{t}$ pair production in the denominator, the normalized $t \bar{t}$ distributions show correlations that are basically the same for each variable, and are a mirror image of the dominant behavior of the $\cos\phi$ distributions for the absolute data. 

\subsection{Update CT14HERA2 PDFs using ATLAS and CMS 8 TeV $t\bar t$ data}

\begin{figure}[H]	
\centering
\includegraphics[width=0.49\textwidth]{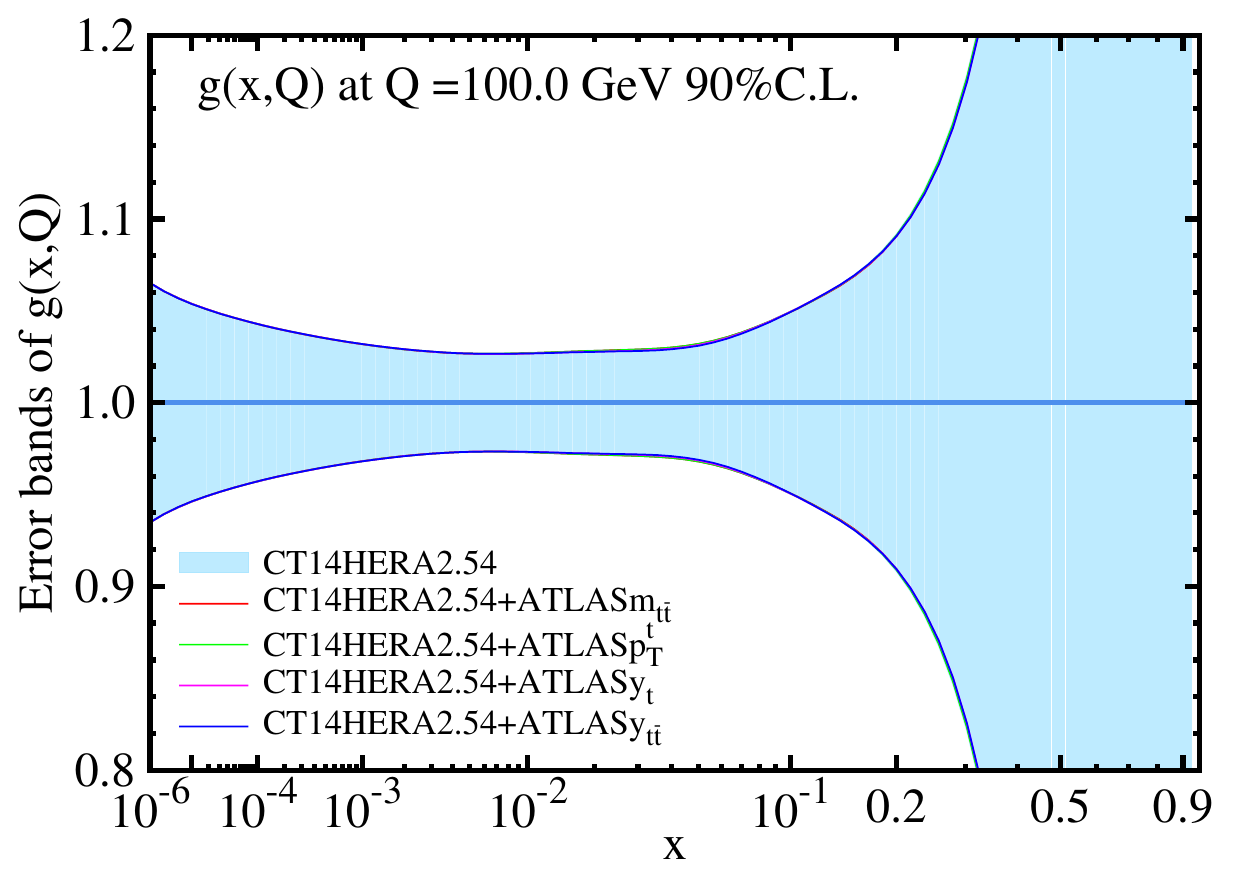}
\includegraphics[width=0.49\textwidth]{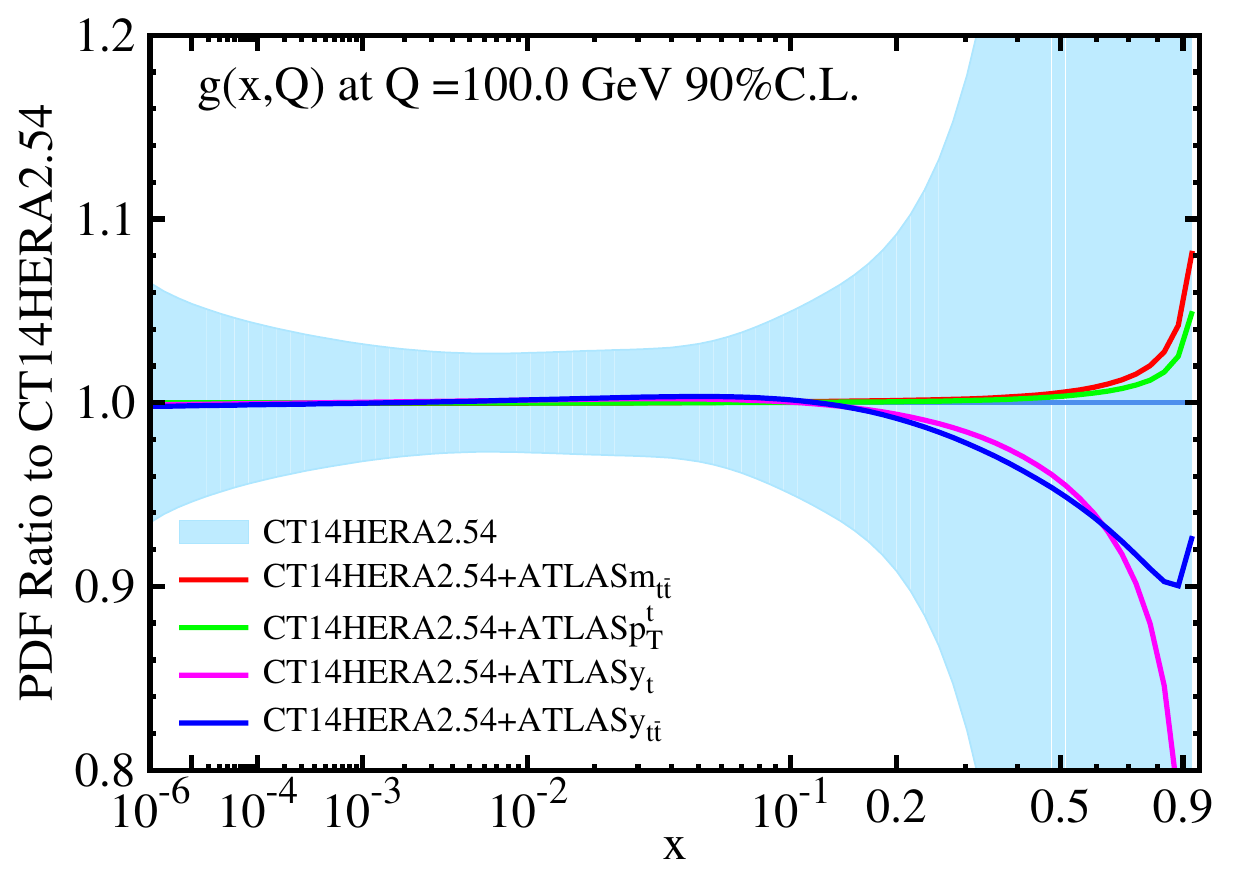}
\includegraphics[width=0.49\textwidth]{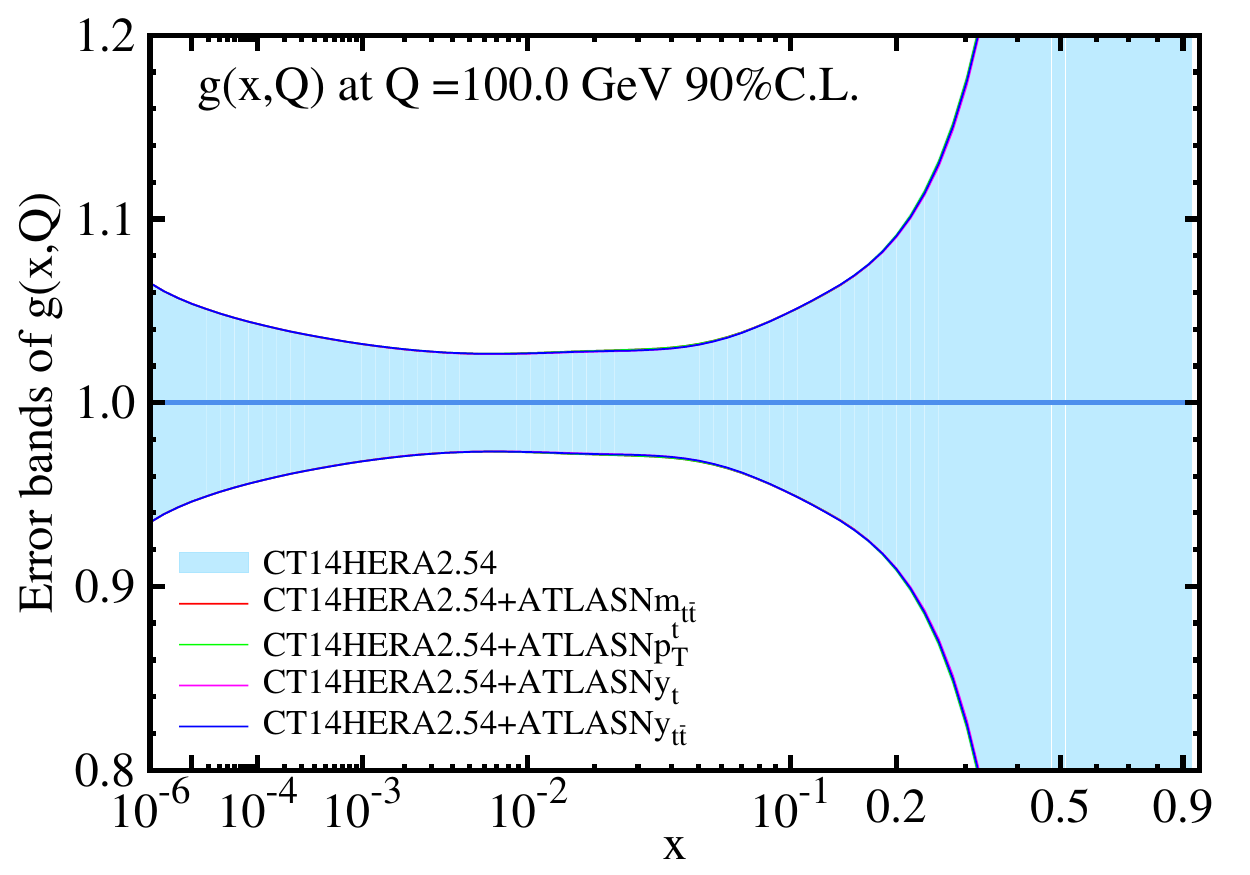}
\includegraphics[width=0.49\textwidth]{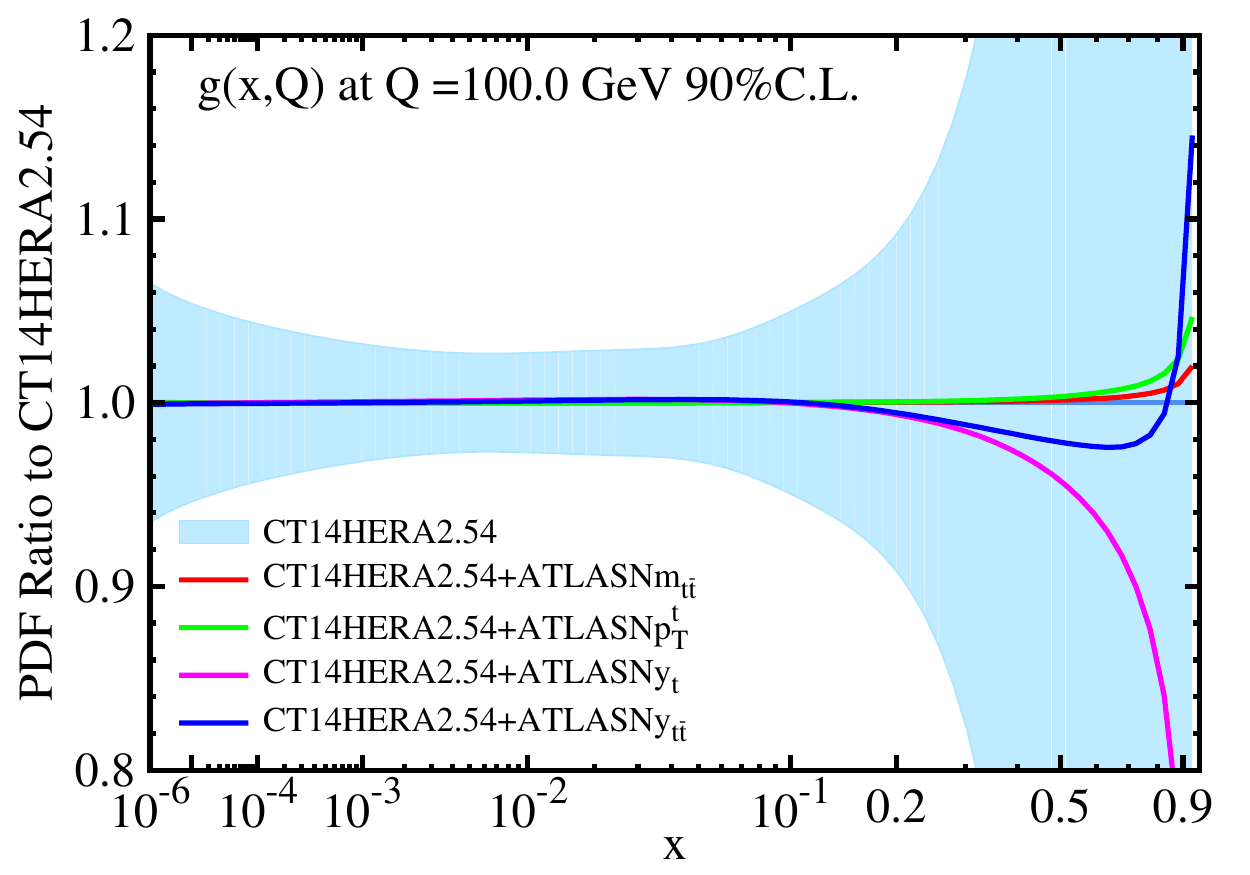}
\includegraphics[width=0.49\textwidth]{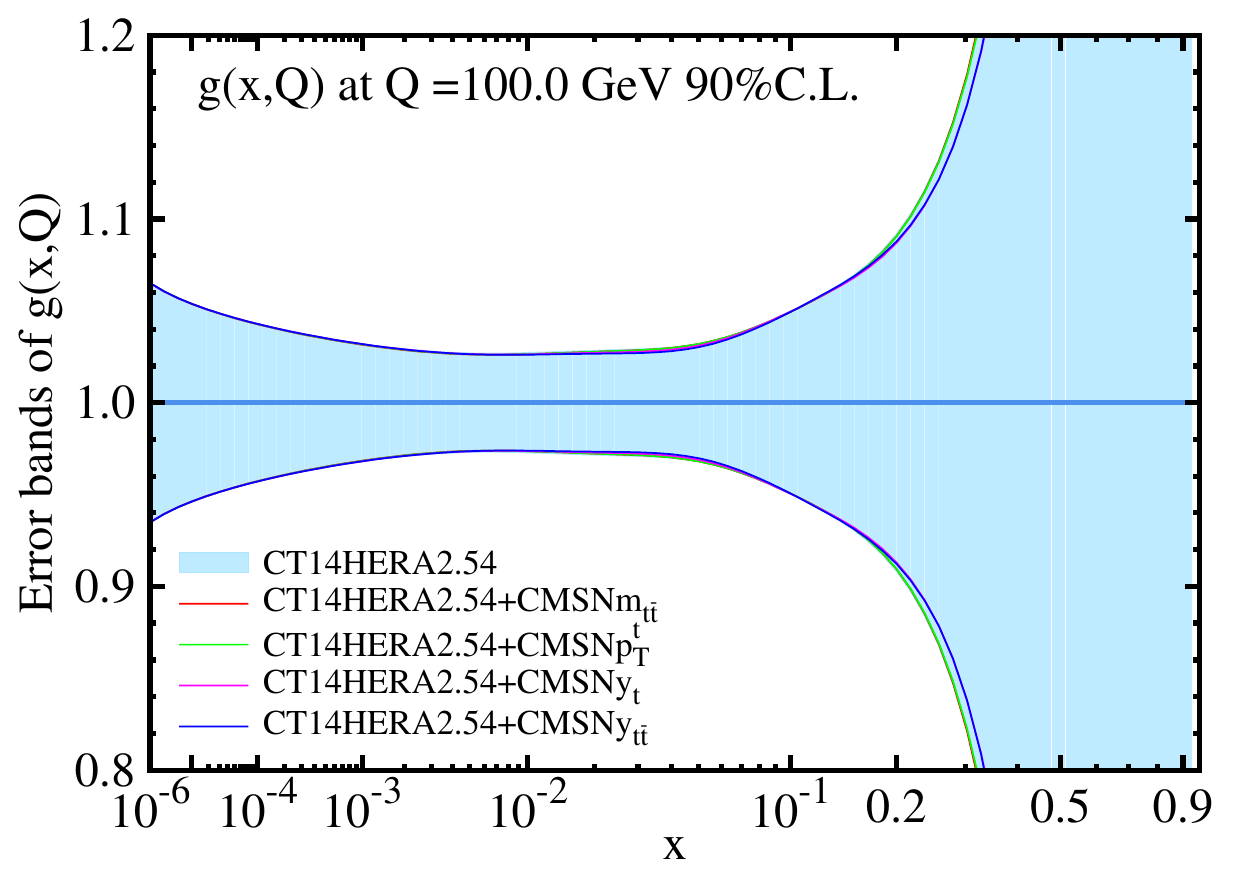}
\includegraphics[width=0.49\textwidth]{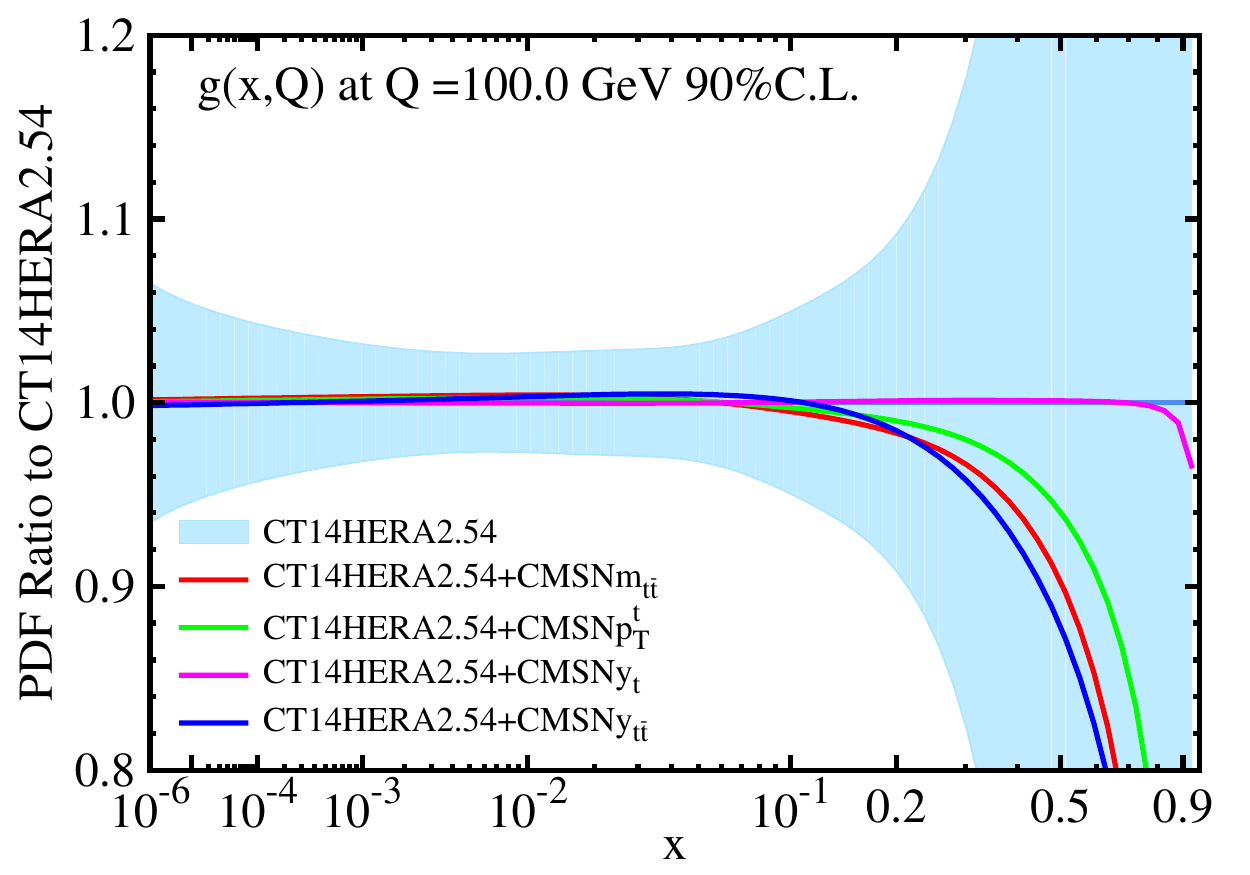}
\caption{The gluon PDF error bands (left) and ratios (right) for {\tt \texttt{ePump}}-updated PDFs  over the best-fit of the base CT14HERA2.54 gluon PDF.}
\label{Fig:CT14HERA2pttb-sing-dis}.
\end{figure}

In this section, using the CT14HERA2.54 PDFs as a basis, we study the impact of the the ATLAS (absolute and normalized) and CMS (normalized)  8 TeV $t\bar t $ full phase-space differential cross-sections as a function of the $y_{t\bar t}$, $m_{t\bar t}$, $p^t_T$, and $y_t$ variables, on the gluon PDF.
The ATLAS and CMS $t\bar{t}$ data are included individually using {\tt \texttt{ePump}}.  The results are shown in Fig.~\ref{Fig:CT14HERA2pttb-sing-dis}. The impact  on both the central gluon distribution and on the gluon uncertainty band (with respect to the CT14HERA2.54 gluon PDF) is shown. It is evident that there is no notable impact on the central gluon from either the absolute or normalized ATLAS 8 TeV  $t\bar{t}$ data  for the $m_{t\bar t}$ and $p^t_T$ distributions. However,  both the absolute and normalized  $|y_{t\bar t}|$ and $|y_t|$ distributions have a relatively minor impact on the best fit gluon PDF $x > 0.2$. It is also evident that none of the distributions result in a significant reduction of the gluon PDF uncertainty at any $x$ value. 
This implies that, either the ATLAS $t\bar{t}$ single differential data are in strong tension with the other data included in CT14HERA2, or the gluon PDF is well constrained by other data, or both.
In contrast to the ATLAS data, we observe that the CMS normalized $y_{t\bar t}$, $m_{t\bar t}$ and $p^t_{T}$ data provide relatively larger impacts on both the central predictions and the uncertainty bands of the CT14HERA2.54 gluon PDF at high $x$, 
while the $y_t$ distribution does not.
In the region $x>0.1$, the inclusion of the $y_{t\bar t}$, $m_{t\bar t}$ and $p^t_{T}$ data leads to a decrease in the gluon PDF, but still well within the PDF error band. 
It is known that, in general,  the gluon PDF is mainly constrained by the DIS and jet data.

\subsection{Optimizing CT14HERA2 PDFs}
\begin{figure}[H]	
	\centering
	\includegraphics[width=0.49\textwidth]{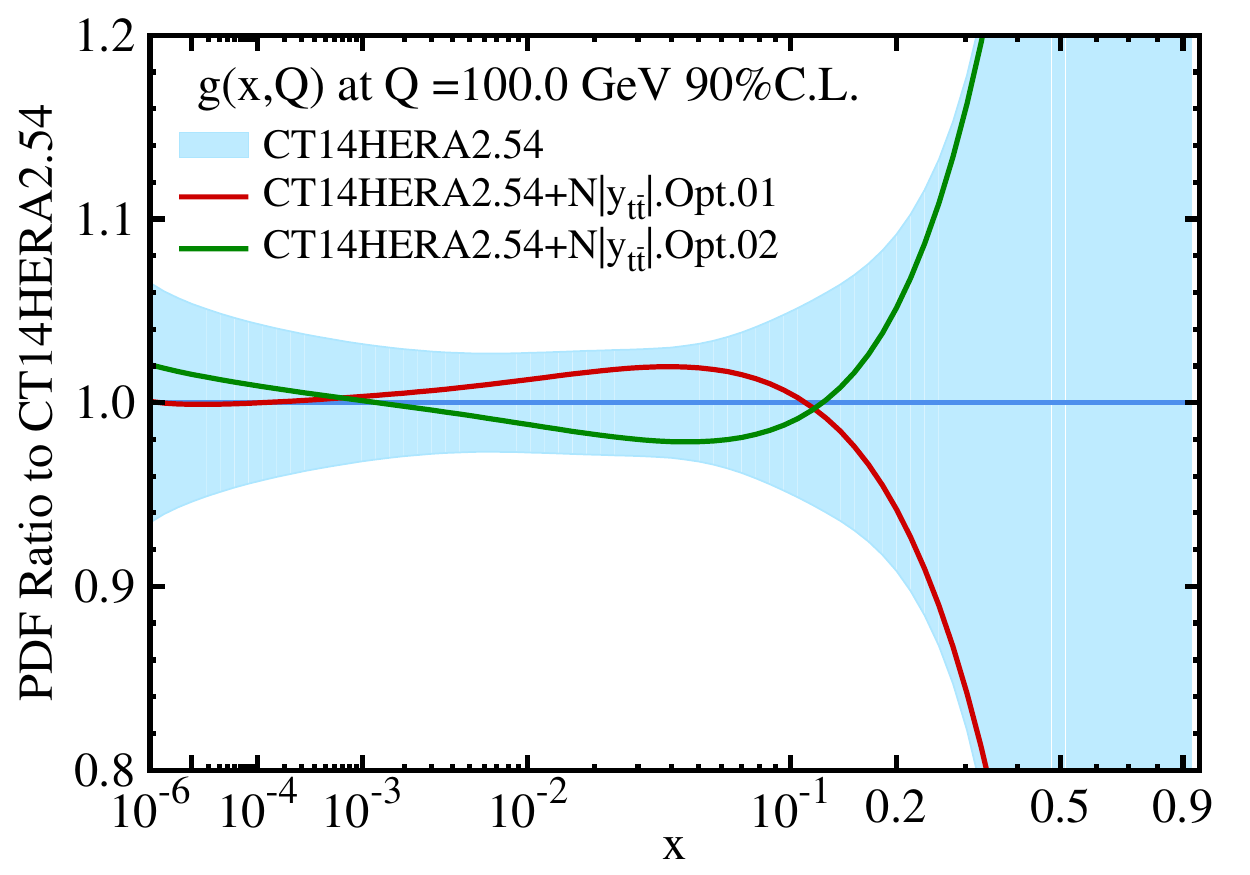}
	\includegraphics[width=0.49\textwidth]{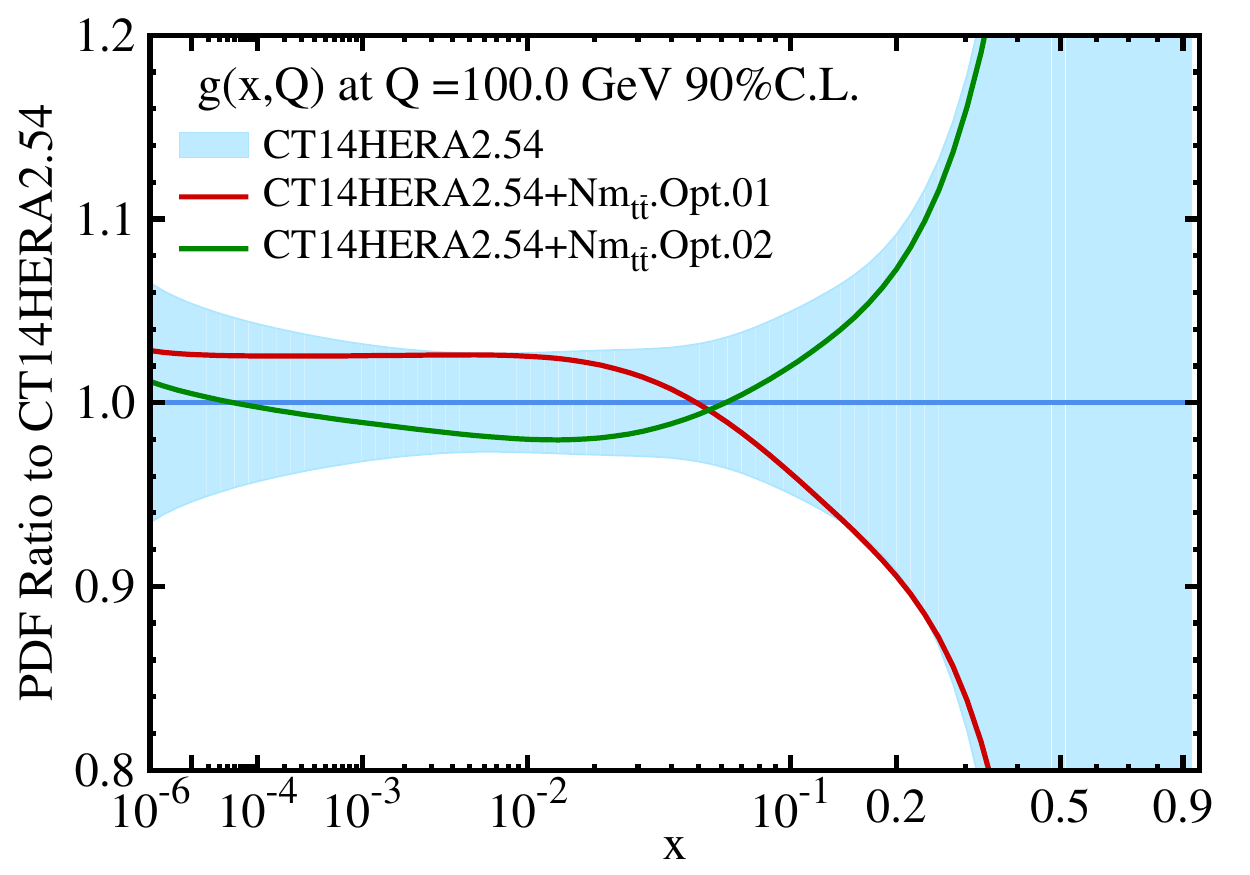}
	\includegraphics[width=0.49\textwidth]{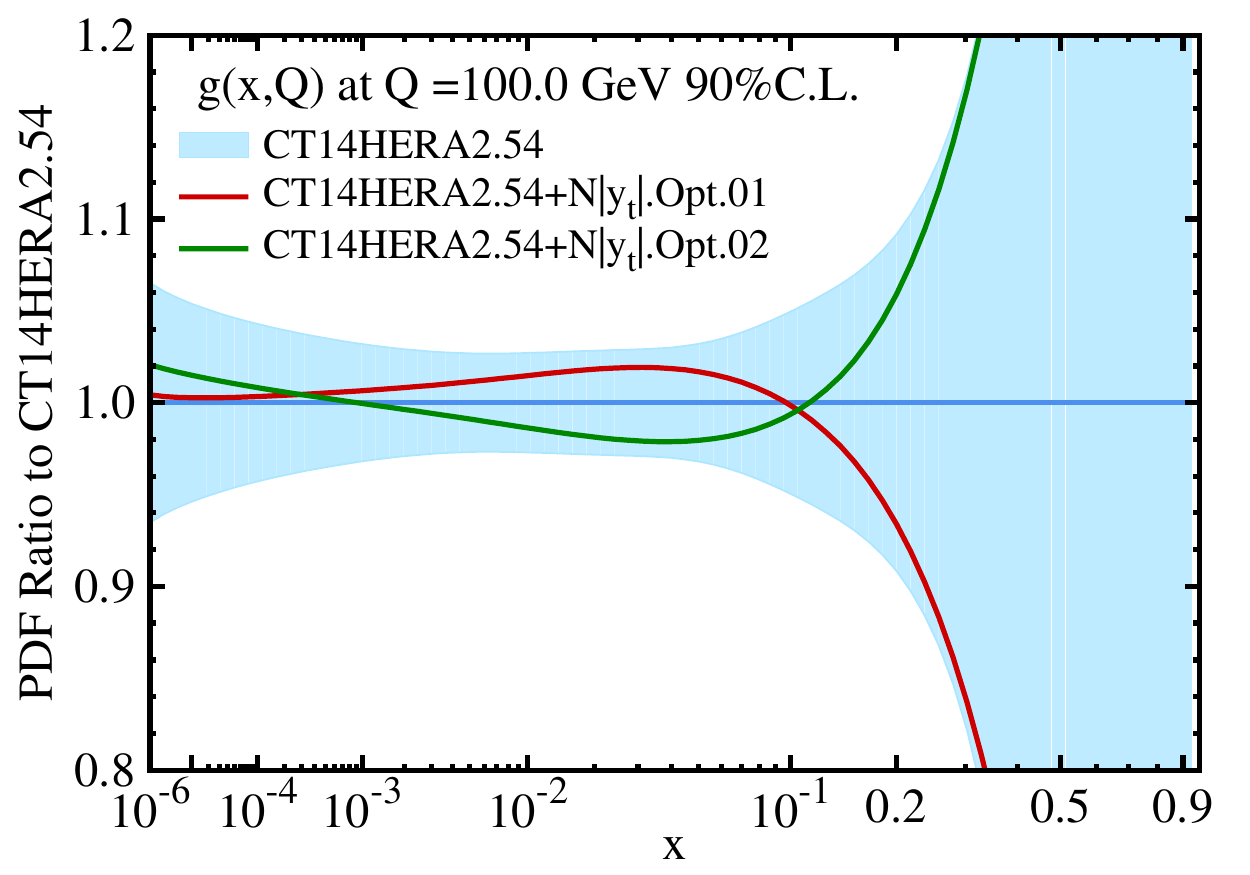}
	\includegraphics[width=0.49\textwidth]{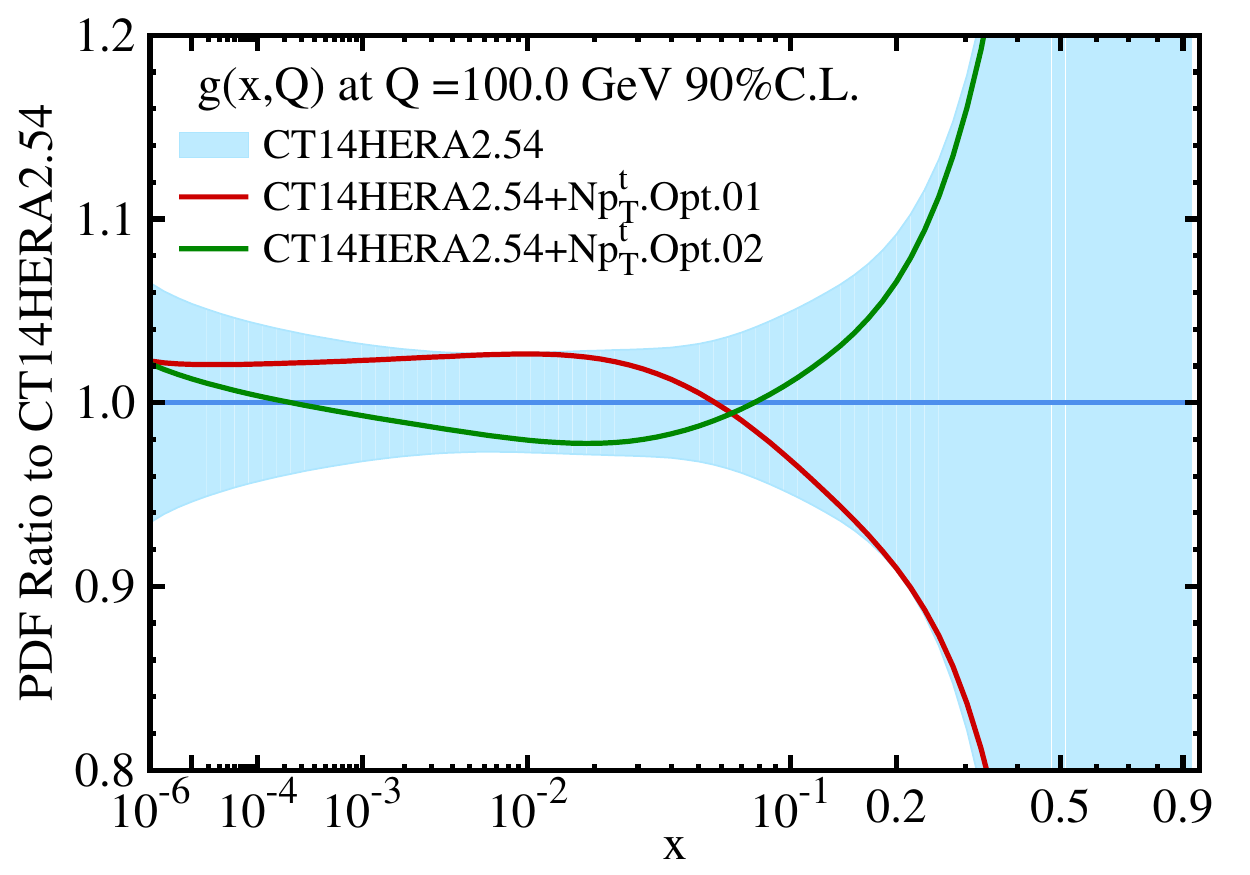}
	\caption{Ratio of the first pair of updated error PDFs and  original CT14HERA2.54 error PDFs to the CT14HERA2 central value of gluon PDF at $Q=100$ GeV.}\label{opt.norm}
\end{figure}

\begin{figure}[H]	
	\centering
	\includegraphics[width=0.49\textwidth]{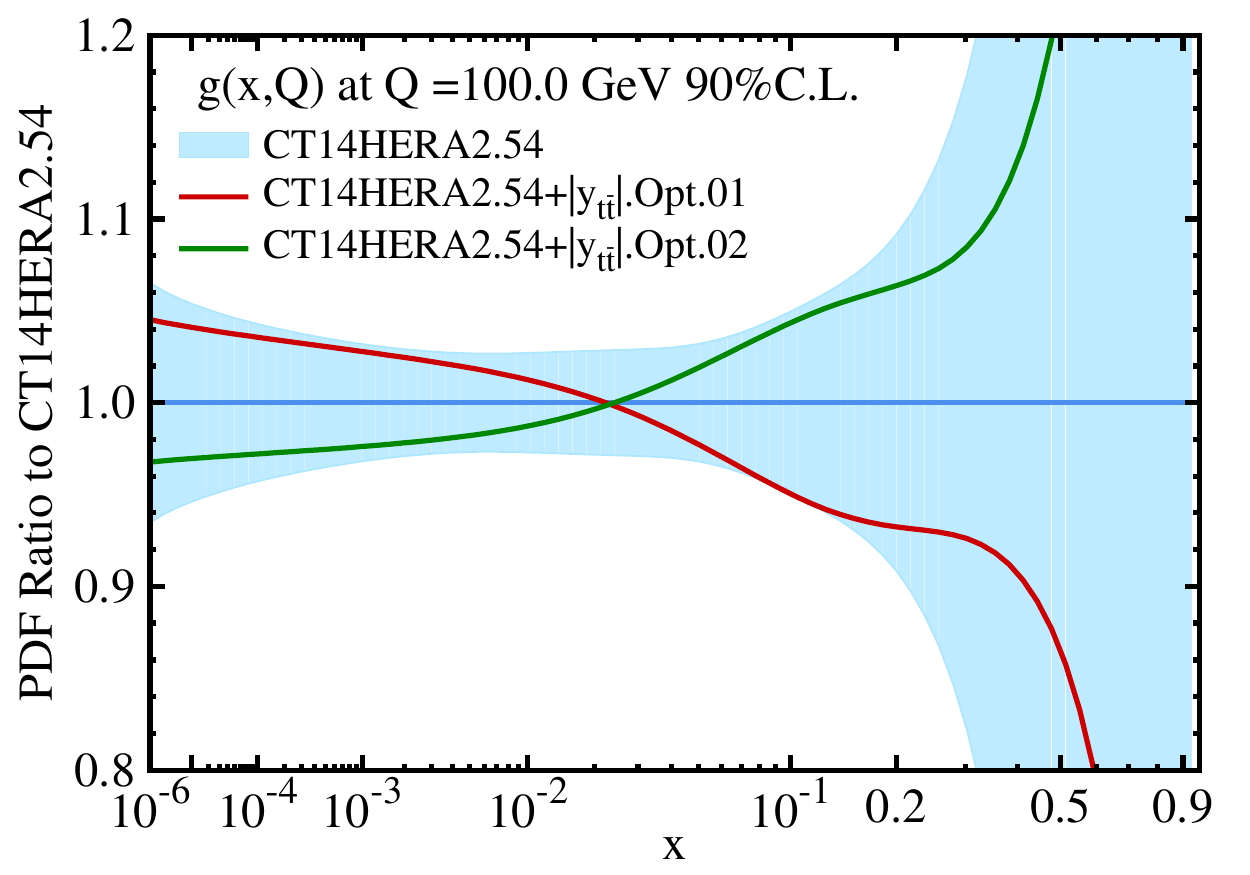}
	\includegraphics[width=0.49\textwidth]{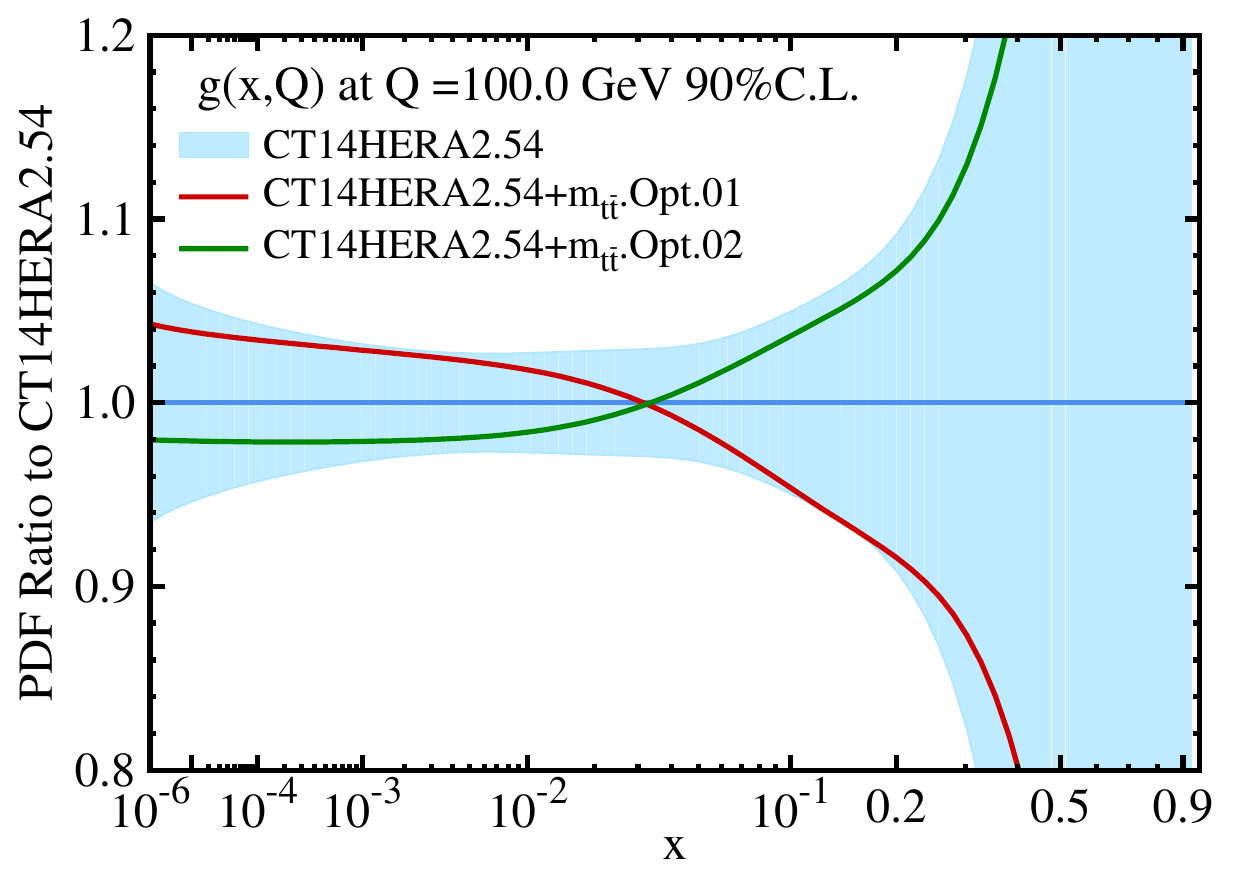}
	\includegraphics[width=0.49\textwidth]{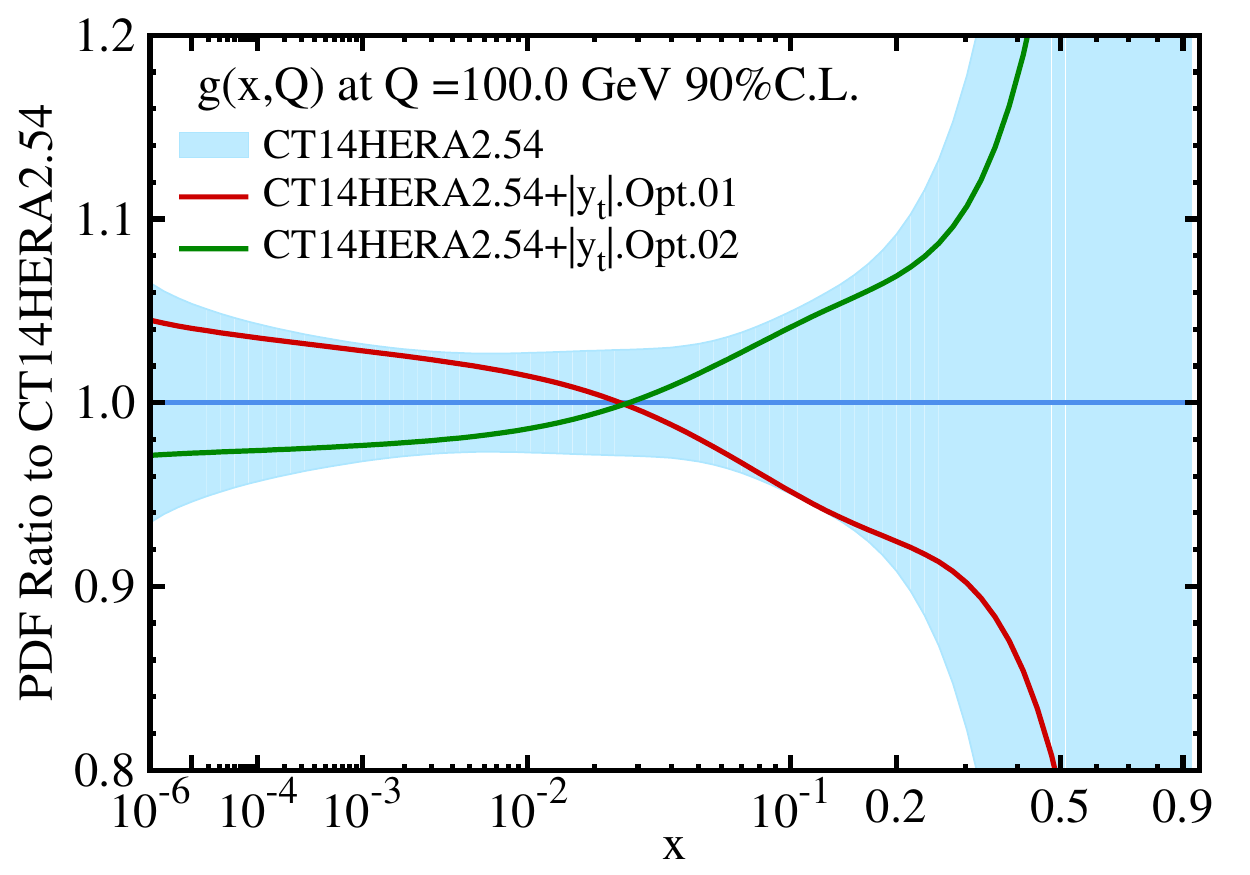}
	\includegraphics[width=0.49\textwidth]{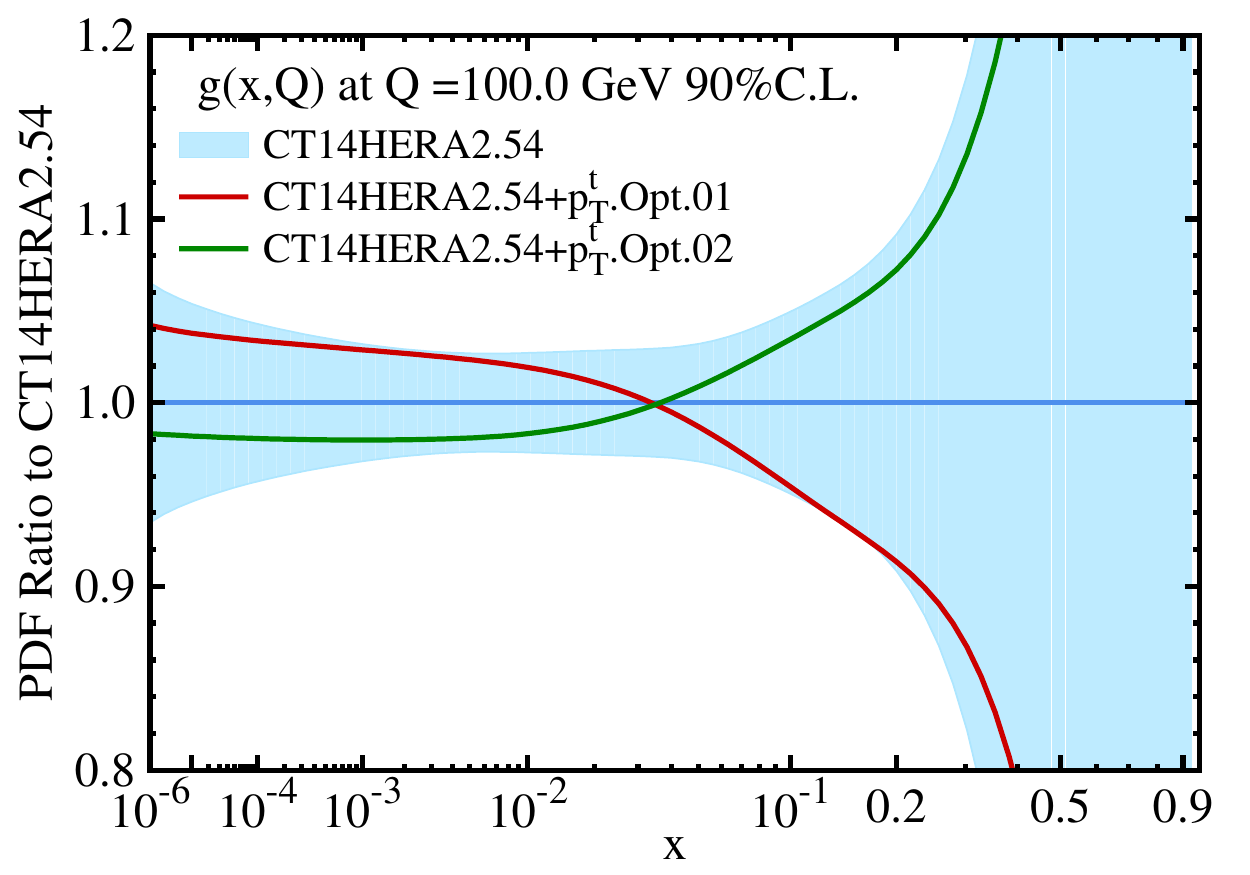}
	\caption{Ratio of the first pair of updated error PDFs and original CT14HERA2.54 error PDFs to the CT14HERA2 central value of gluon PDF at $Q=100$ GeV.}\label{opt.abs}
\end{figure}

In this section we apply the \texttt{ePump} optimization method in order to further explore the impact of the ATLAS and CMS 8 TeV  $t\bar t$ differential cross section data to the gluon PDF 
uncertainty. The \texttt{ePump} optimization method is similar to the data set 
diagonalization method~\cite{Pumplin:2009nm}.
For the optimization, we use CT14HERA2.54 error PDFs that have 27 error set pairs,  the absolute and normalized 8 TeV $t\bar t$ differential cross section 
from  NNLO $t\bar t$ fastNLO tables~\cite{Czakon:2016dgf,Czakon:2017dip}.
The optimized error PDFs are ordered by the size of their eigenvalues, and  one can determine how many error PDFs are necessary to obtain the  dependence of the observables on the PDFs.
The sensitivity of each $t\bar t$ data  to the gluon PDF at relevant  x-range    can be illustrated by comparing the pair of gluon error PDFs (two for each eigenvector) with the original CT14HERA2 gluon error PDFs, relative to the CT14HERA2 best fit values.
Therefore in Figs.~\ref{opt.norm} and \ref{opt.abs}, we have plotted ratios of the first
 pair of gluon error  PDFs (red and green lines) and the original CT14HERA2.54 gluon error PDFs (blue band) at $Q = 100$ GeV, to the CT14HERA2 best fit value of the gluon PDF.
Also, in Table \ref{tab:Opt-1-4}  we provide the maximal amount of gluon error bands covered by the first and second eigenvectors pairs for four $t\bar t$ differential distributions. 
From this table, we see that the first and second  eigenvectors pairs of gluon error  PDFs are almost completely cover the CT14HERA2.54 gluon PDF error bands for the whole x range x-range, indicating the dependence of the  $t\bar t$ differential cross section on the gluon PDF. Since the other eigenvector pairs, after ePump-Optimization, have negligible effect, the total error band of the gluon-PDF, in the relevant $x$-region, can be approximated by taking the quadrature sum of the error bands from the first two leading eigenvector sets. 
 And the contribution from the remaining 23 eigenvectors pairs are almost identically zero.
The first eigenvector pair  gives  the largest contribution  to the PDF uncertainty  for each  $t\bar t$ distributions, 
especially, it is more than $90\%$ for absolute  $m_{t\bar t}$ and $p^t_T$  as well as normalized $|y_t|$ distributions. 
As we see  PDF uncertainty for each $t\bar t$ distribution depends mostly on the first and second eigenvector pairs, we may use only these four eigenvector PDFs to study the PDF-induced uncertainty related to the $t\bar t$ production, instead of using the full 54 error sets of CT14HERA2+$t\bar t$ PDFs.

\begin{table}[H] \centering
	\caption{The eigenvalues of the first and second eigenvectors pairs}\label{tab:Opt-1-4} \vspace{0.4cm}
	\setlength{\abovecaptionskip}{20pt}
	\begin{spacing}{1.5}
    \begin{tabular}{|c|c|c|c|c|c|c|} \hline
distributions &  first eig.vec. & normalized & absolute & second eig.vec. & normalized & absolute \\ \hline
$y_{t\bar t}$ & \multirow{4}{*}{(01,02)} & 82.4 \% & 71.9 \%  & \multirow{4}{*}{(03,04)} & 17.4 \% & 25.6 \% \\ \cline{1-1} \cline{3-4} \cline{6-7} 
$m_{t\bar t}$ &                          & 84.1 \% & 92.5 \%  &                          & 15.8 \% & 7.3  \% \\ \cline{1-1} \cline{3-4} \cline{6-7} 
$y_t$       &                            & 90.6 \% & 83.5 \%  &                          & 9.2  \% & 16.0 \% \\ \cline{1-1} \cline{3-4} \cline{6-7} 
$p^t_T$  &                               & 86.1 \% & 93.2 \%  &                          & 13.7 \% & 6.7  \%  \\ 
          \hline
        \end{tabular}
	\end{spacing}
\end{table}

\subsection{Comparison between 8 TeV $t\bar t$ data and theory from original and new CT14HERA2 }

\begin{figure}[H]	
\centering
\includegraphics[width=0.49\textwidth]{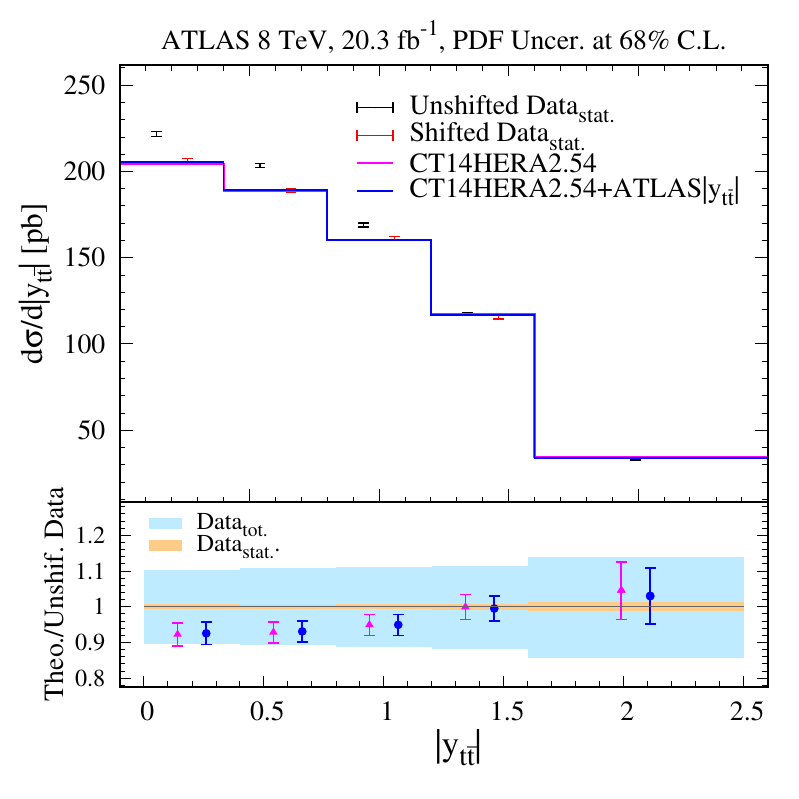}
\includegraphics[width=0.49\textwidth]{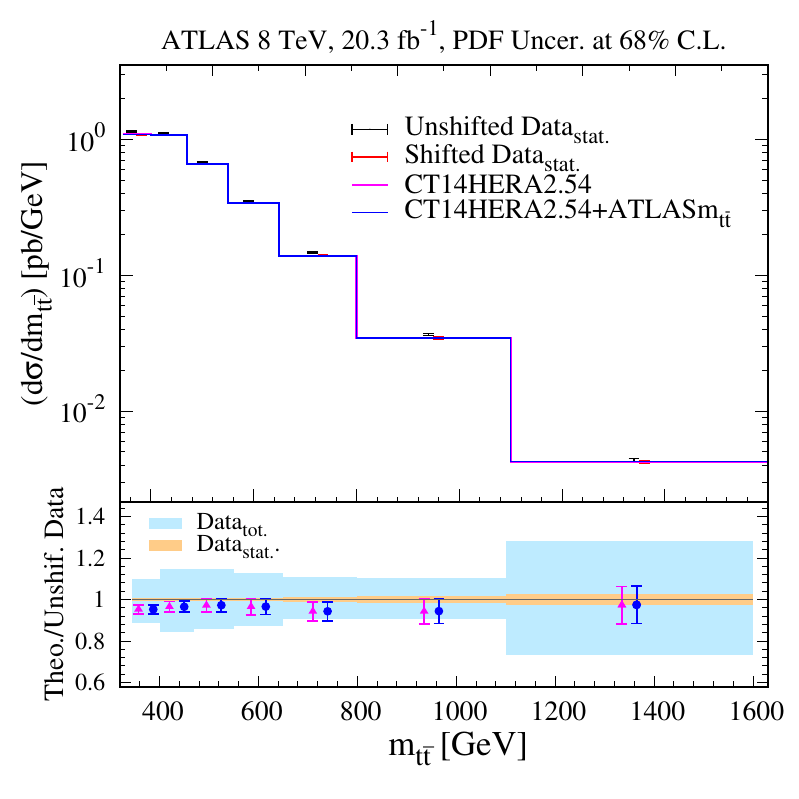}
\includegraphics[width=0.49\textwidth]{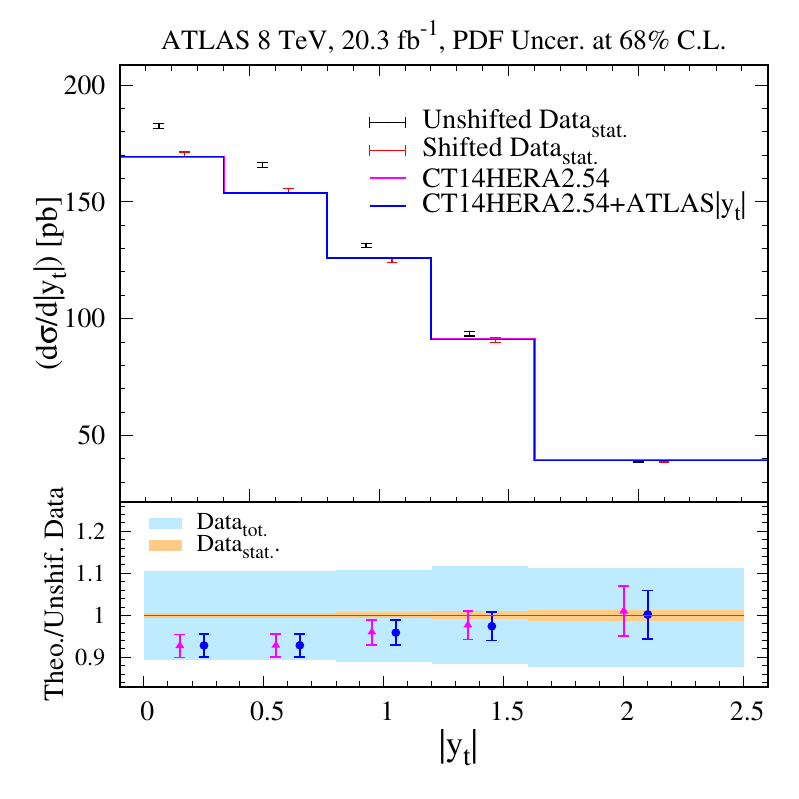}
\includegraphics[width=0.49\textwidth]{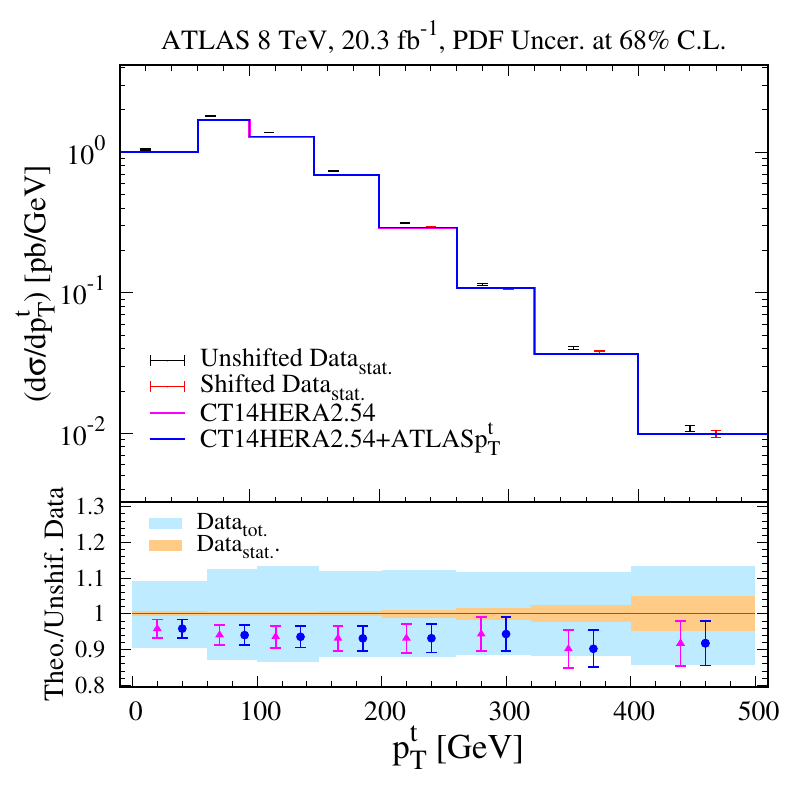}
\caption{Comparison of differential cross sections $d\sigma/d|y_{t\bar t}|$, $d\sigma/dm_{t\bar t}$,  $d\sigma/d|y_t|$, $d\sigma/dp^t_T$ from  CT14HERA2.54 PDFs and from {\tt \texttt{ePump}} updated 	
	CT14HERA2.54+ATLAS$|y_{t\bar t}|$, 
	CT14HERA2.54+ATLAS$m_{t\bar t}$, 
	CT14HERA2.54+ATLAS$|y_t|$, 
	CT14HERA2.54+ATLAS$p^t_T$ PDFs and differential ATLAS 8 TeV $t\bar t$ production cross sections data as a function of the $|y_{t\bar t}|$,  $m_{t\bar t}$, $|y_t|$ and $p^t_{T}$. 
}\label{ATL-abs-data-theory}
\end{figure}

\begin{figure}[H]	
\centering
\includegraphics[width=0.49\textwidth]{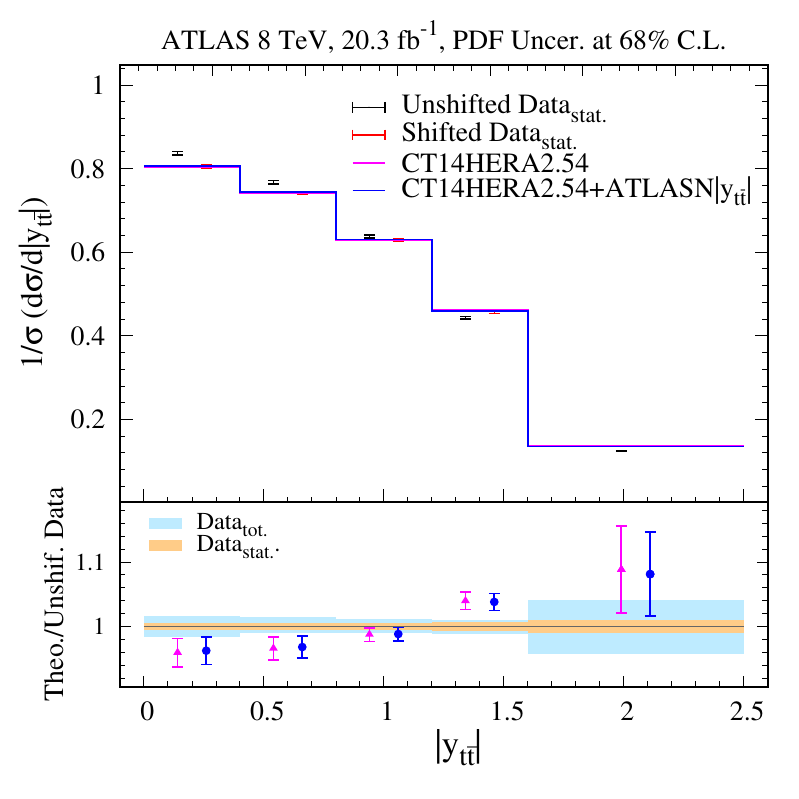}
\includegraphics[width=0.49\textwidth]{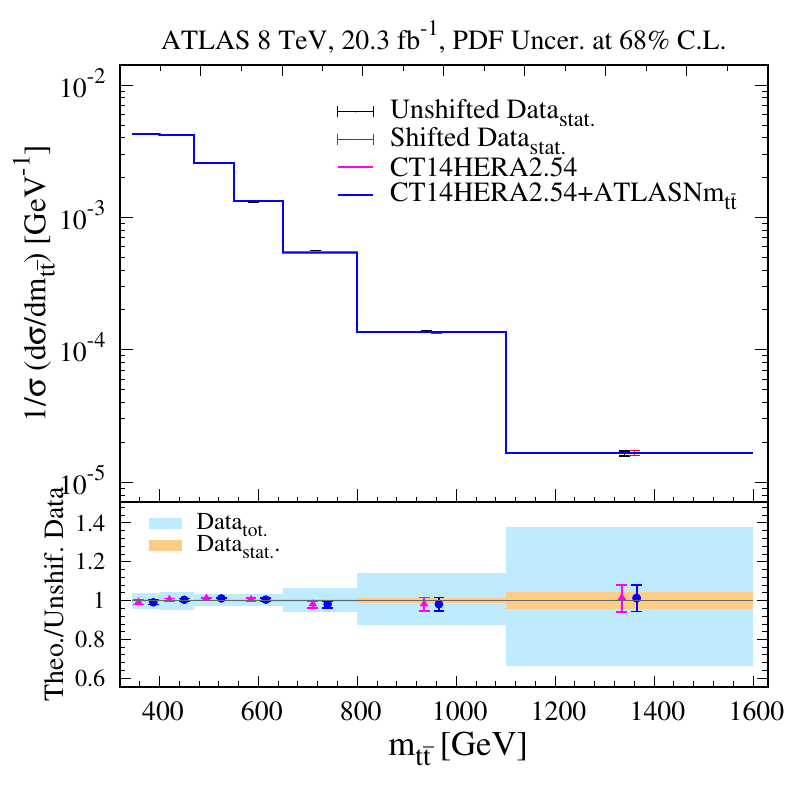}
\includegraphics[width=0.49\textwidth]{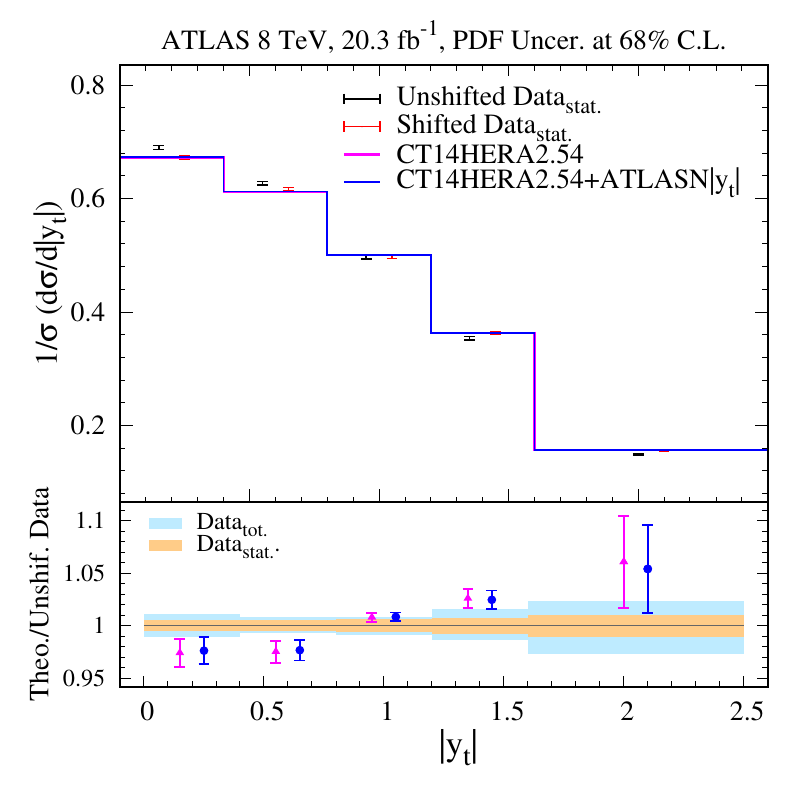}
\includegraphics[width=0.49\textwidth]{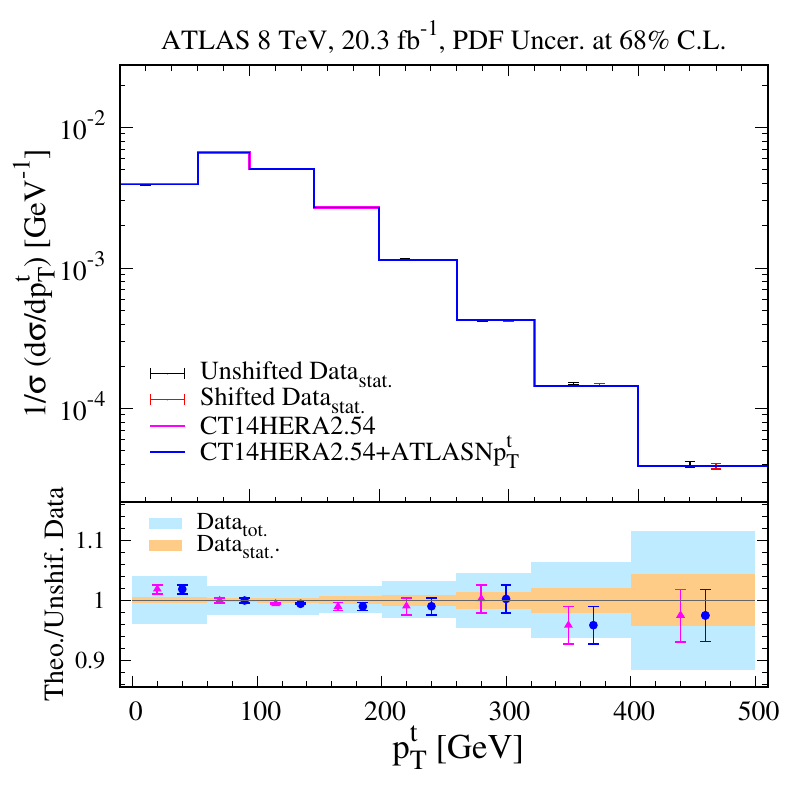}
\caption{Comparison normalized differential cross sections $1/\sigma \;  d\sigma/d|y_{t\bar t}|$, $1/\sigma \; d\sigma/dm_{t\bar t}$, $1/\sigma \;  d\sigma/dp^t_T$, $1/\sigma \; d\sigma/d|y_t|$ from  CT14HERA2.54 PDFs and from {\tt \texttt{ePump}} updated  
	CT14HERA2.54+ATLASN$|y_{t\bar t}|$, 
	CT14HERA2.54+ATLASN$m_{t\bar t}$, 
	CT14HERA2.54+ATLASN$|y_t|$, 
	CT14HERA2.54+ATLASN$p^t_T$ PDFs and 
normalized differential ATLAS 8 TeV $t\bar t$
production cross sections data as a function of the  $|y_{t\bar t}|$,  $m_{t\bar t}$, $|y_t|$ and $p^t_{T}$. 
}\label{ATL-norm-data-theory}
\end{figure}

\begin{figure}[H]	
	\centering		
	\includegraphics[width=0.49\textwidth]{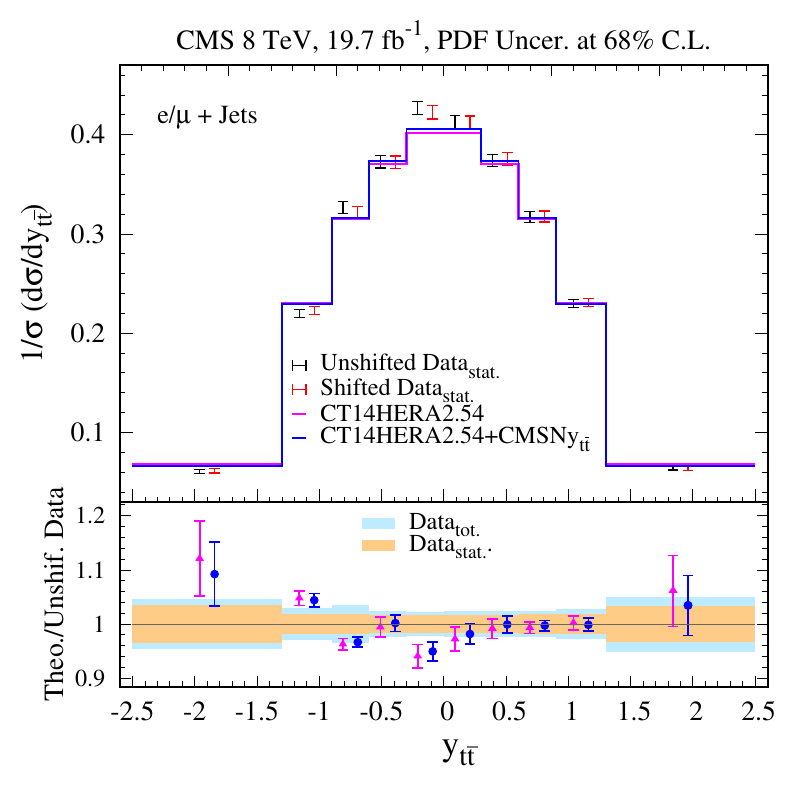}
	\includegraphics[width=0.49\textwidth]{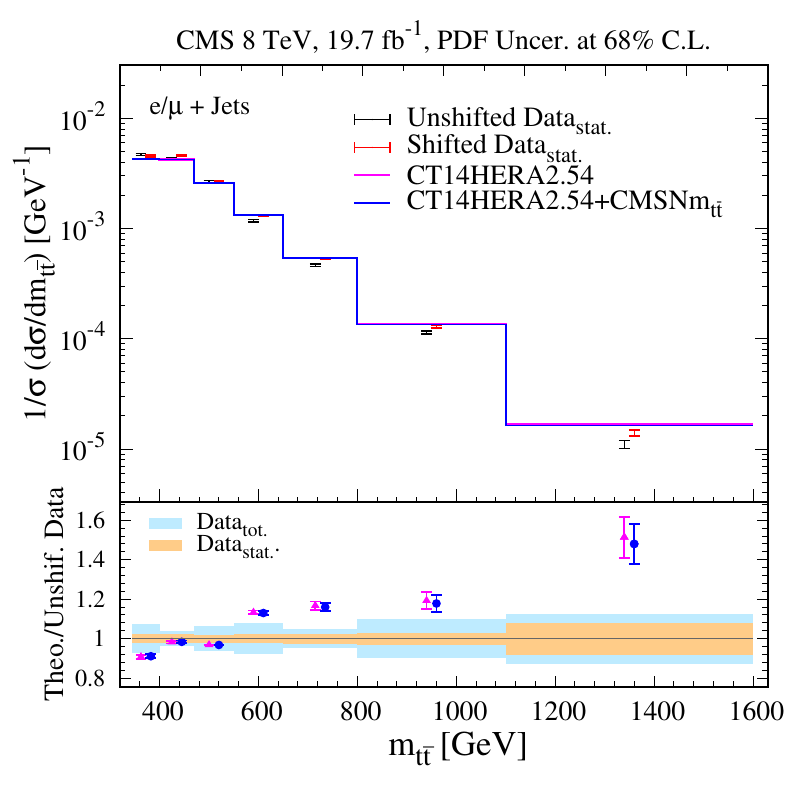}
	\includegraphics[width=0.49\textwidth]{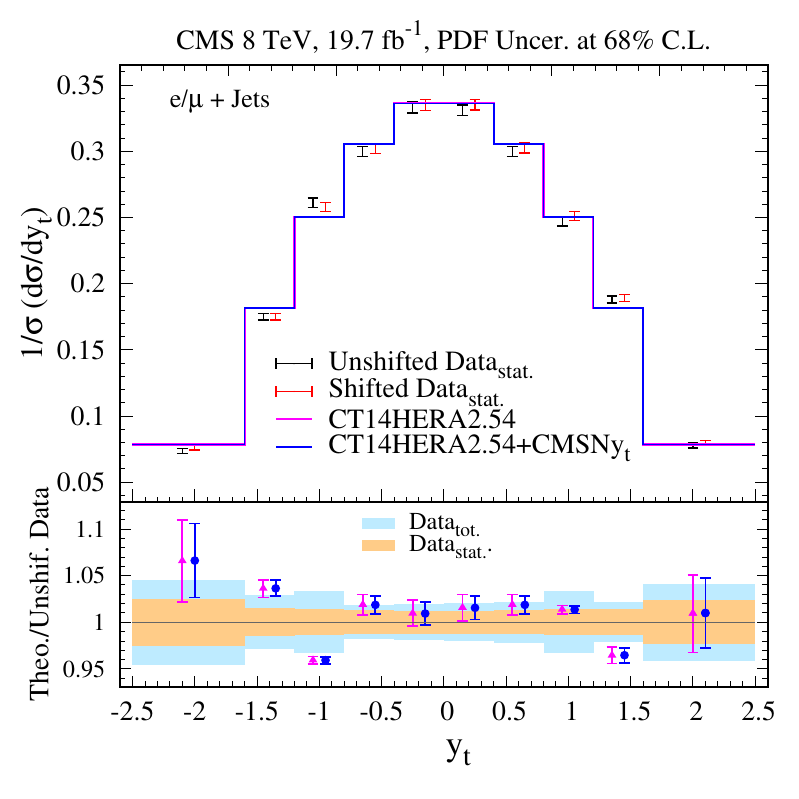}
	\includegraphics[width=0.49\textwidth]{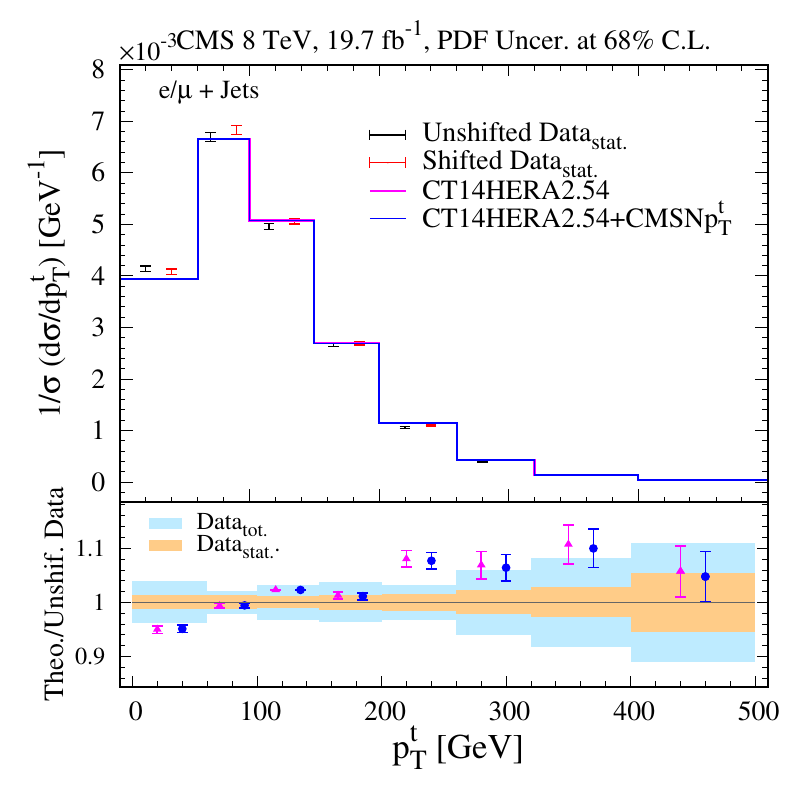}
	\vspace{0.4cm}
	\caption{Comparison normalized differential cross sections $1/\sigma \;  d\sigma/d|y_{t\bar t}|$, $1/\sigma \; d\sigma/dm_{t\bar t}$, $1/\sigma \;  d\sigma/dp^t_T$, $1/\sigma \; d\sigma/d|y_t|$ from  CT14HERA2.54 PDFs and from {\tt \texttt{ePump}} updated  
		CT14HERA2.54+CMSN$y_{t\bar t}$, 
		CT14HERA2.54+CMSN$m_{t\bar t}$, 
		CT14HERA2.54+CMSN$y_t$, 
		CT14HERA2.54+CMSN$p^t_T$ PDFs and  CMS 8 TeV normalized differential  $t\bar t$ production cross sections data as a function of the  $y_{t\bar t}$,  $m_{t\bar t}$, $y_t$ and $p^t_{T}$. 
	} \label{CMS-N-data-theory}
\end{figure}

In this subsection we show the theory predictions after considering the $t\bar t$ data  and compare with the measurements.
The comparisons between the theory predictions from before and after updated CT14HERA2.54 PDFs and the ATLAS 8 TeV absolute and normalized  differential $t \bar t$ data as well as CMS 8 TeV normalized  differential $t \bar t$ data are presented in Figs.~\ref{ATL-abs-data-theory}, \ref{ATL-norm-data-theory} and \ref{CMS-N-data-theory}.
In those figures, the magenta solid lines correspond to the theoretical predictions from CT14HERA2.54, the blue solid lines are the theoretical predictions from 
updated CT14HERA2.54 PDFs.	
The black and red error bars on each data and shifted data point (upper part of the figure) include only statistical error. 
Shifted data $D_k^{sh}$ is defined as, 
$$
D_k^{sh}\equiv D_k -
\sum_{\alpha=1}^{N_{\lambda}} \lambda_{\alpha}(a_0) \beta_{k\alpha}
$$
where $D_k$ is an $k$-th data point (value); $\lambda_\alpha$ is know as nuisance parameter;$\sum_{\alpha=1}^{N_{\lambda}} \beta_{k\alpha}$ are the correlated systematic errors for the k-th data point.
The blue bands in ratio plots indicate the total uncertainty, that are  the quadratic sum of statistical and systematic uncorrelated uncertainties, on the data in each bin. The yellow bands in ratio plots indicate the
statistical uncorrelated uncertainties on the data in each bin.
The error bars on the theoretical predictions show the 68\% C.L..
We see that there is an overall shift for all the raw data points. This means that the correlated systematic errors, weighted by their corresponding nuisance parameters, play an important role in the fitting.
We find that there is little  impovement in agreement with the measurements after calculating theoretical predictions evaluated with the new PDFs that obtained adding the ATLAS and CMS 8 TeV $t\bar t$ production differential cross sections data.

\section{Consistency between $t\bar t$ data and data in CT14HERA2}\label{sec:tension}

\begin{figure}[H]
	\begin{center}
		\includegraphics[width=0.49\textwidth]{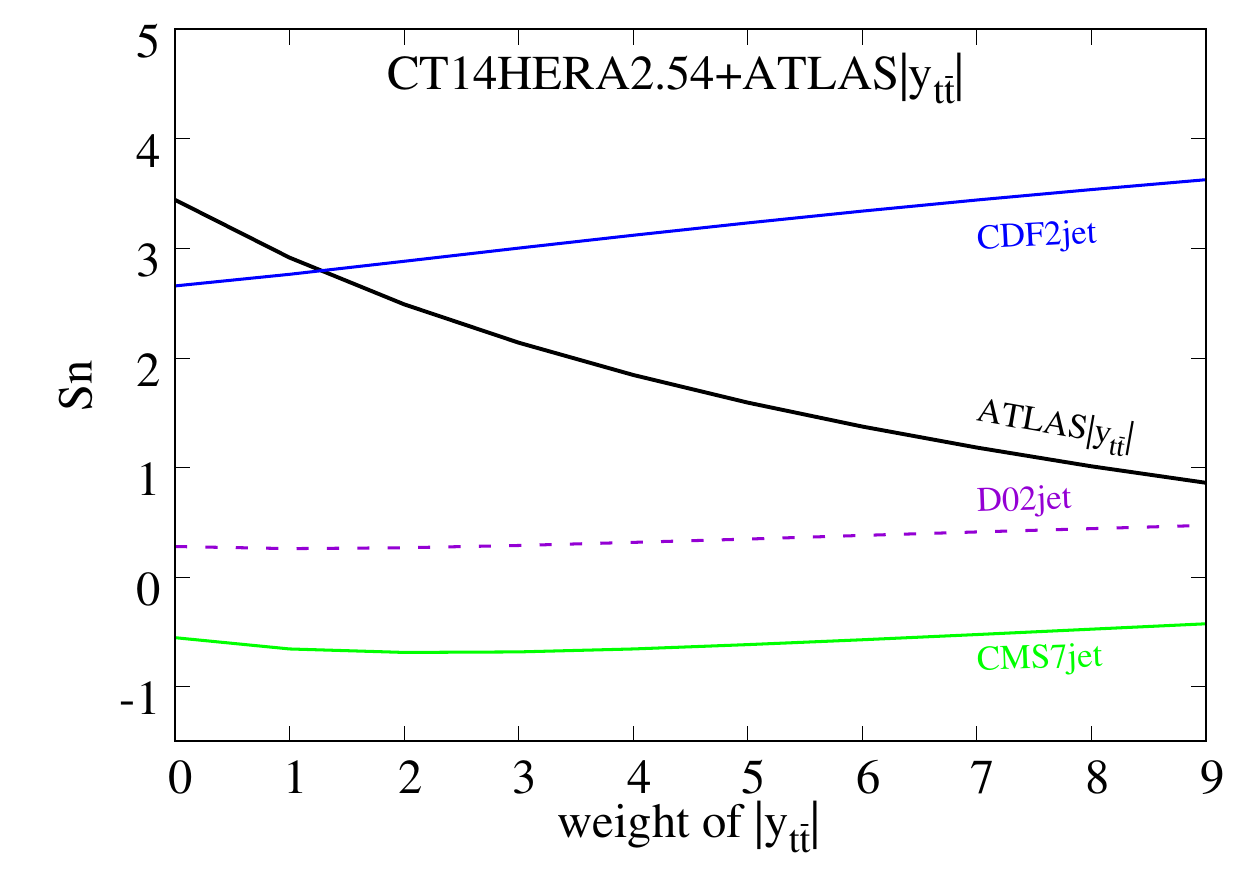}	
		\includegraphics[width=0.49\textwidth]{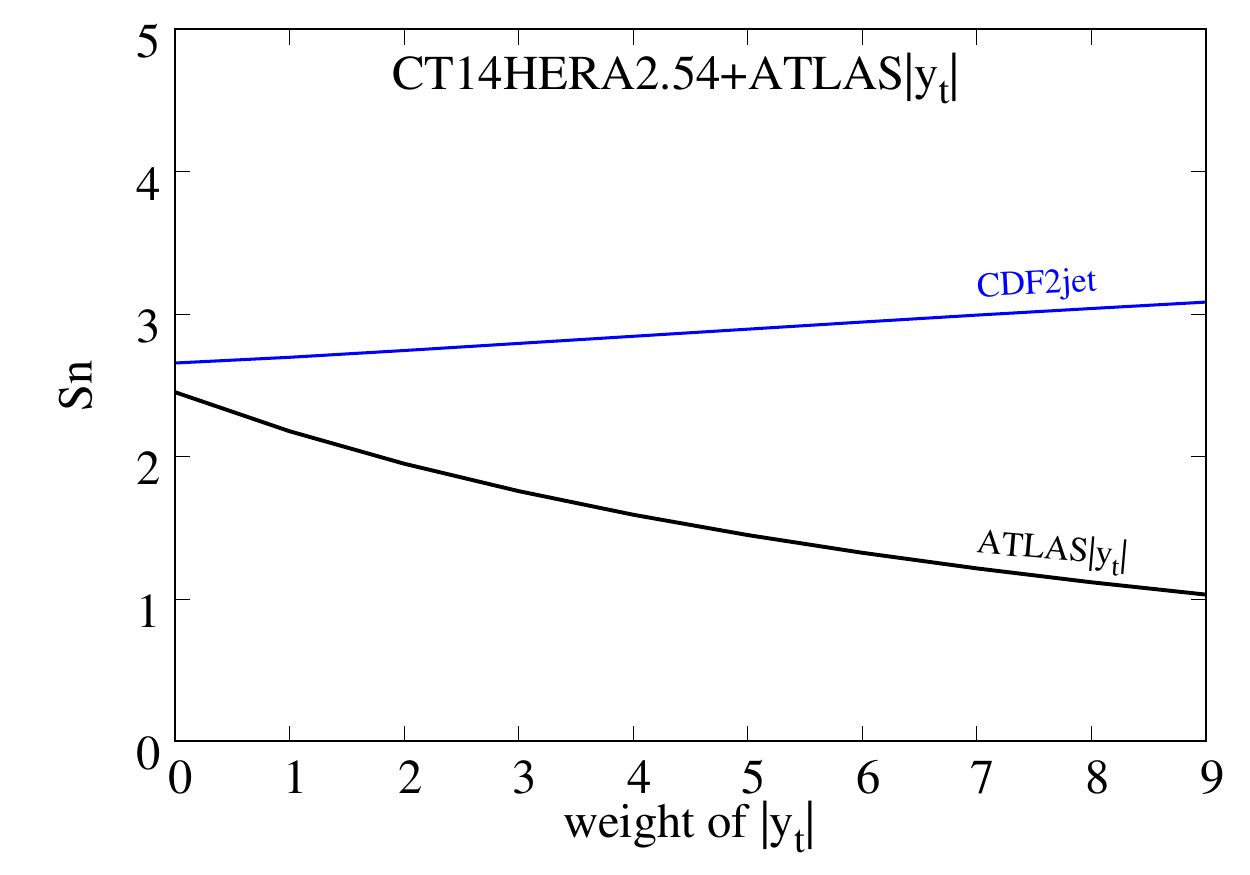}	
		\includegraphics[width=0.49\textwidth]{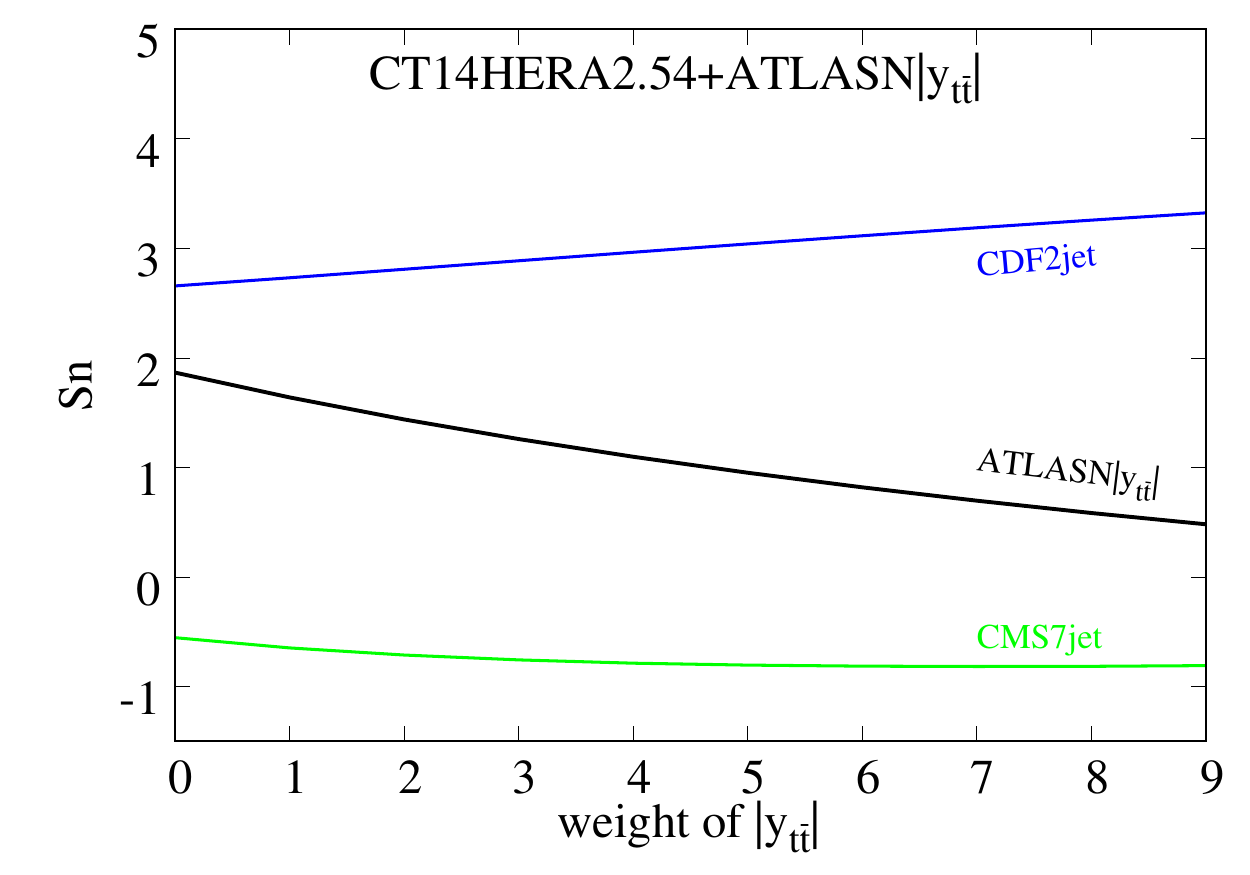}	
		\includegraphics[width=0.49\textwidth]{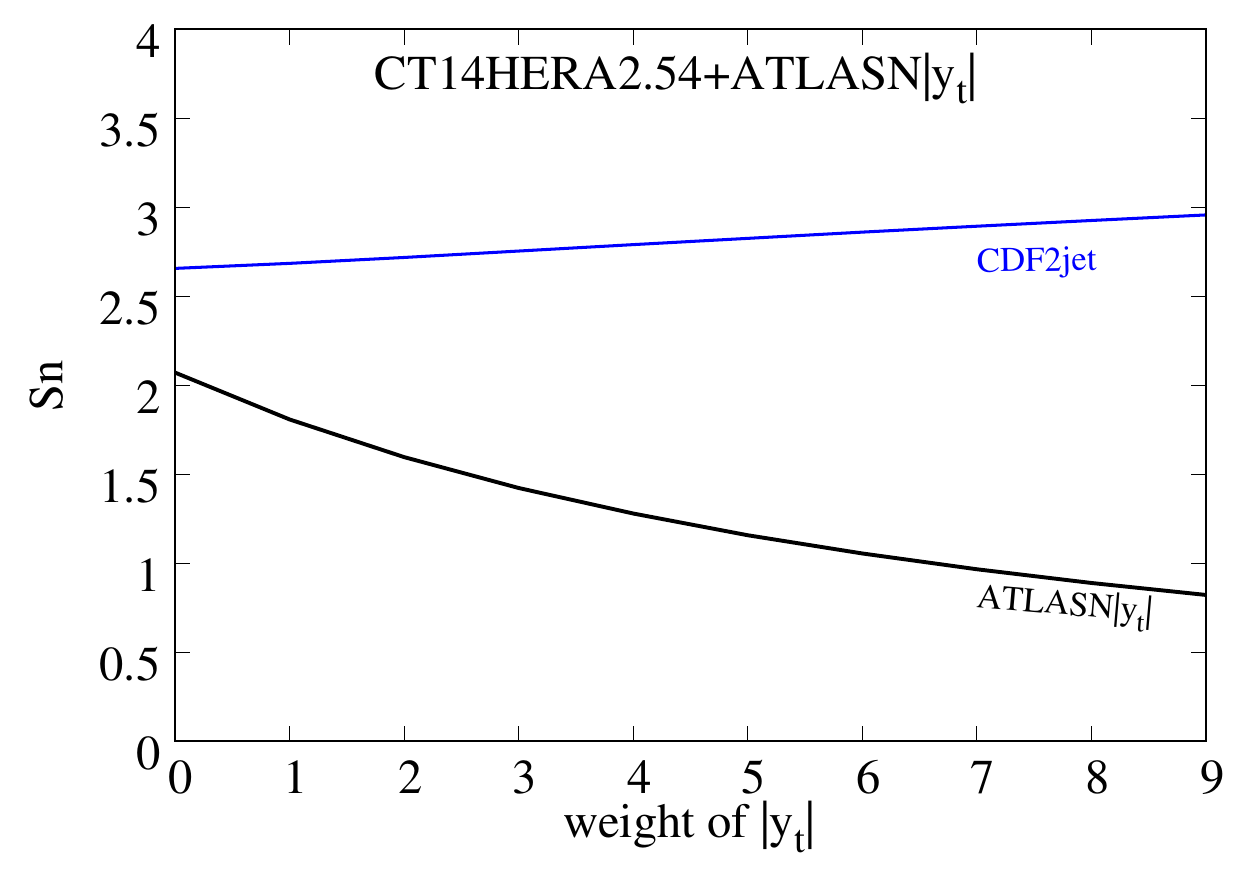}
	\end{center}
	\caption{The equivalent Gaussian variable Spartyness  $S_n$  of some data in {\tt \texttt{ePump}} updated CT14HERA2.54 versus weight of the ATLAS absolute and normalized data for the absolute value of the rapidity $|y_t|$ of the top quark distribution  and top-quark pair rapidity $|y_{t\bar t}|$ distribution  at 8 TeV.  
	}\label{tension-1}
\end{figure}

\begin{figure}[H]
	\begin{center}
		\includegraphics[width=0.49\textwidth]{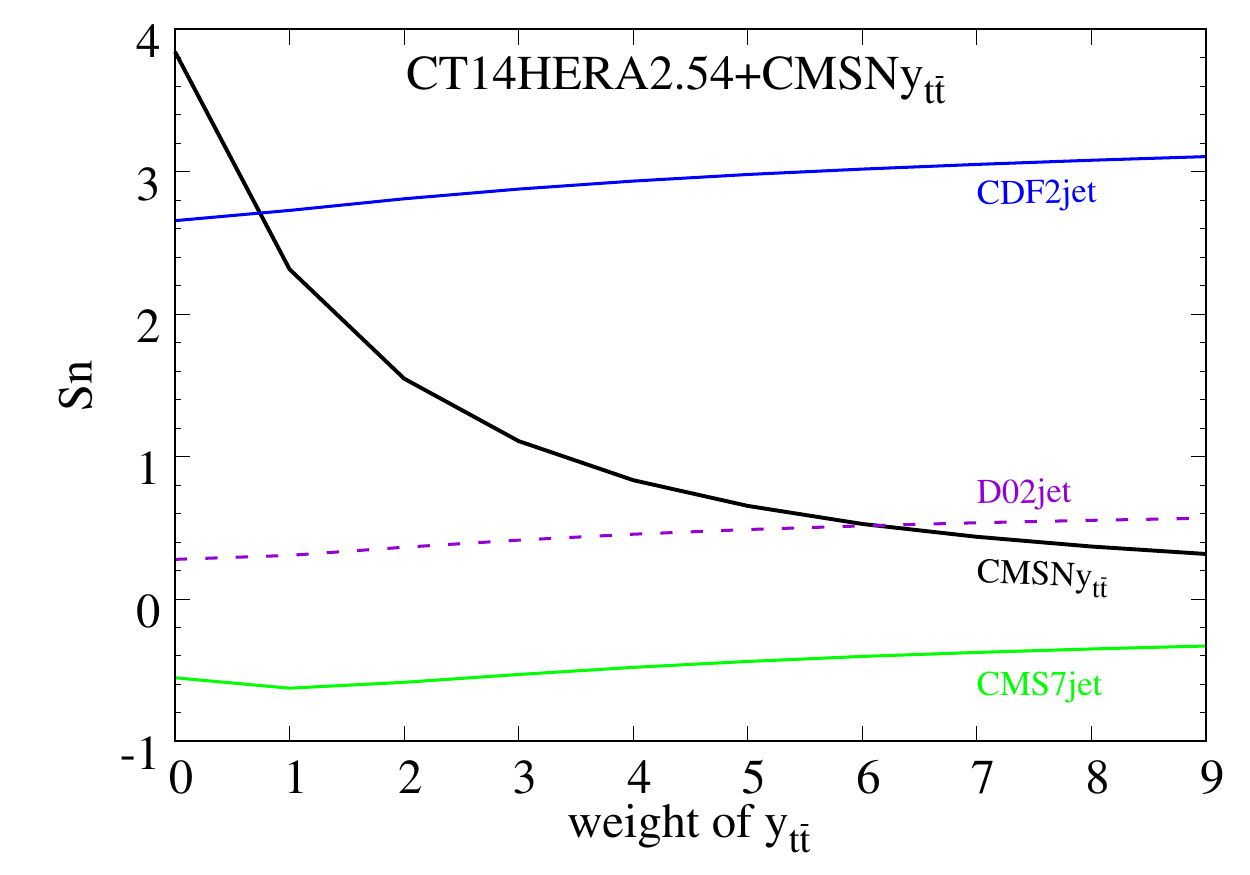}
		\includegraphics[width=0.49\textwidth]{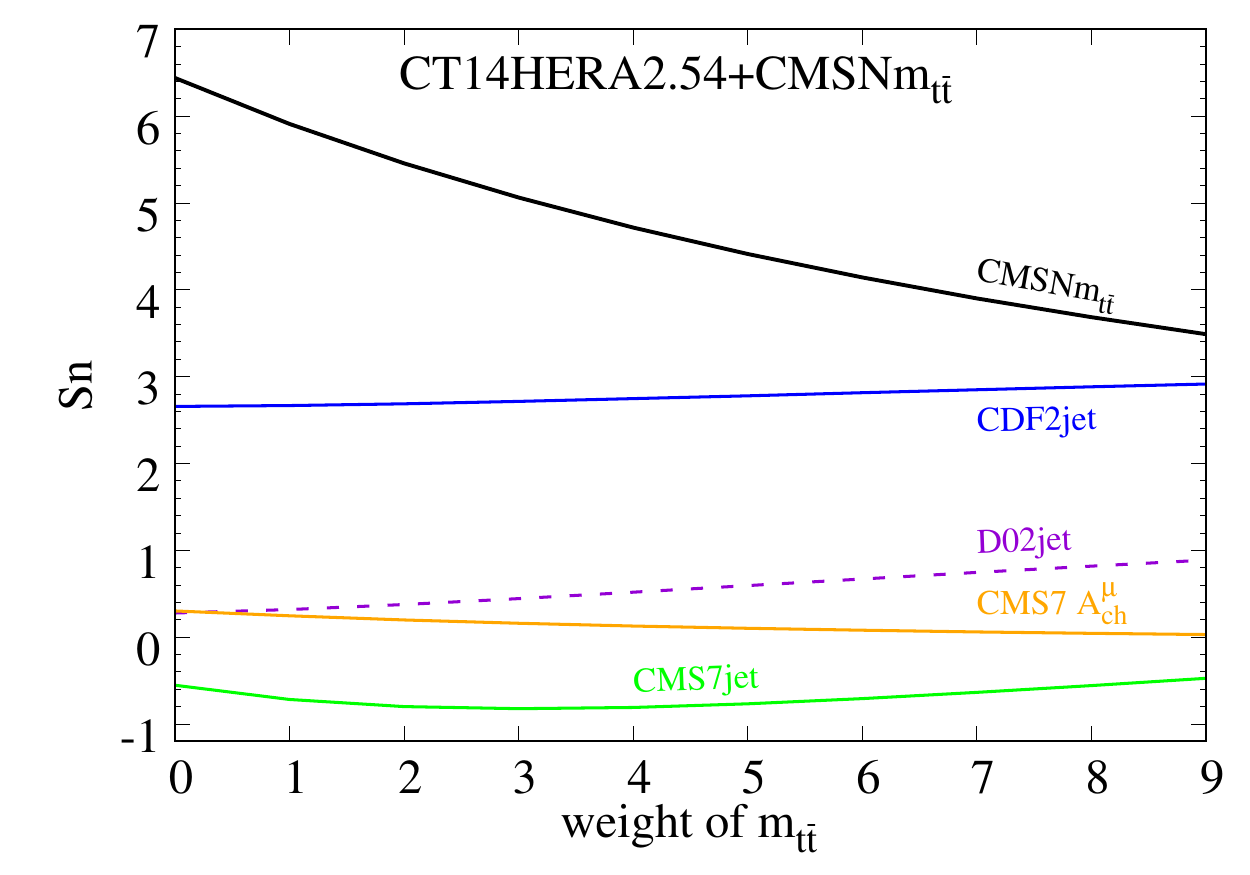}
		\includegraphics[width=0.49\textwidth]{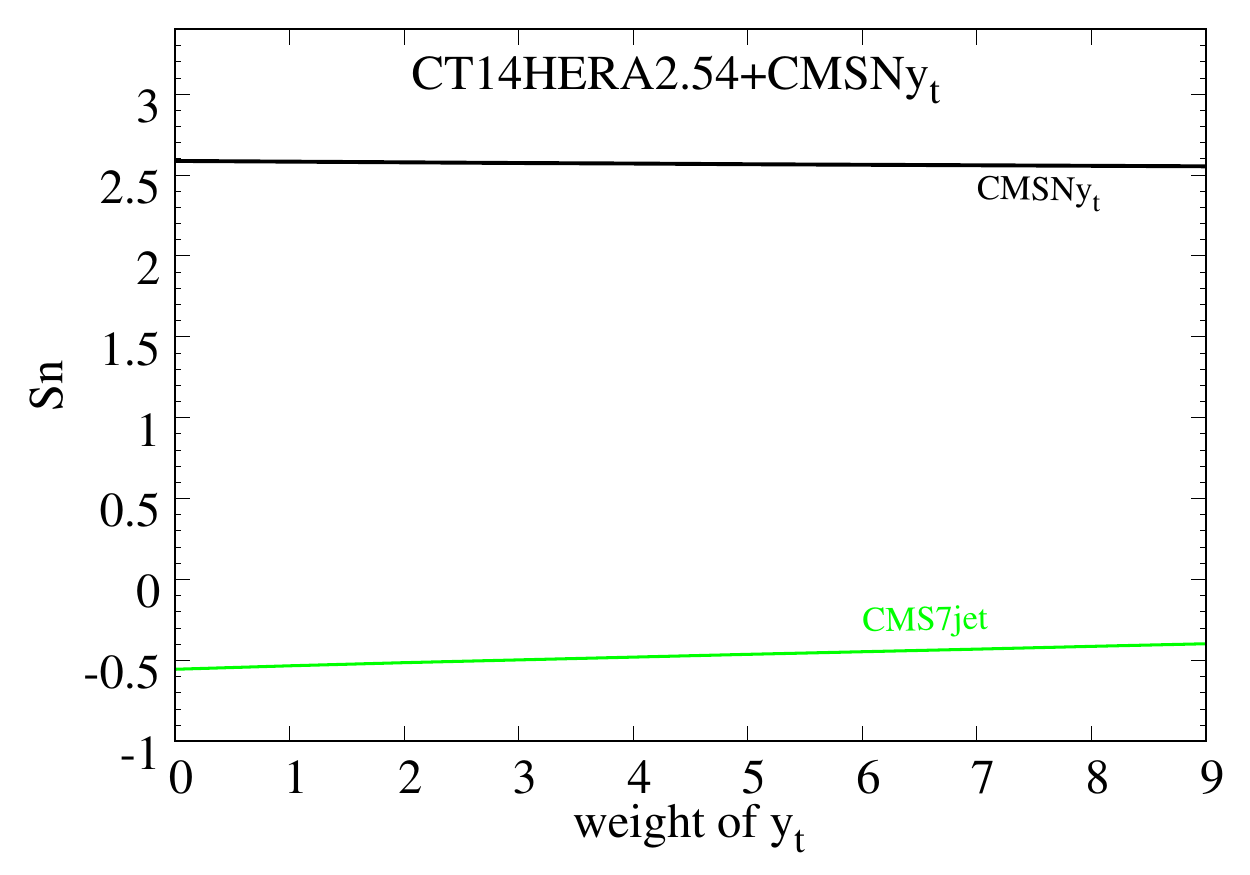}
		\includegraphics[width=0.49\textwidth]{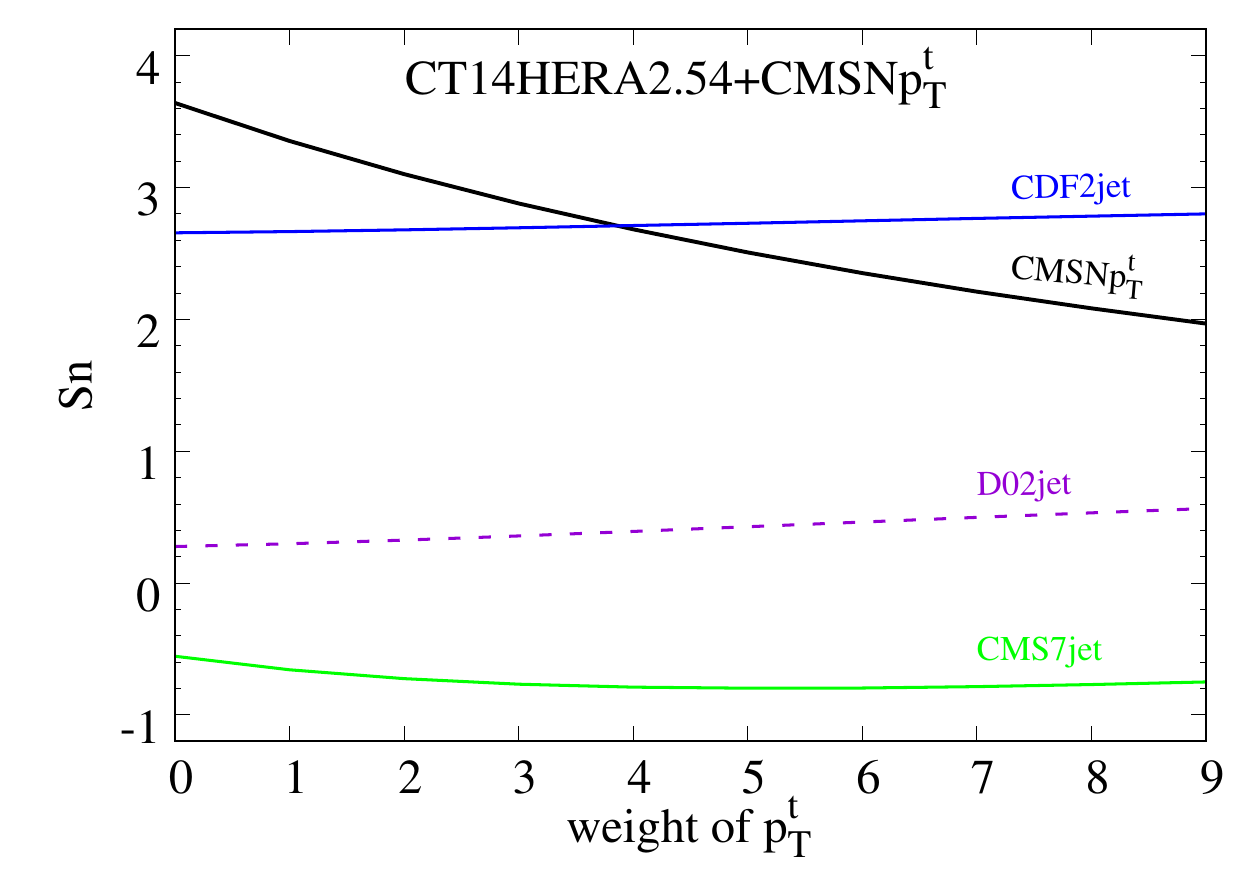}
	\end{center}
	\caption{The equivalent Gaussian variable Spartyness  $S_n$  of some data in {\tt \texttt{ePump}} updated CT14HERA2.54 versus weight of the CMS normalized  $1/\sigma \; d\sigma/dy_t$ and $1/\sigma \; d\sigma/dy_{t\bar t}$ data  at 8 TeV. }\label{tension-2}
\end{figure}

As we present in the last two sections, we observe the CT14HERA2 gluon PDF receive minor impact after including the 8 TeV ATLAS absolute and normalized and CMS normalized $t \bar{t}$ data. This can be a result of strong tension from the data included in the CT14HERA2 PDF.
In order to study for possible tensions between the 
single differential $t \bar{t}$ data from ATLAS and CMS
and the data sets included in the CT14HERA2 PDFs, we increase the weight of the $t\bar t$ data when we updating the CT14HERA2 PDFs using the {\tt \texttt{ePump}}.
We consider weight from zero to nine for the single differential 
$t\bar t$ data individually for testing the tension among the $t\bar t$ data and other data included in the CT14HERA2 PDF.
Weight zero case is just the CT14HERA2 fit without any change.
Weight one case corresponds to CT14HERA2 fit with $t\bar t$ data included individually. 
Weight larger than one is equivalent to having more $t\bar t$ data points with the same experimental uncertainties~\cite{Hou:2019gfw}.
Instead of $\chi^2$, we present the change of goodness-of-fit of each data by the variable $S_n$~\cite{Dulat:2013hea}, which can be treated as a rescale of $\chi^2$ based on the number of data points of the data. Values of $S_n$ between $-1$ and $1$ correspond to a good fit (at the $68\%$ C.L.);  large positive values of $S_n$($\gtrsim 2$) correspond to a poor fit; while large negative values ($\lesssim -2$) mean that it fit unusually well.
If we increase the weight of the ATLAS and CMS 8 TeV $t \bar t $ data in the fit, the $S_n$ of the $t \bar t $ data decreases with its reduced $\chi^2$, as it should be; when the weight of the $t \bar t $ data is becoming large, the $S_n$ of some particular data in CT14HERA2 may increase by noticeable amount.
If some of the data in CT14HERA2 have tension with the $t \bar t $ data,
the $S_n$ of those data will become larger when the weight of the $t \bar t $ data increases.
We find that most of the data in CT14HERA2 do not show significant tension with
the 8 TeV single differential ATLAS and CMS $t\bar t $ data. However, we observe that some data in CT14HERA2 do show some tension with the $t\bar t$ data. 
In Figs.~\ref{tension-1}-\ref{tension-2}, we show the change of $S_n$ for some data in CT14HERA2 as the weight of the $t\bar t$  data increases from 0 to 9. 
We observe that some of the data in CT14HERA2 has minor change in $S_n$ as the weight of the $t \bar t $ data increases. For example, we see that the $S_n$ of the CDF jet data~\cite{Aaltonen:2008eq}
and the D0 jet data~\cite{Abazov:2008ae} increases the most; while the $S_n$ of the CMS 7 TeV jet data~\cite{Chatrchyan:2012bja} reduce mildly when the weight of the ATLAS normalized $|y_{t\bar t}|$ data or CMS normalized $p^t_T$ increase.
	As a result, we did not observe strong tension on the ATLAS and CMS single differential $t \bar{t}$ data from the data included in the CT14HERA2 PDF. But we do observe that, the jet data is relatively more sensitive to the includsion of the $t \bar{t}$ data. It is quite reasonable, because the jet data provide constraint on gluon PDF as the $t \bar{t}$ data do. The inclusion of the $t \bar{t}$ data would forming a "competitive" relationship with the jet data on constraining gluon PDF.

\section{The impact of ATLAS and CMS 8 TeV $t\bar{t}$ data on  CT14HERA2mJ PDFs}\label{sec:ATLAS8updateHERA2mJ}

As we learn from the last section that, before the $t \bar{t}$ data, the gluon PDF of the CT14HERA2 receive well constraint from the jet data, namely 
	CDF~\cite{Aaltonen:2008eq}, D0~\cite{Abazov:2008ae}, ATLAS~\cite{Aad:2011fc} and CMS~\cite{Chatrchyan:2012bja}.
In order to see the impact of the $t\bar t$ data on gluon PDF,
we need to suppress the contribution from jet data.
For this purpose, first, we generated the Hessian eigenvector sets "CT14HERA2mJ"
("mJ" here means "minus jet")
by global analysis after removing the four inclusive
jet production data from Tevatron and LHC Run I in the CT14HERA2 fit.
And then we update CT14HERA2mJ PDFs  using 
{\tt \texttt{ePump}} by including  $t\bar t$ data one by one.
In this section, we provide comparisons of the CT14HERA2mJ 
before and after the
{\tt \texttt{ePump}} updating by adding the ATLAS and CMS 8 TeV $t\bar t$ singgle differential cross section data as a function of 
$y_{t\bar t}$, $m_{t\bar t}$, $y_t$ and $p^t_T$.

\subsection{Correlation between CT14HERA2mJ gluon PDF and $t\bar t$ data}

\begin{figure}[H]	
	\centering
	\includegraphics[width=0.49\textwidth]{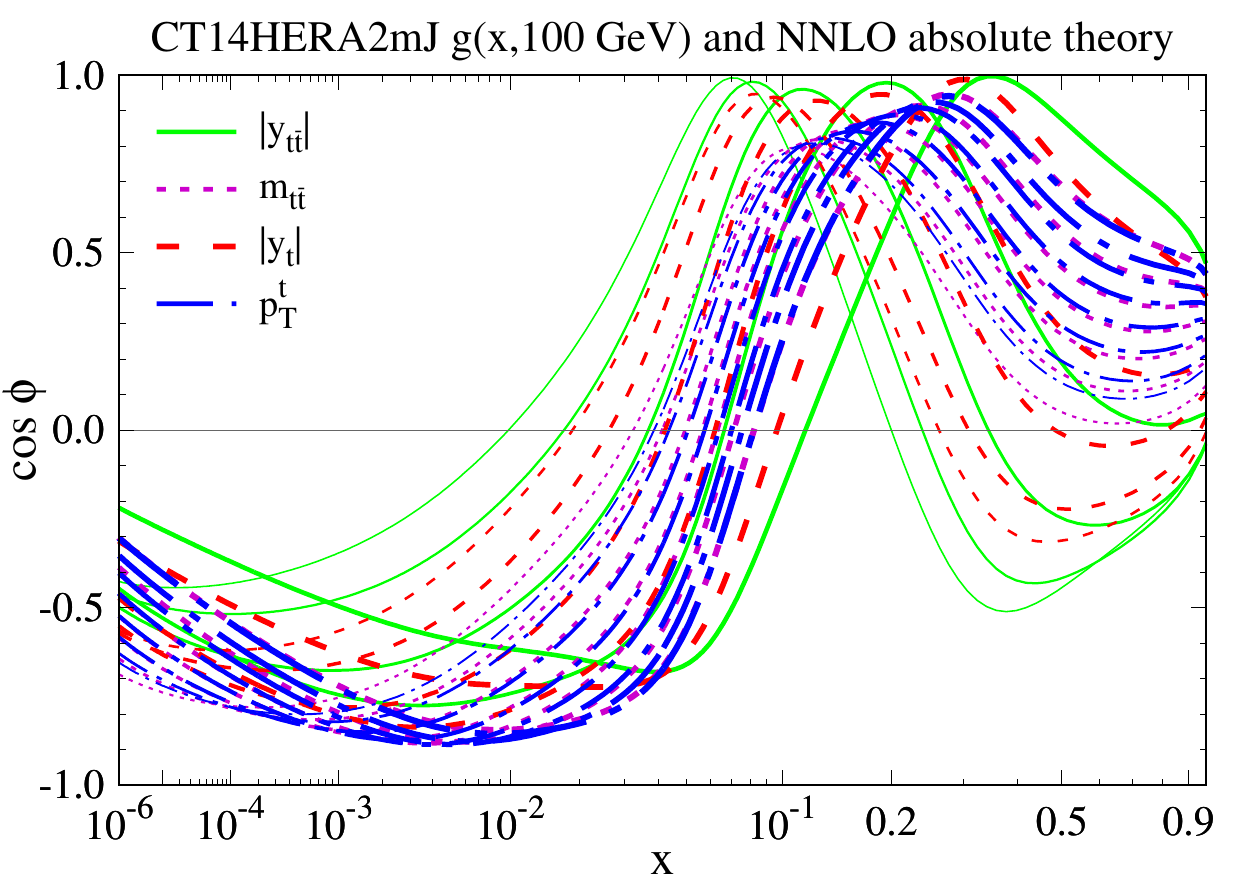}
	\includegraphics[width=0.49\textwidth]{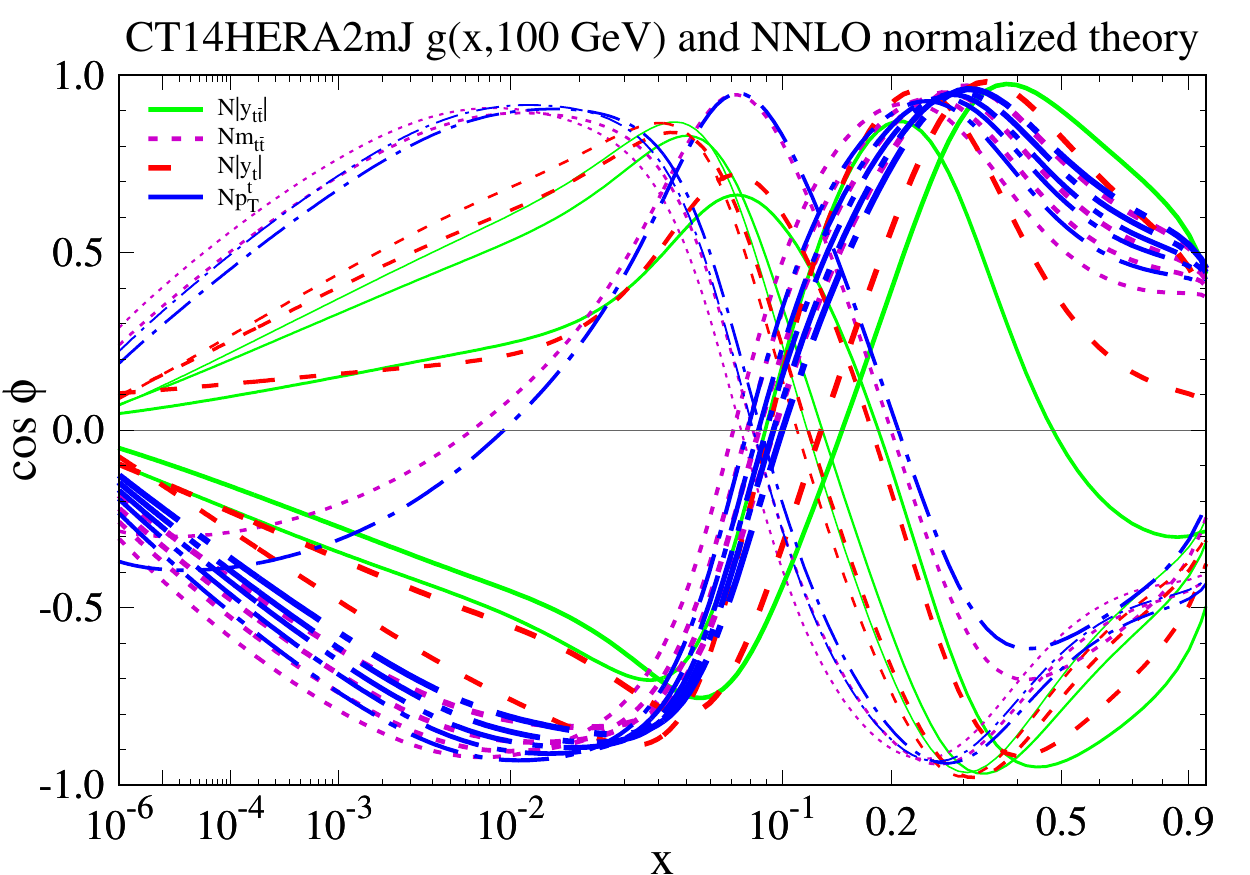}
	\vspace{0.4cm}
	\caption{Correlation cosine $\cos \phi$ between CT14HERA2mJ $g(x,Q= 100 GeV)$ PDF and fastNLO predictions for each bins of the  absolute (left) and normalized (right) $|y_{t\bar t}|$ (solid green), $m_{t\bar t}$ (dark magenta), $|y_t|$ (red) and $p^t_{T}$ (blue) top-quark differential distributions.}\label{plot:ATL-gluonmJ}
\end{figure}

We first check the
correlation between the absolte  and normalized  differential $t\bar t$ data and the CT14HER2mJ g($x, Q = 100$ GeV) PDF.
Without the inclusion of the jet data in the CT14HER2mJ, the gluon PDF receive constraints mostly from the deep inelastic scattering (DIS) data, and have different behavior as the gluon in the CT14HERA2 PDF. As showing in Fig.~\ref{plot:ATL-gluonmJ}, the correlation between $t\bar t$ data and the gluon CT14HER2mJ PDF keep the main features as that for the gluon PDF in CT14HERA2 PDF shown in Fig.~\ref{cosphi-CT14HERA2-gluon-ttbarTheory}. 

\subsection{Update CT14HERA2mJ PDFs using ATLAS 8 TeV $t\bar t$ data}

\begin{figure}[H]
	\begin{center}
		\includegraphics[width=0.45\textwidth]{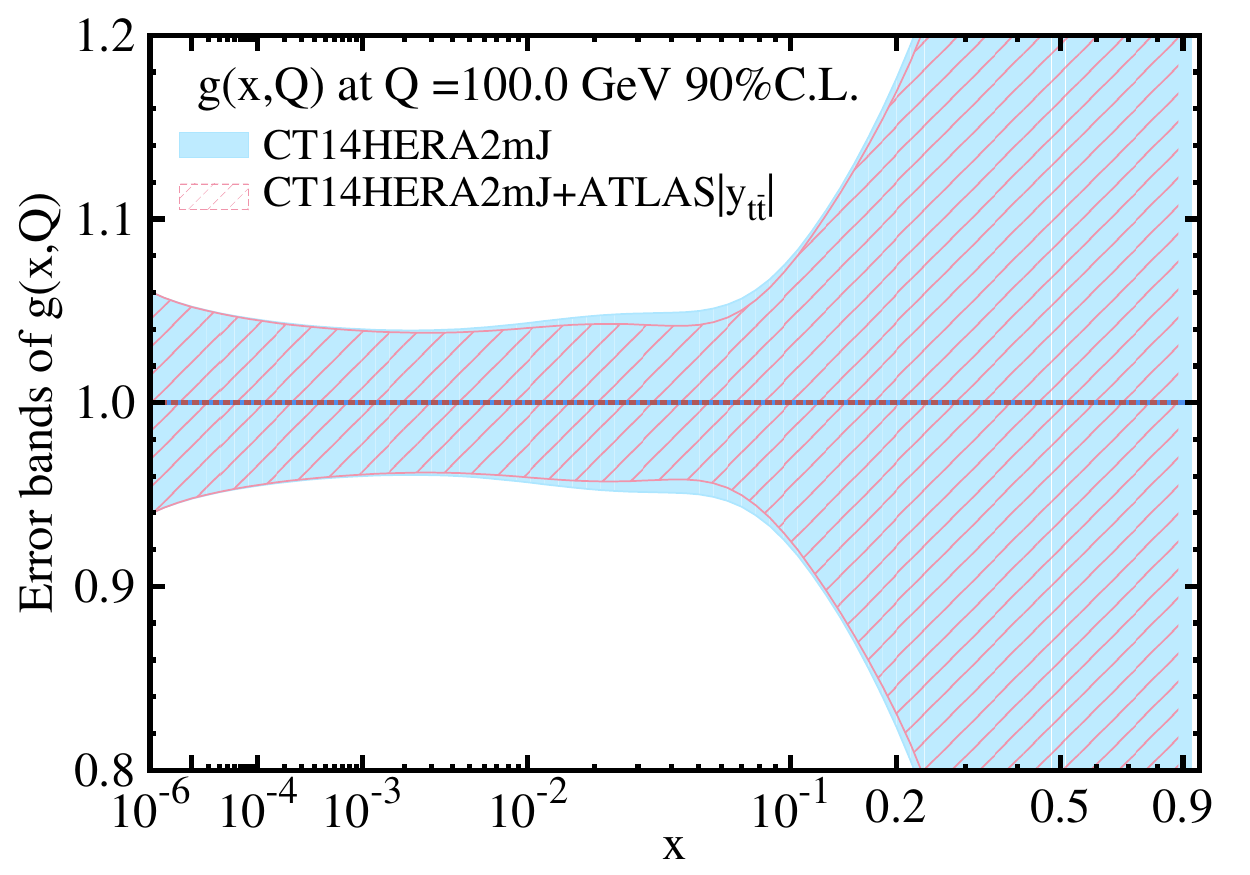}
		\includegraphics[width=0.45\textwidth]{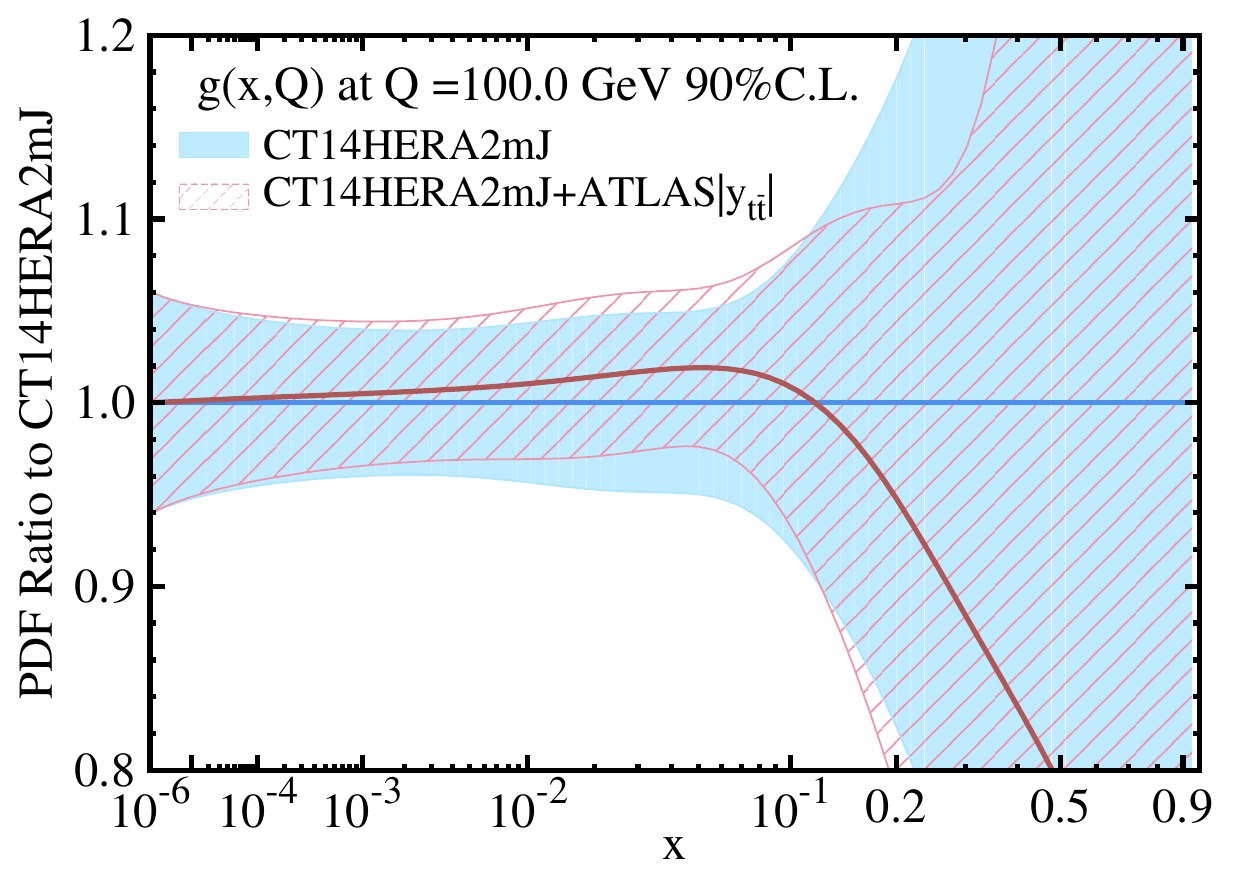}
		\includegraphics[width=0.45\textwidth]{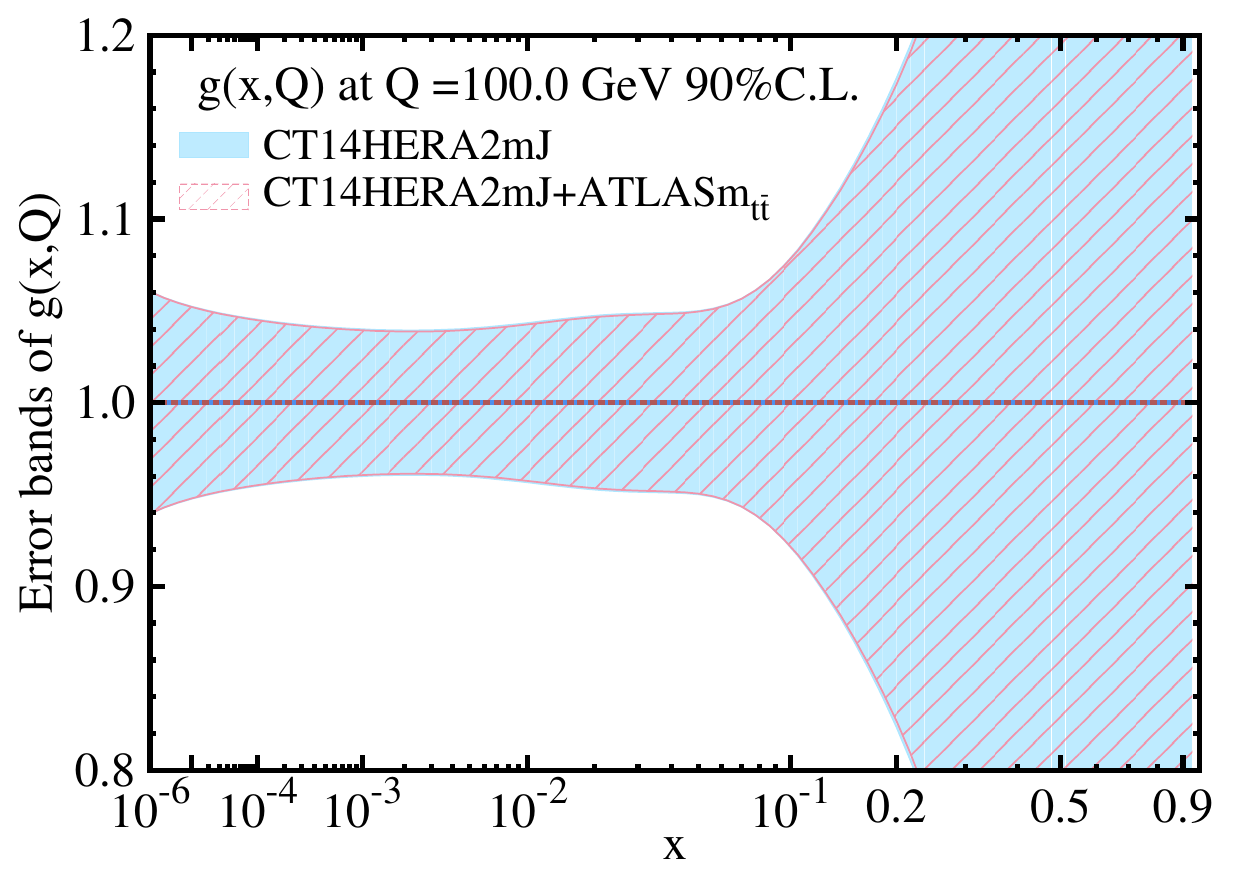}
		\includegraphics[width=0.45\textwidth]{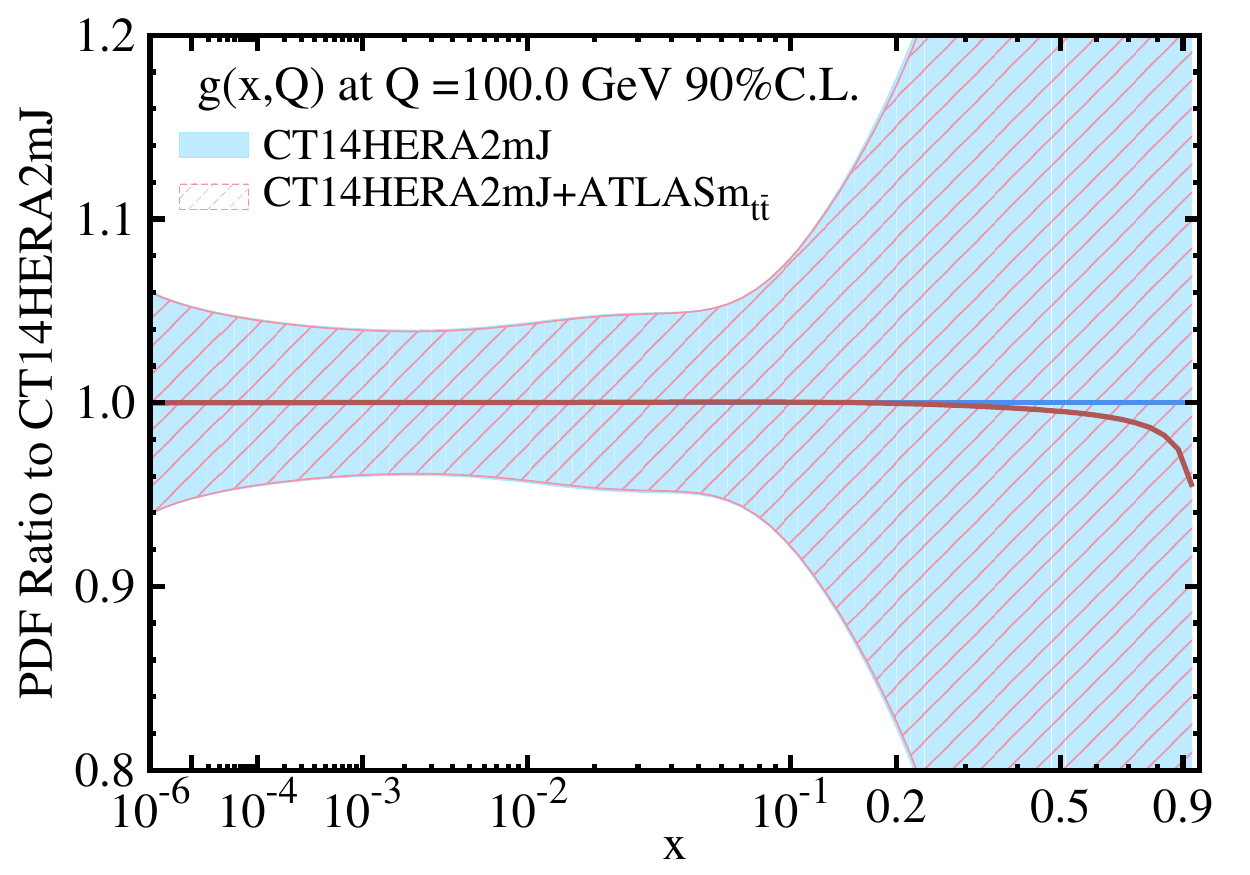}
		\includegraphics[width=0.45\textwidth]{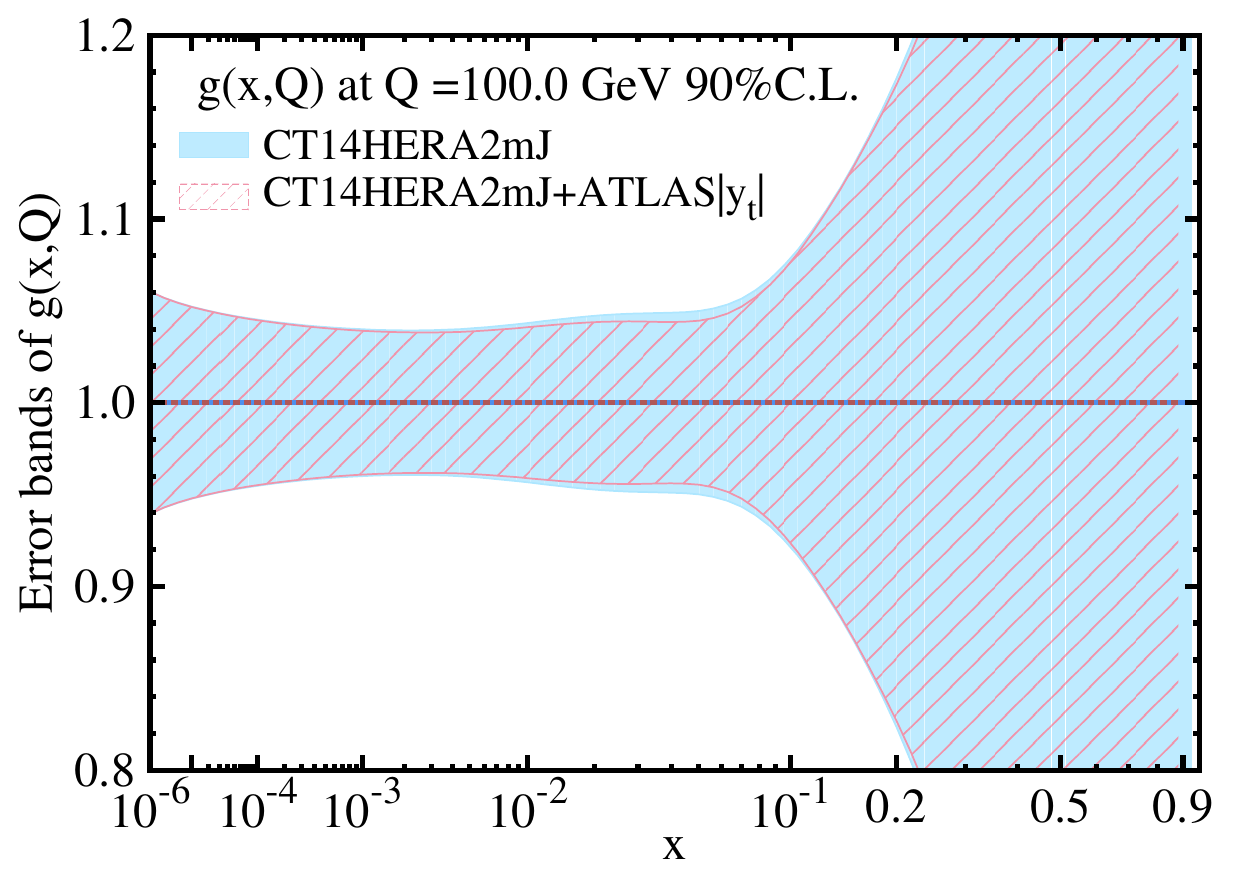}
		\includegraphics[width=0.45\textwidth]{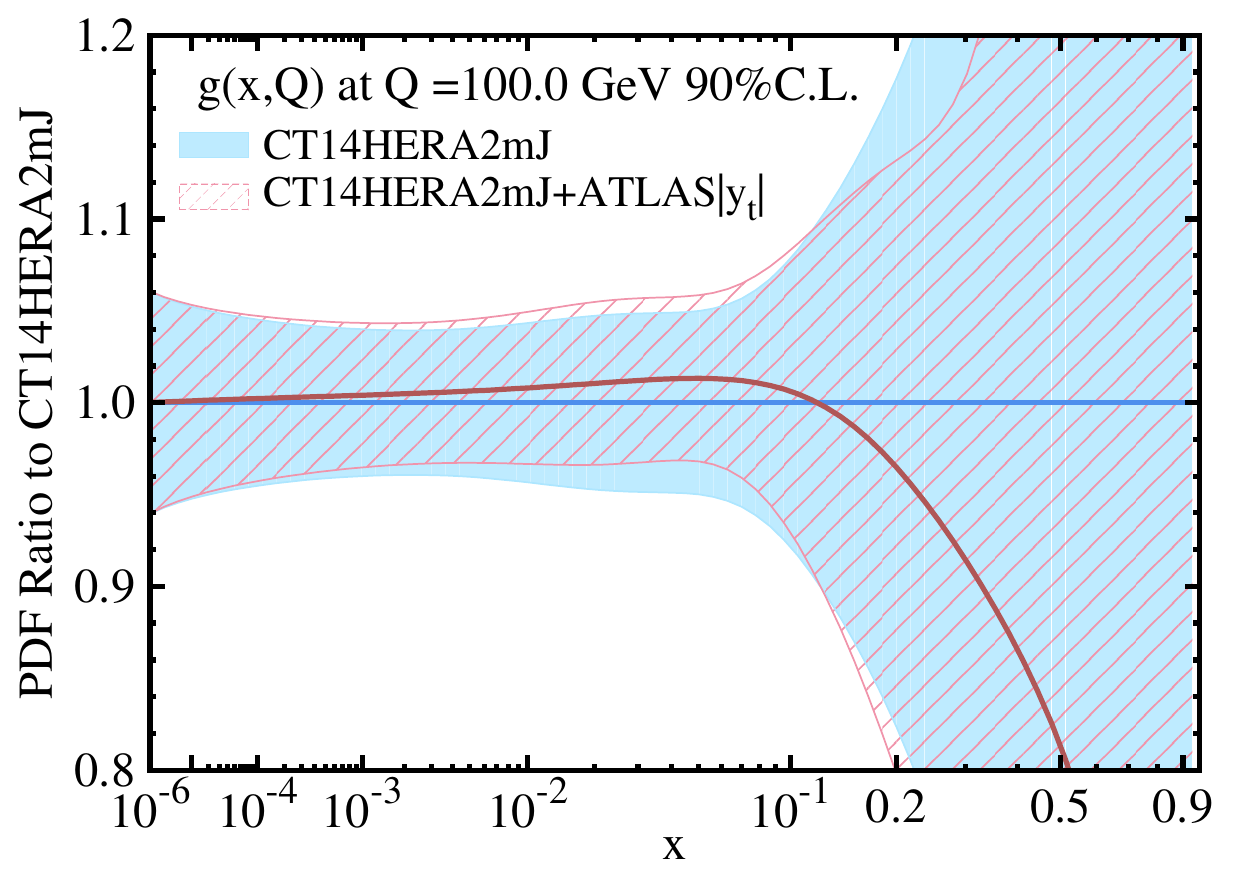}
		\includegraphics[width=0.45\textwidth]{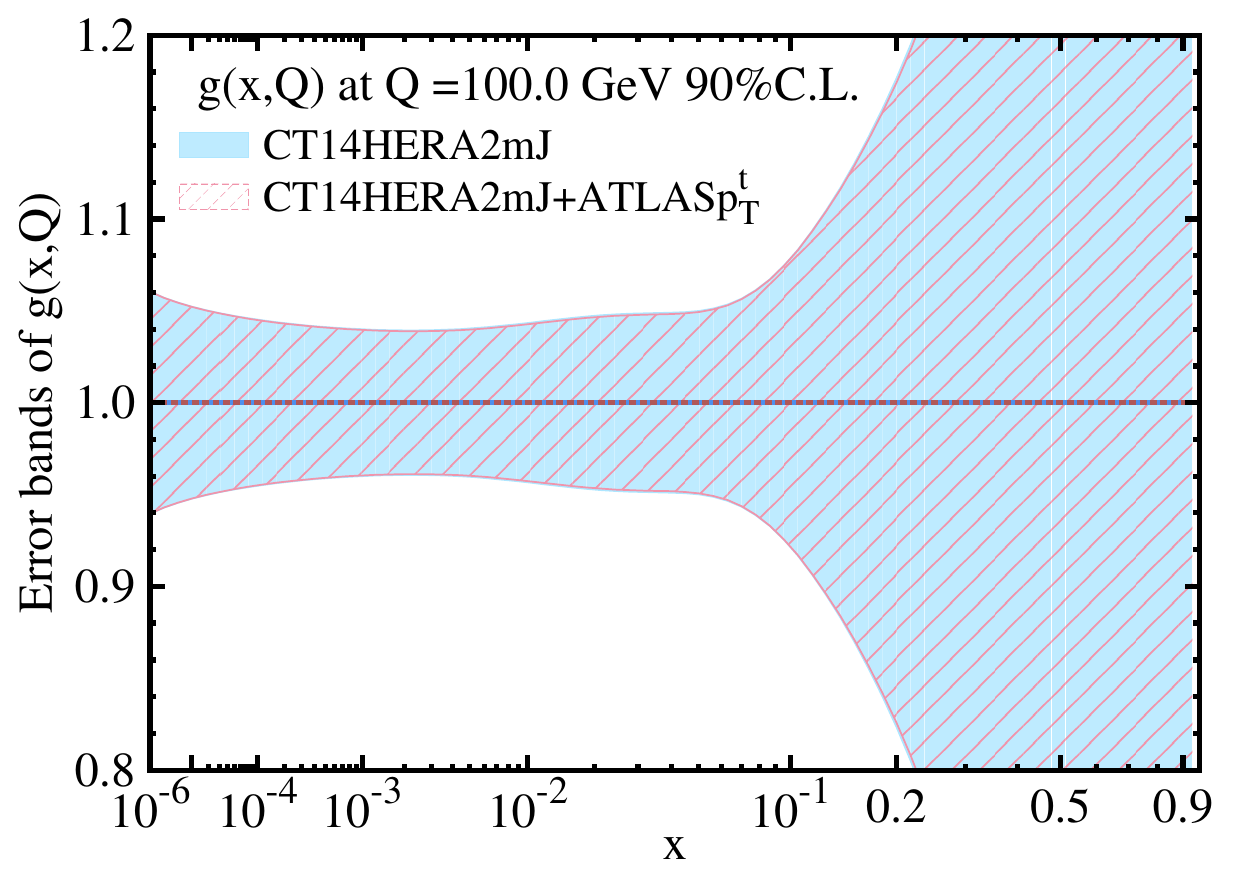}
		\includegraphics[width=0.45\textwidth]{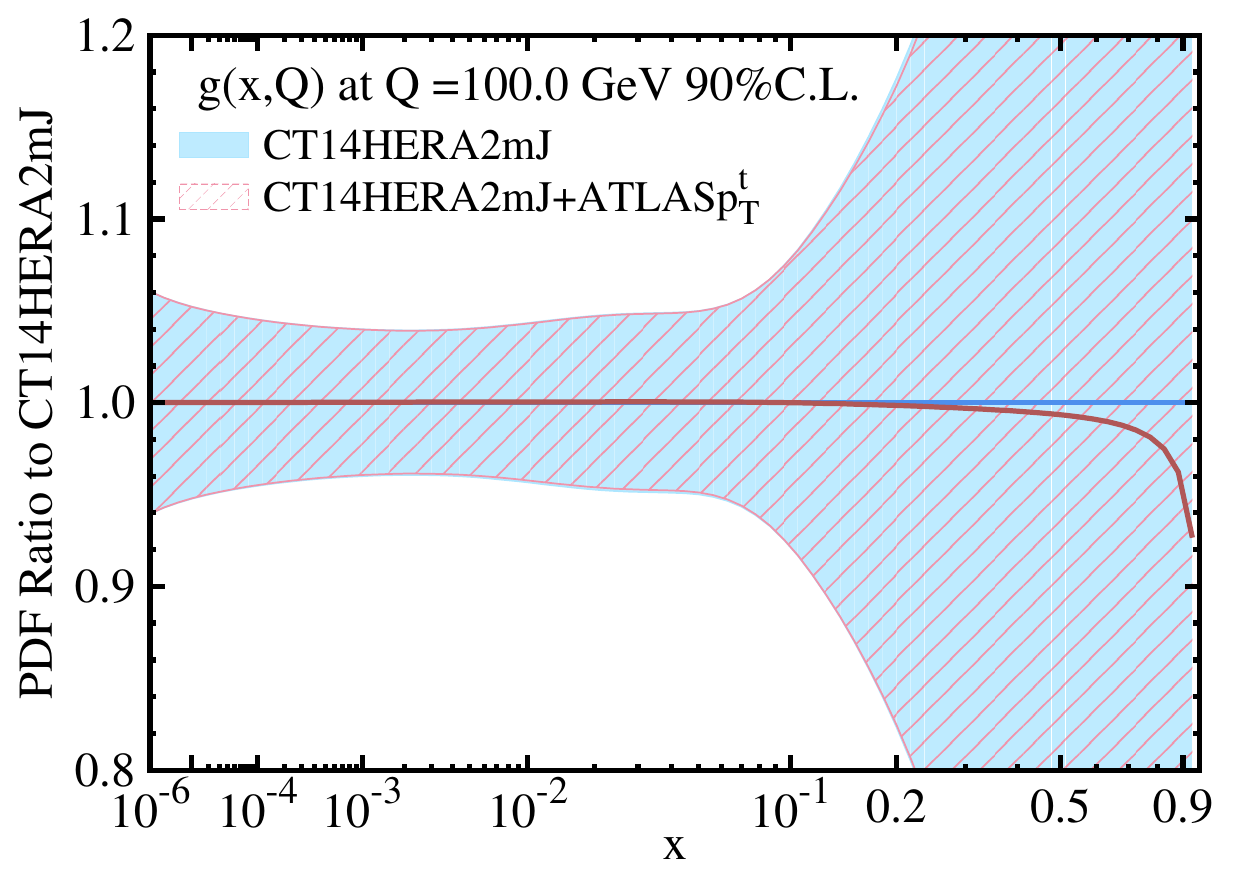}
	\end{center}
	\caption{ The gluon PDF ratios for {\tt \texttt{ePump}}-updated 
		CT14HERA2mJ+ATLAS$|y_{t\bar t}|$, 
		CT14HERA2mJ+ATLAS$m_{t\bar t}$, 
		CT14HERA2mJ+ATLAS$|y_t|$, 
		CT14HERA2mJ+ATLAS$p^t_T$ PDFs,
		which are obtained by including ATLAS 8 TeV absolute  $d\sigma/d|y_{t\bar t}|$, $d\sigma/dm_{t\bar t}$, $d\sigma/d|y_t|$, and $d\sigma/dp^t_T$  data,
		over the best-fit of the base  CT14HERA2.54 gluon PDFs.
	}\label{Fig:CT14HERA2mJetpATLAS8Att}
\end{figure}

\begin{figure}[H]
	\begin{center}
		\includegraphics[width=0.45\textwidth]{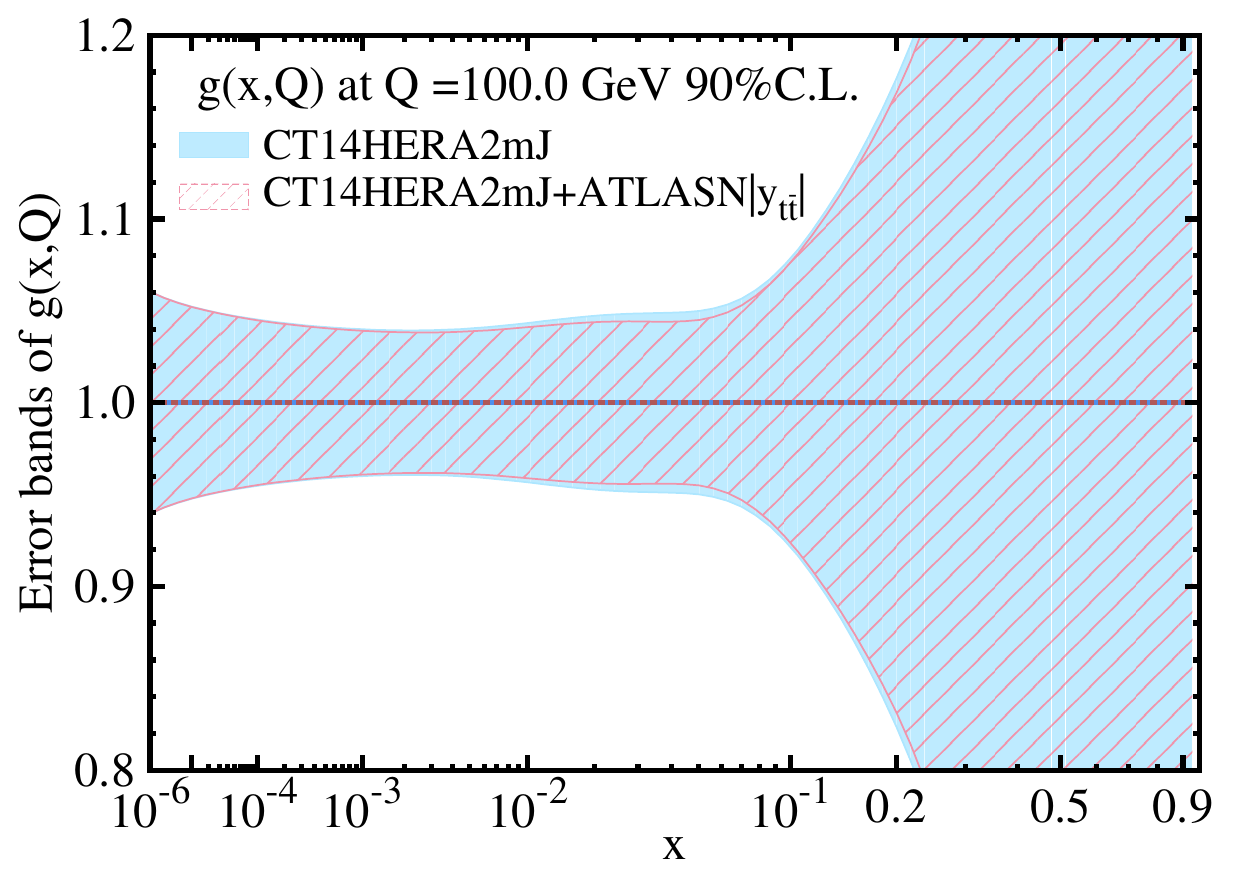}
		\includegraphics[width=0.45\textwidth]{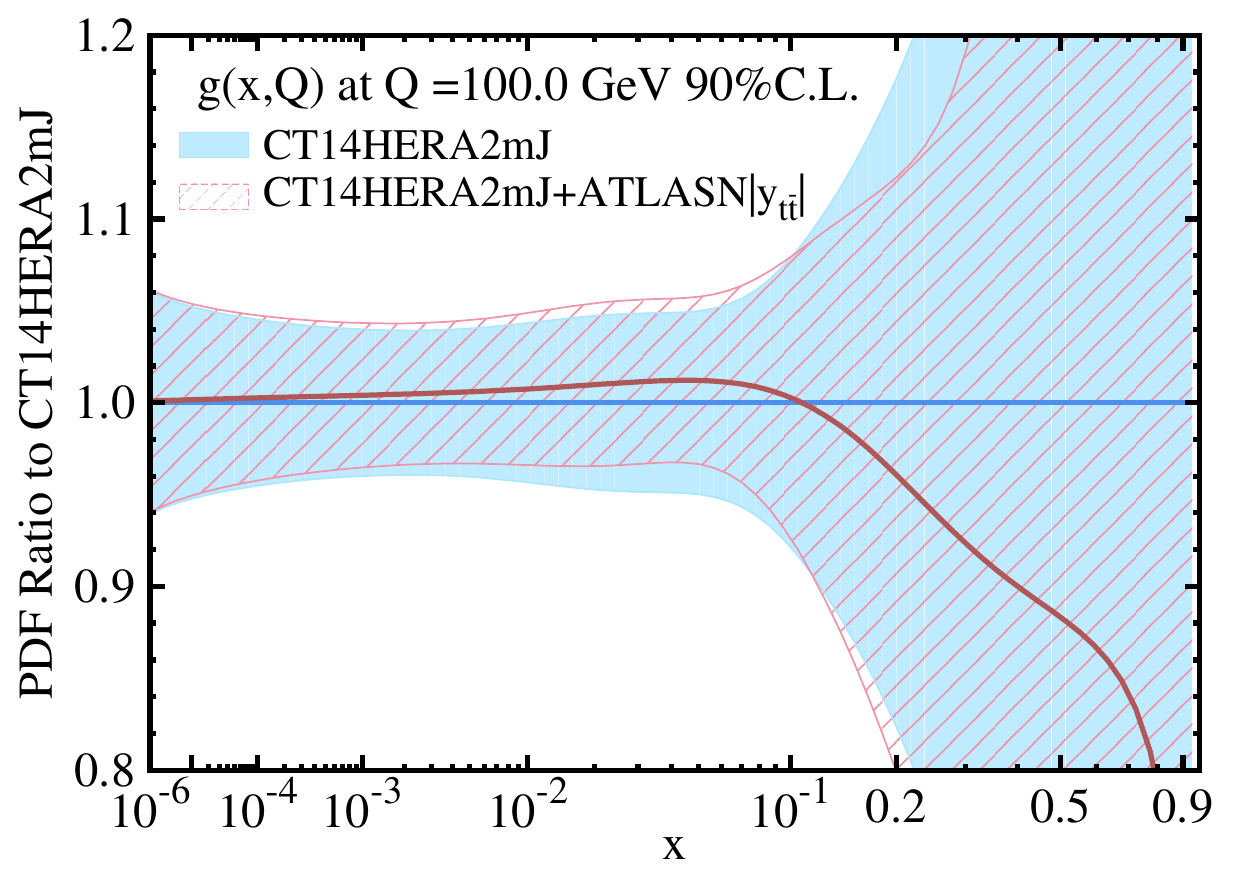}
		\includegraphics[width=0.45\textwidth]{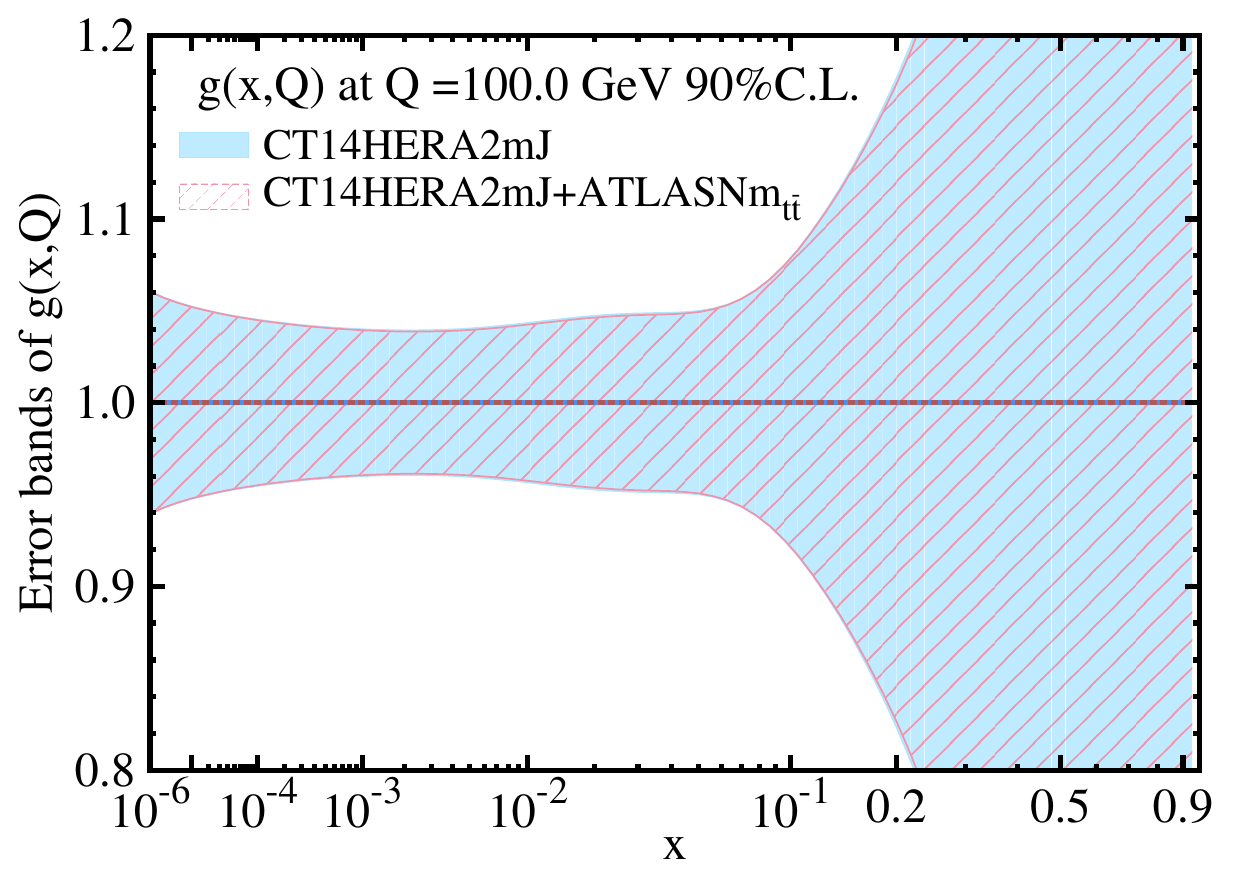}
		\includegraphics[width=0.45\textwidth]{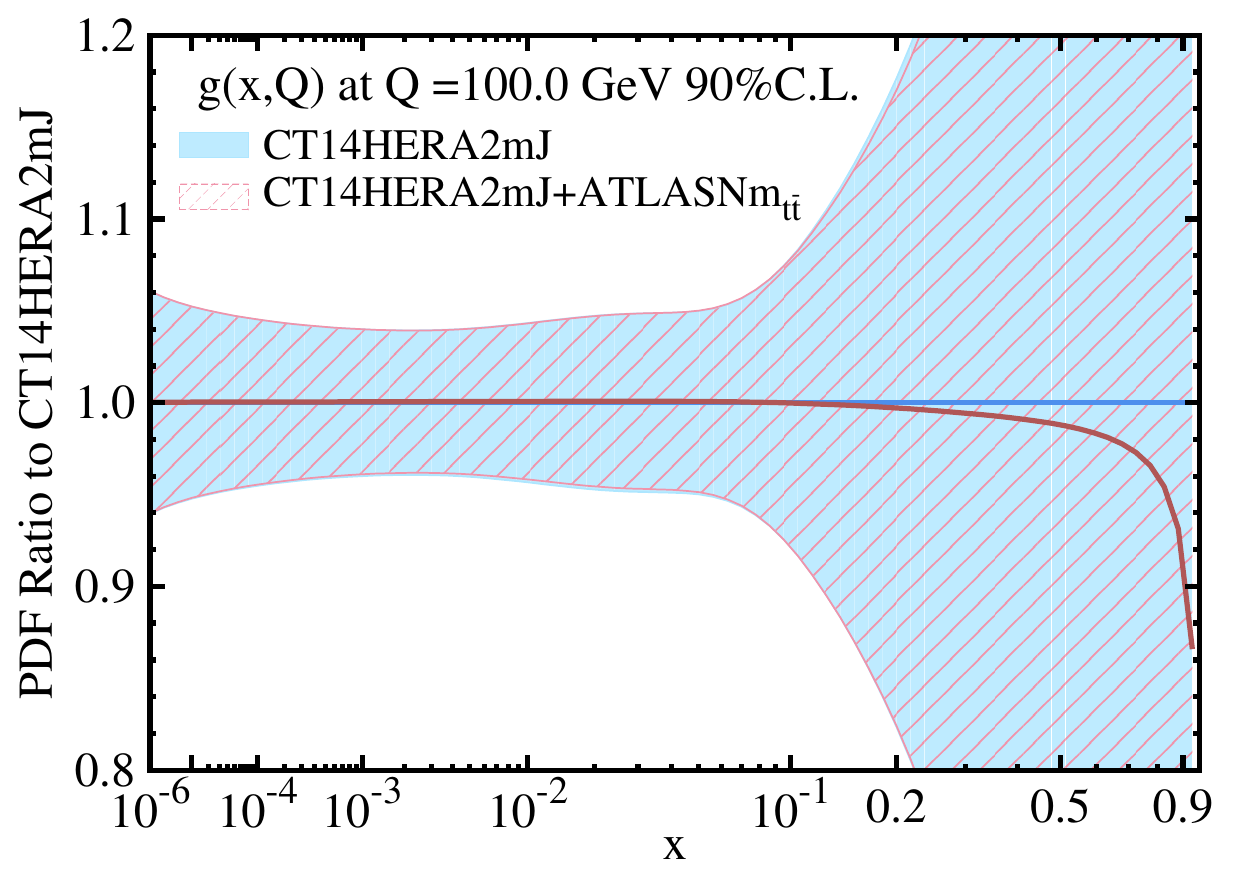}
		\includegraphics[width=0.45\textwidth]{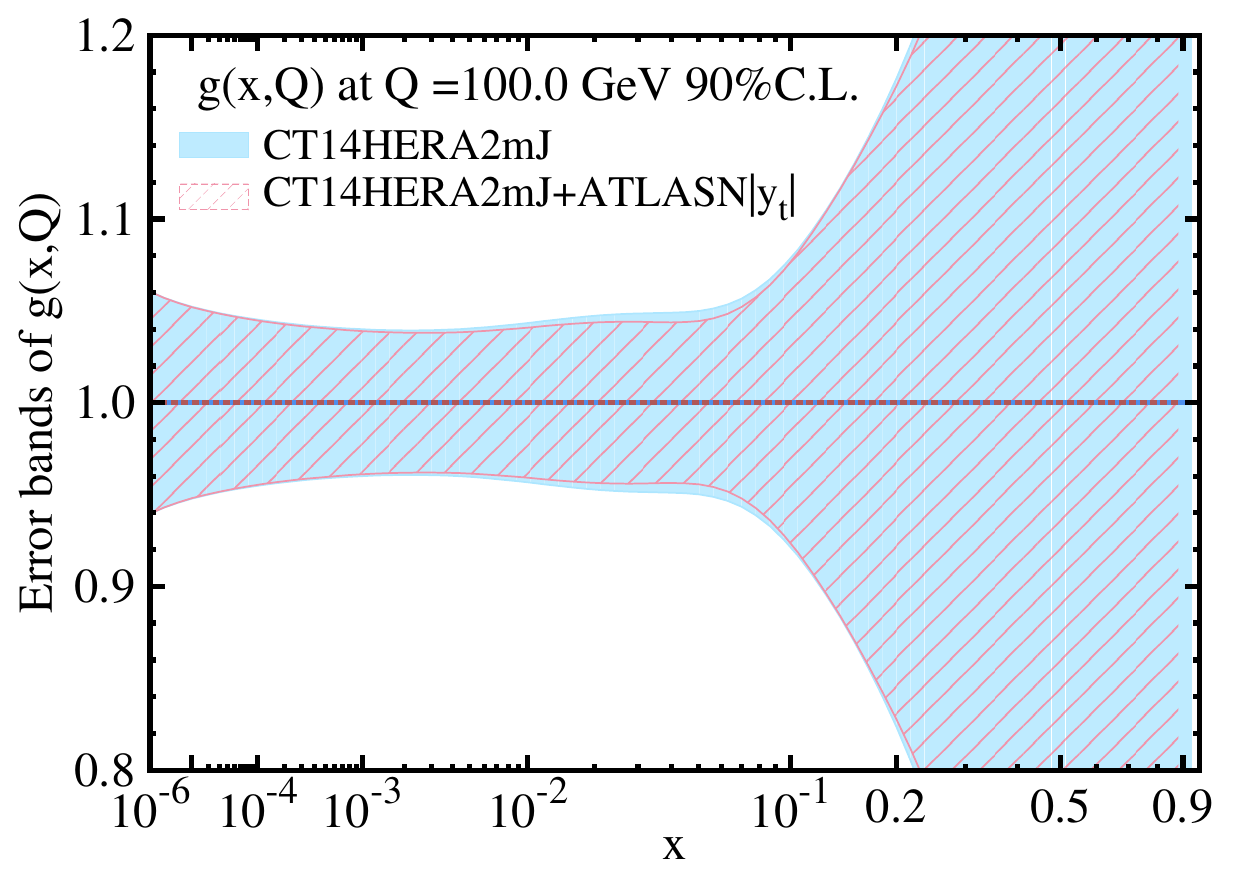}
		\includegraphics[width=0.45\textwidth]{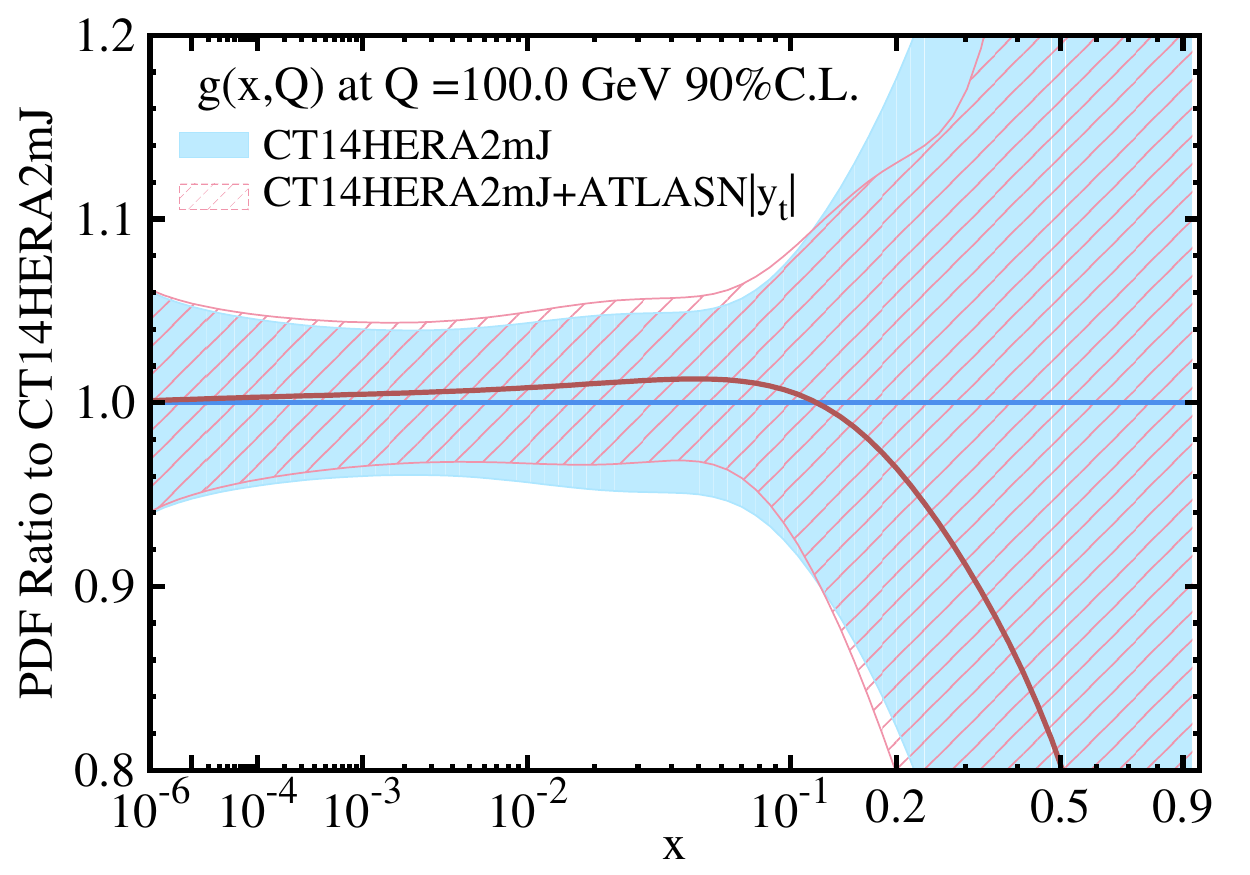}
		\includegraphics[width=0.45\textwidth]{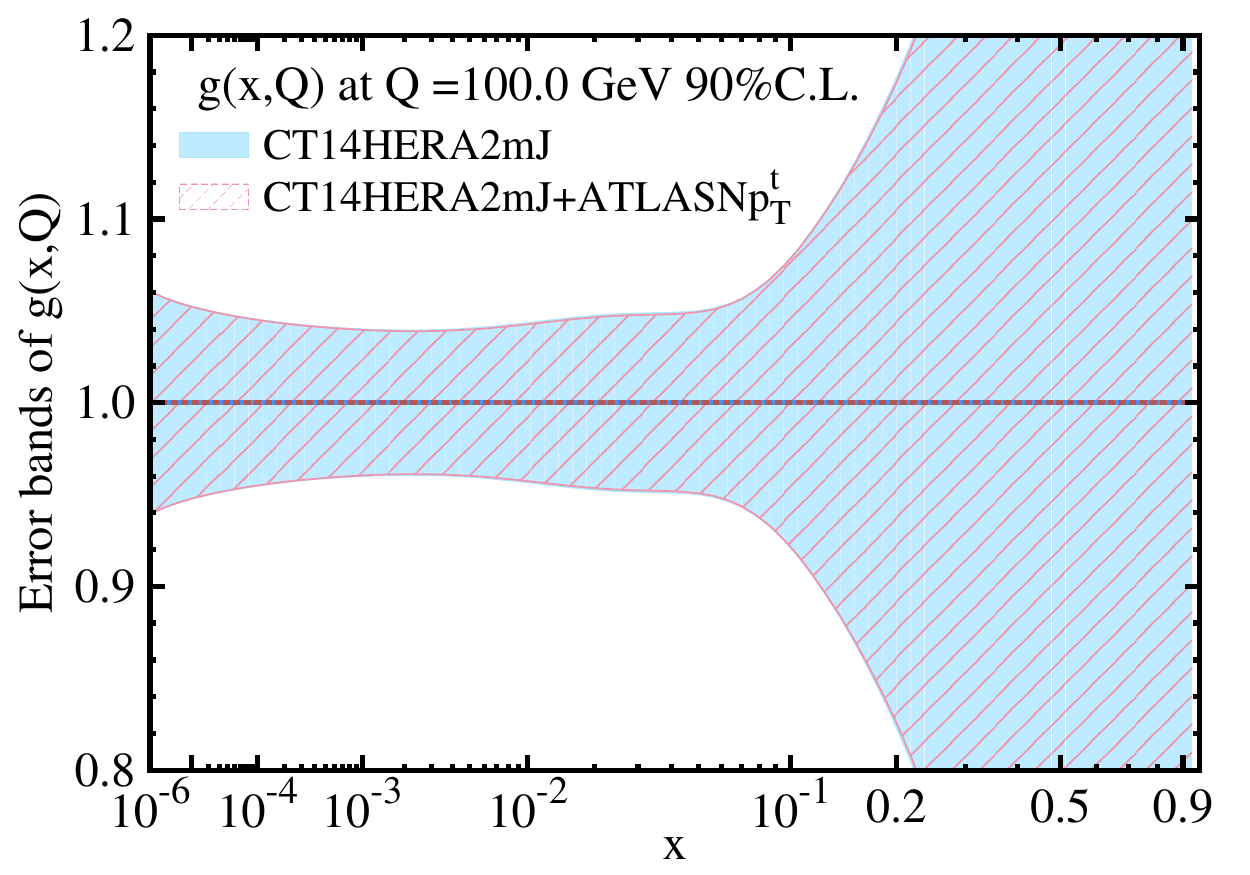}
		\includegraphics[width=0.45\textwidth]{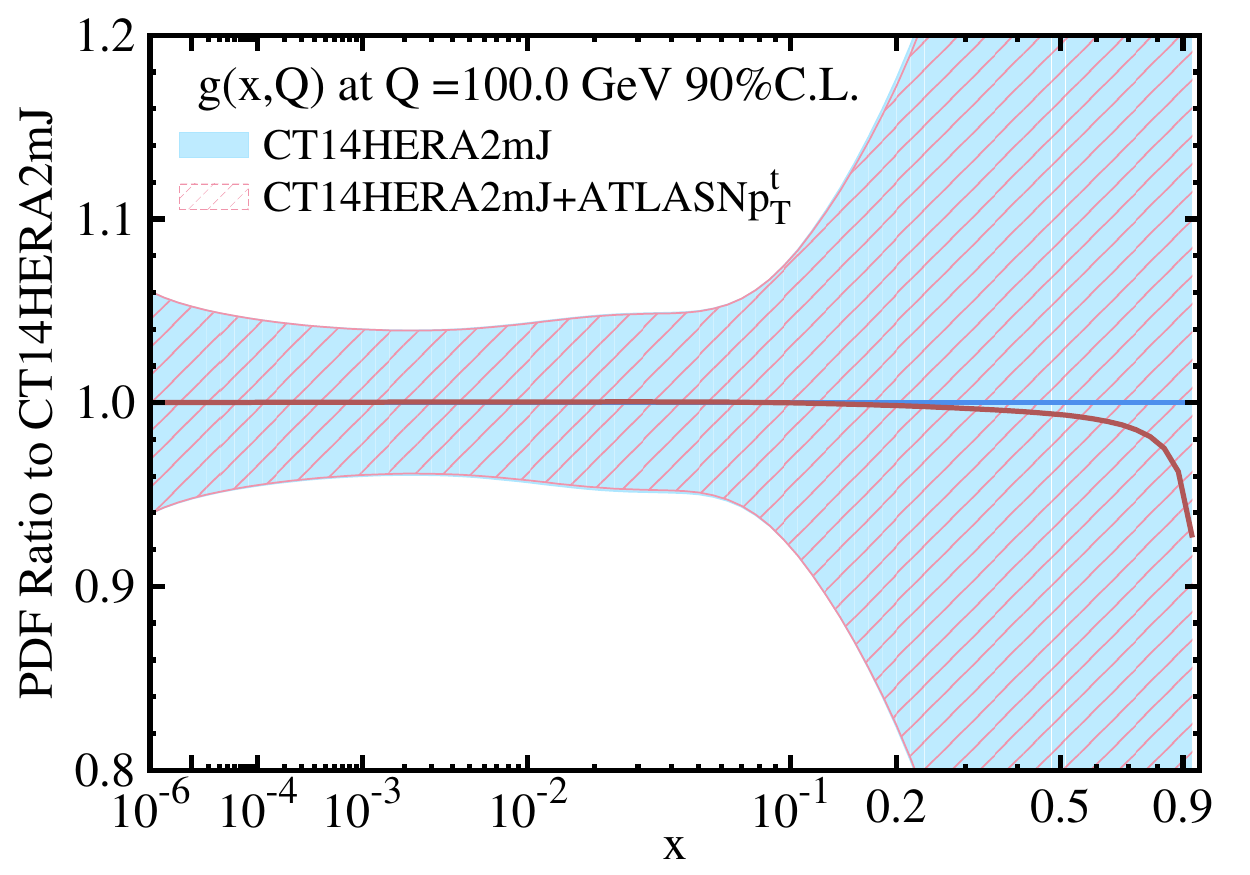}
	\end{center}
	\caption{The gluon PDF ratios for {\tt \texttt{ePump}}-updated 
		CT14HERA2mJ+ATLAS$|y_{t\bar t}|$, 
		CT14HERA2mJ+ATLAS$m_{t\bar t}$, 
		CT14HERA2mJ+ATLAS$|y_t|$, 
		CT14HERA2mJ+ATLAS$p^t_T$ PDFs,
		which are obtained by including ATLAS 8 TeV
		normalized $1/\sigma \; d\sigma/d|y_{t\bar t}|$, $1/\sigma \; d\sigma/dm_{t\bar t}$, $1/\sigma \; d\sigma/d|y_t|$, $1/\sigma \; d\sigma/dp^t_T$ data, over the best-fit of the base  CT14HERA2.54 gluon PDFs. }\label{Fig:CT14HERA2mJetpATLAS8Ntt}
\end{figure}

In Figs.~\ref{Fig:CT14HERA2mJetpATLAS8Att} and \ref{Fig:CT14HERA2mJetpATLAS8Ntt}, we show  the {\tt \texttt{ePump}} updating PDFs, starting from the CT14HERA2mJ PDFs, and including the
absolute and normalized ATLAS 8 TeV $t\bar t$ data one by one.
The impact on the gluon PDF from  those $t\bar t$ data can be seen by comparing the difference between the gluon PDF before and
after  {\tt \texttt{ePump}} updating.
It is apparent that, without the jet data in the fit, the CT14HERA2mJ uncertainty band is larger than the CT14HERA2 band, 
the absolute  $d\sigma/d|y_{t\bar t}|$,  $d\sigma/d|y_t|$  and 
normalized  $1/\sigma \; d\sigma/d|y_{t\bar t}|$,  $1/\sigma \; d\sigma/d|y_t|$ data  have a larger impact on the central PDF (and to a smaller extent on the uncertainty band) of the CT14HERA2mJ gluon PDF.

More specifically, we observe an increase of the CT14HERA2mJ best fit gluon PDF for $10^{-4} \lesssim x \lesssim 0.15$ and a decrease for $x \gtrsim 0.15$ after including $d\sigma/d|y_{t\bar t}|$ and $d\sigma/d|y_t|$ data. We also observe a small reduction of the gluon PDF uncertainty bands for $x \sim 0.05$.
However, there is still no obvious impact on the CT14HERA2mJ gluon
PDF after including the absolute
$d\sigma/dm_{t\bar t}$, $d\sigma/dp^t_T$
and normalized $1/\sigma \; d\sigma/dm_{t\bar t}$, $1/\sigma \; d\sigma/dp^t_T$ data.
The results shown in the Figs.~\ref{Fig:CT14HERA2mJetpATLAS8Att} and \ref{Fig:CT14HERA2mJetpATLAS8Ntt} directly confirm our understanding from the last section that, the reason we see only minor impacts on the CT14HERA2 gluon PDF is because the CT14HERA2 gluon PDF is well constrained by the four jet data sets included in the CT14HERA2 PDF. The removal of the jet data increases the constraining power of the $t \bar t$ data, but the impact is still smaller than that of the jet data sets.

\subsection{Update CT14HERA2mJ PDFs using CMS 8 TeV $t\bar{t}$ data }

\begin{figure}[H]
	\begin{center}
		\includegraphics[width=0.45\textwidth]{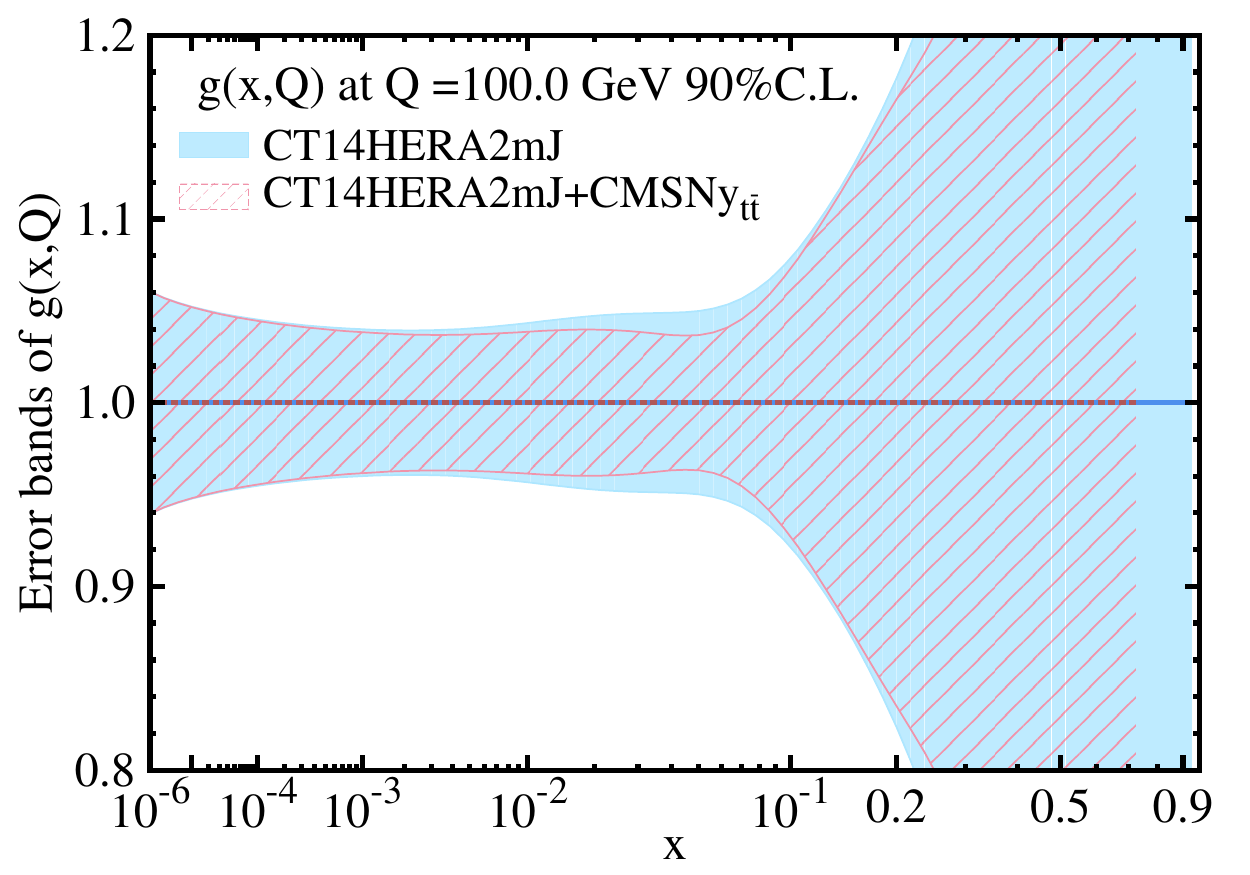}
		\includegraphics[width=0.45\textwidth]{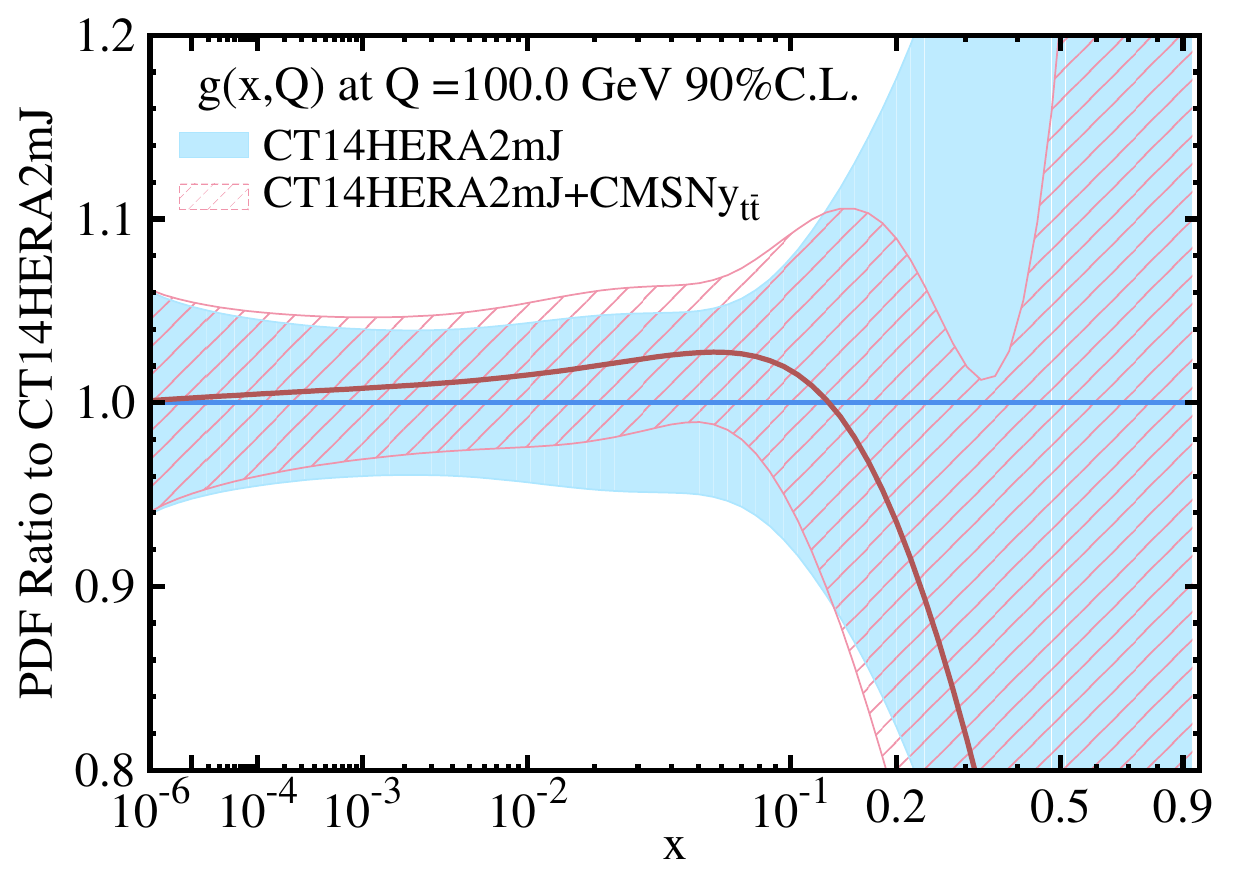}
		\includegraphics[width=0.45\textwidth]{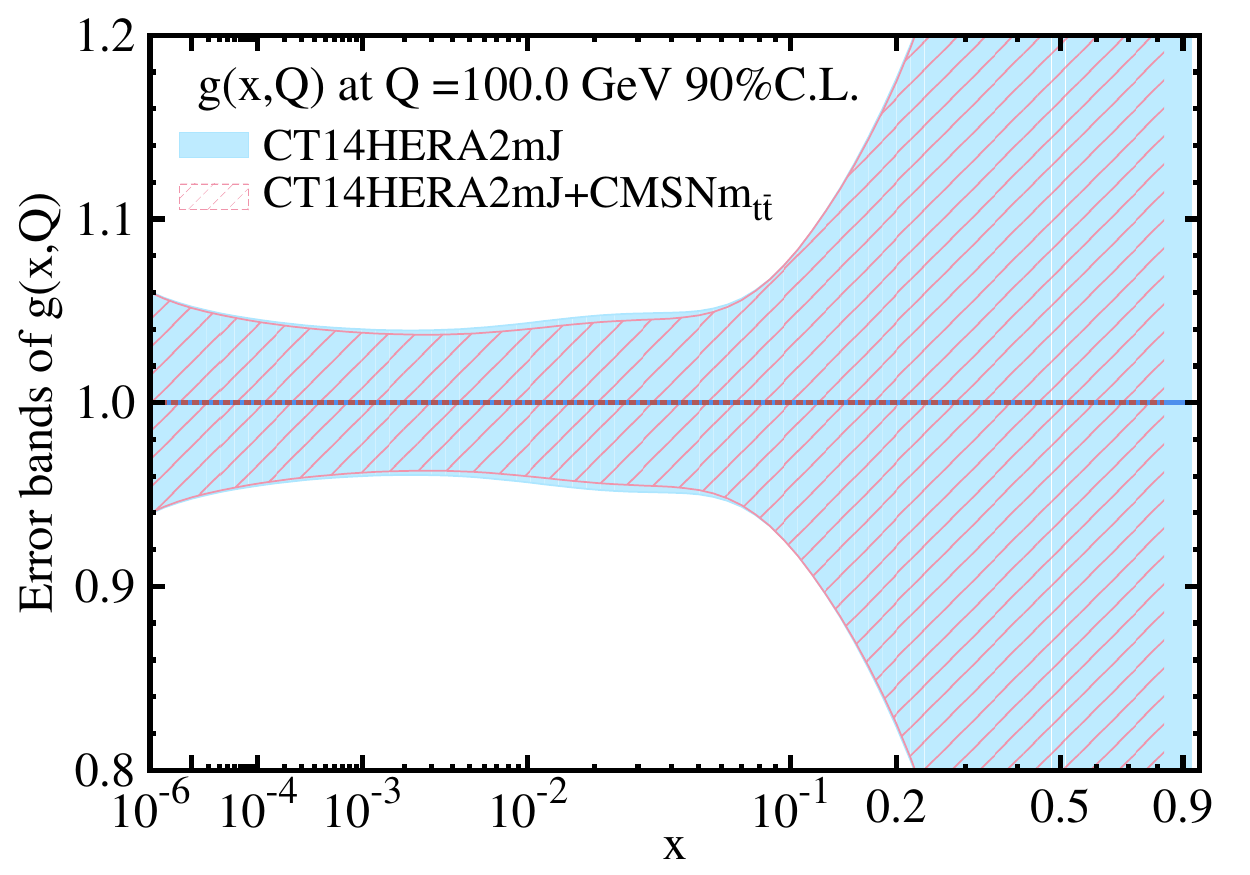}
		\includegraphics[width=0.45\textwidth]{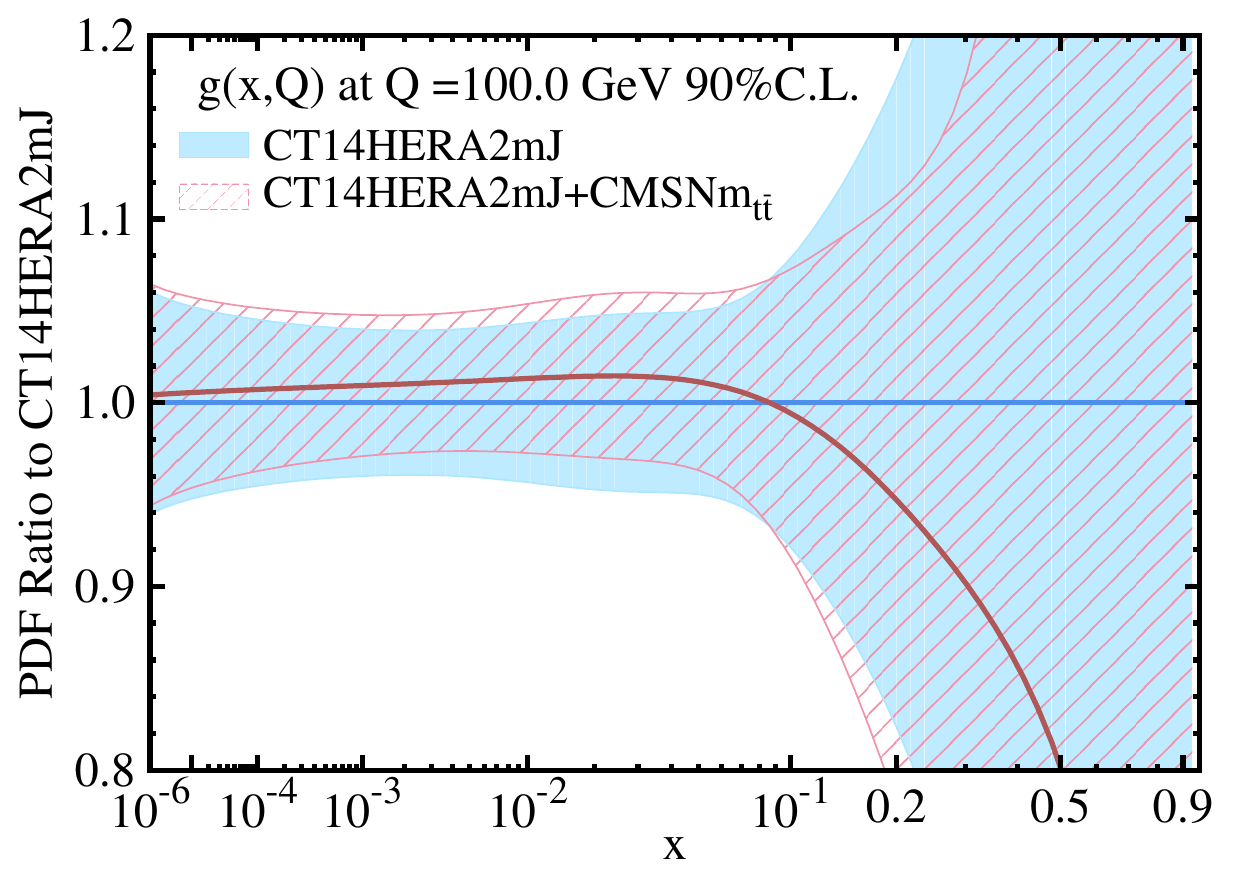}
		\includegraphics[width=0.45\textwidth]{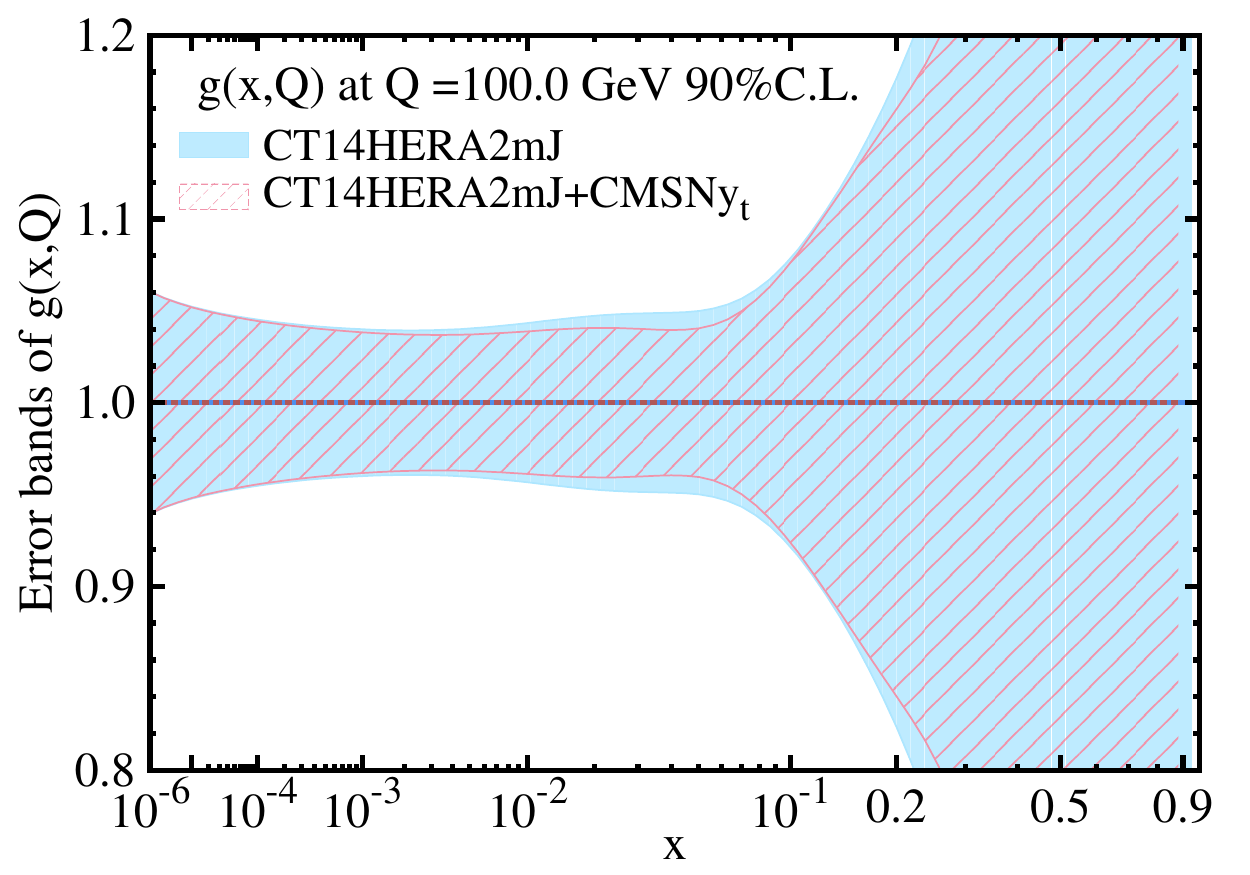}
		\includegraphics[width=0.45\textwidth]{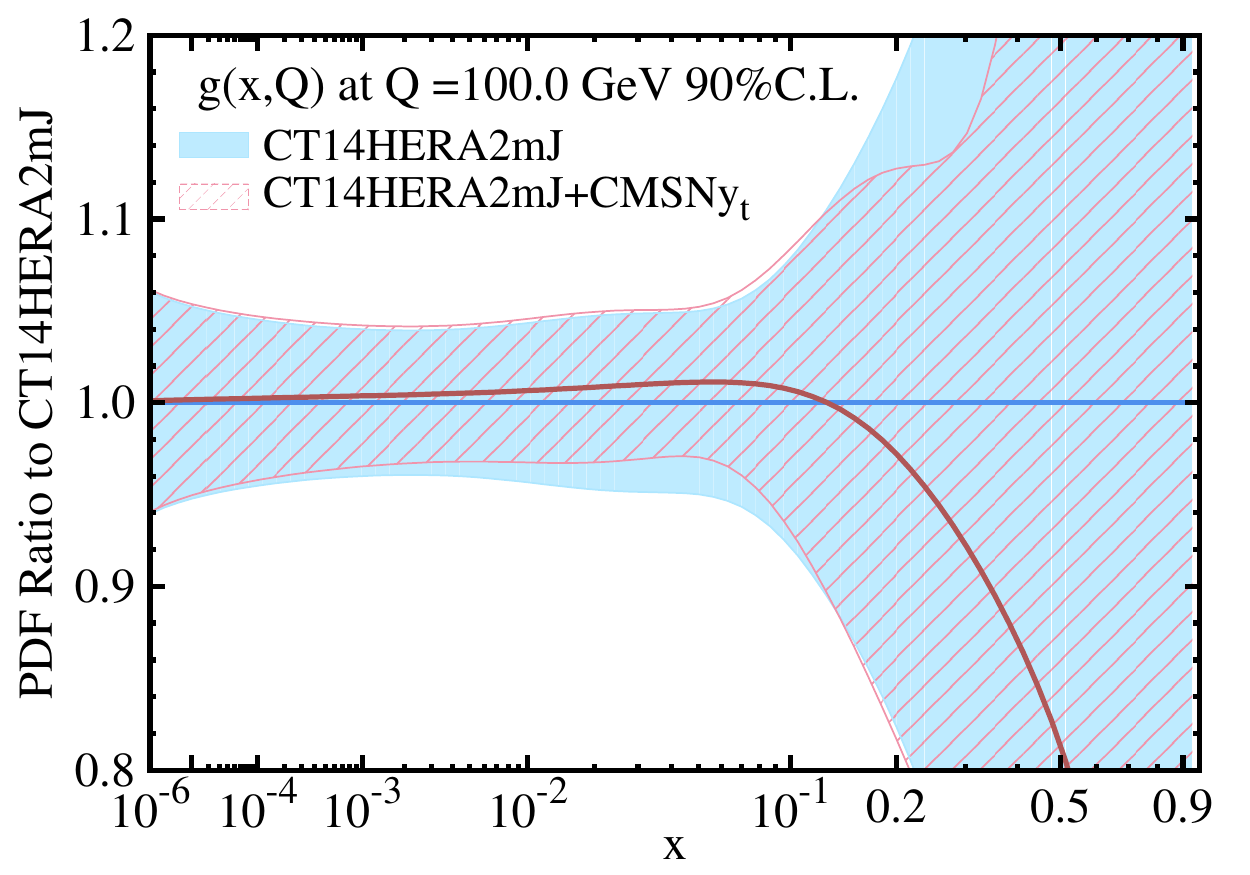}
		\includegraphics[width=0.45\textwidth]{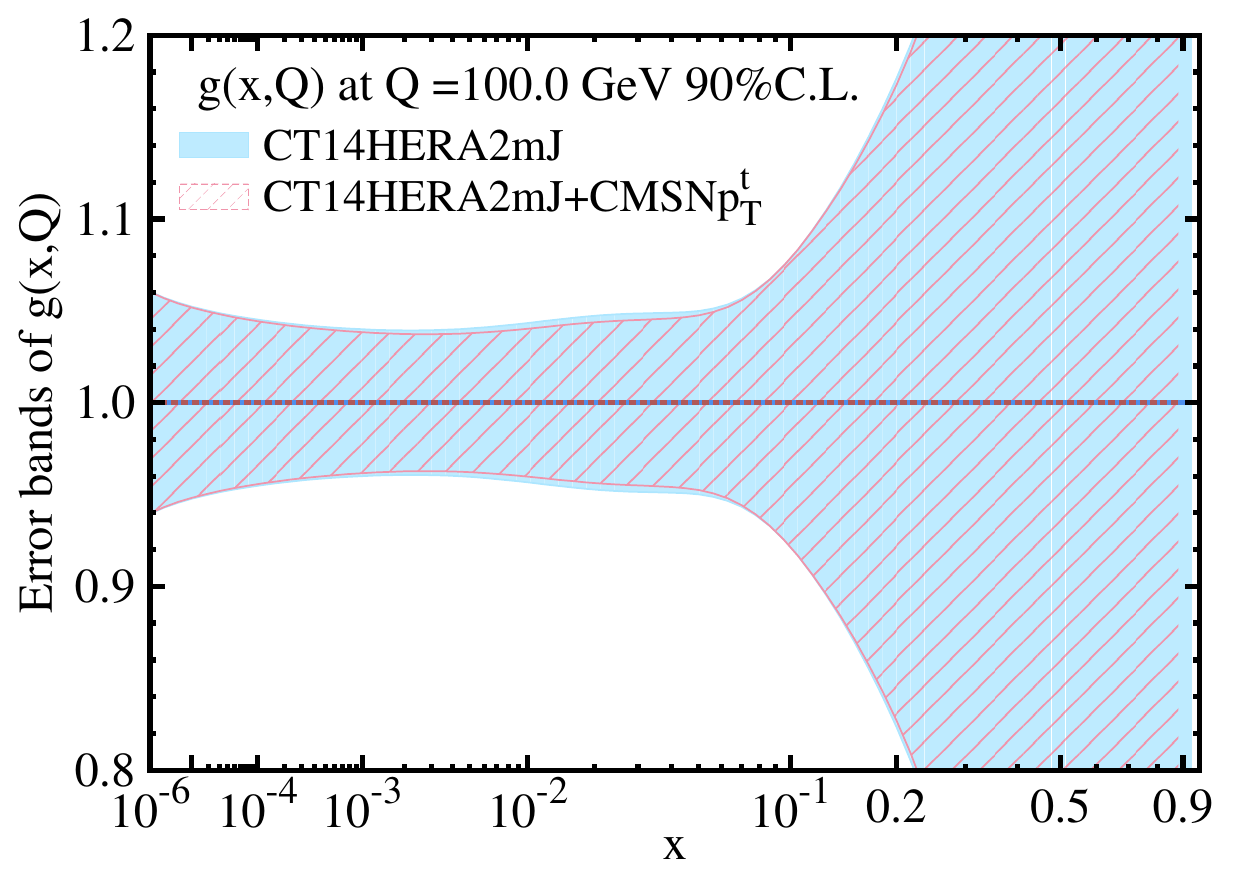}
		\includegraphics[width=0.45\textwidth]{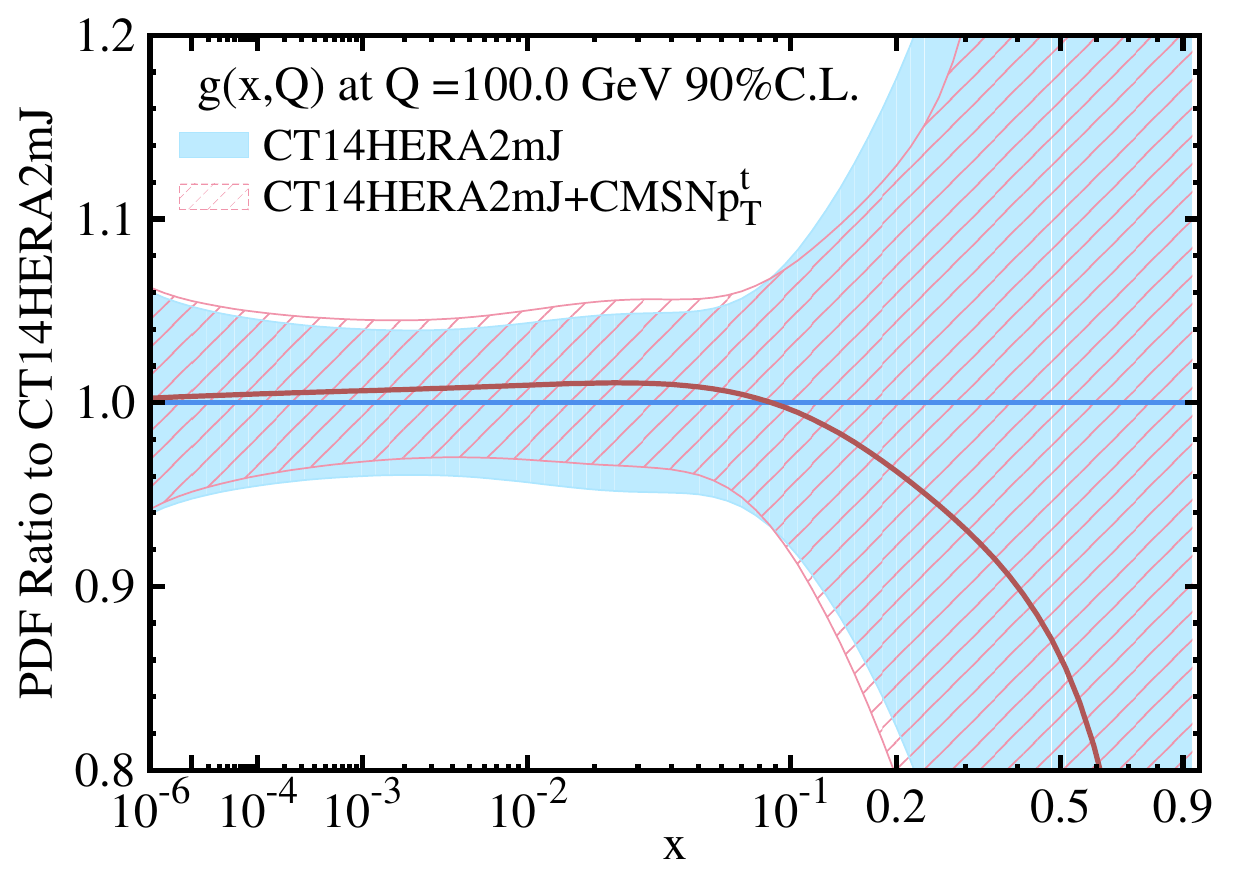}
	\end{center}
	\caption{The gluon PDF ratios for {\tt \texttt{ePump}}-updated 
		CT14HERA2mJ+CMSN$y_{t\bar t}$, 
		CT14HERA2mJ+CMSN$m_{t\bar t}$, 
		CT14HERA2mJ+CMSN$y_t$, 
		CT14HERA2mJ+CMSN$p^t_T$ PDFs over the best-fit of the base CT14HERA2mJ gluon PDFs.
	}\label{Fig:CT14HERA2mJetpCMS8Ntt}
\end{figure}

In this section by including the normalized CMS 8 TeV   $1/\sigma \; d\sigma/dy_{t\bar t}$, $1/\sigma \; d\sigma/dm_{t\bar t}$, $1/\sigma \; d\sigma/dy_t$, and $1/\sigma \; d\sigma/dp^t_T$ data one by one, we
update CT14HERA2mJ PDFs. 
The impact on gluon PDF from $t\bar{t}$ data can be seen by
comparing the difference between the gluon PDF before and
after the {\tt \texttt{ePump}} updating. 
From Fig.~\ref{Fig:CT14HERA2mJetpCMS8Ntt} we see that, without the jet data in the fit, the normalized  differential CMS 8 TeV $t\bar t$  data 
$y_{t\bar t}$, $m_{t\bar t}$,  $y_t$ and $p^t_{T}$  
have rather obvious impact on both the central predictions and uncertainty bands of the CT14HERA2mJ gluon PDF at the 
region $10^{-4} \lesssim x \lesssim 0.6$. namely, each $t\bar t$  data  increases the gluon PDF in the 
region $10^{-4} \lesssim x \lesssim 0.15$ while each $t\bar t$ data deceases it in the region $x \gtrsim 0.15$, 
but the updated gluon PDF in four cases are well within the uncertaity bands of PDFs.

\section{The impact of CMS 7 TeV inclusive jet data on CT14HERA2mJ}\label{sec:jetupdateHERA2mJ}

\begin{figure}[H]
    \begin{center}
\includegraphics[width=0.49\textwidth]{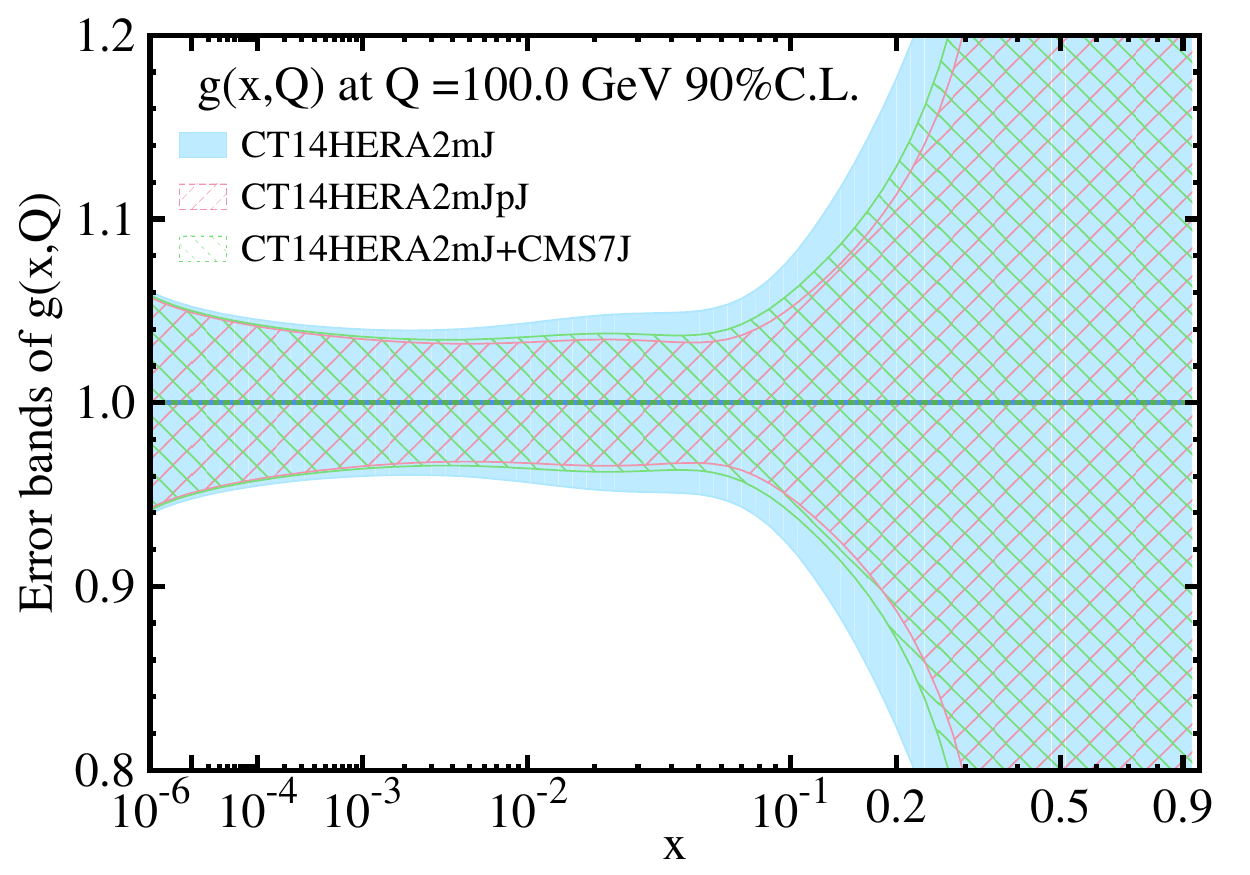}
\includegraphics[width=0.49\textwidth]{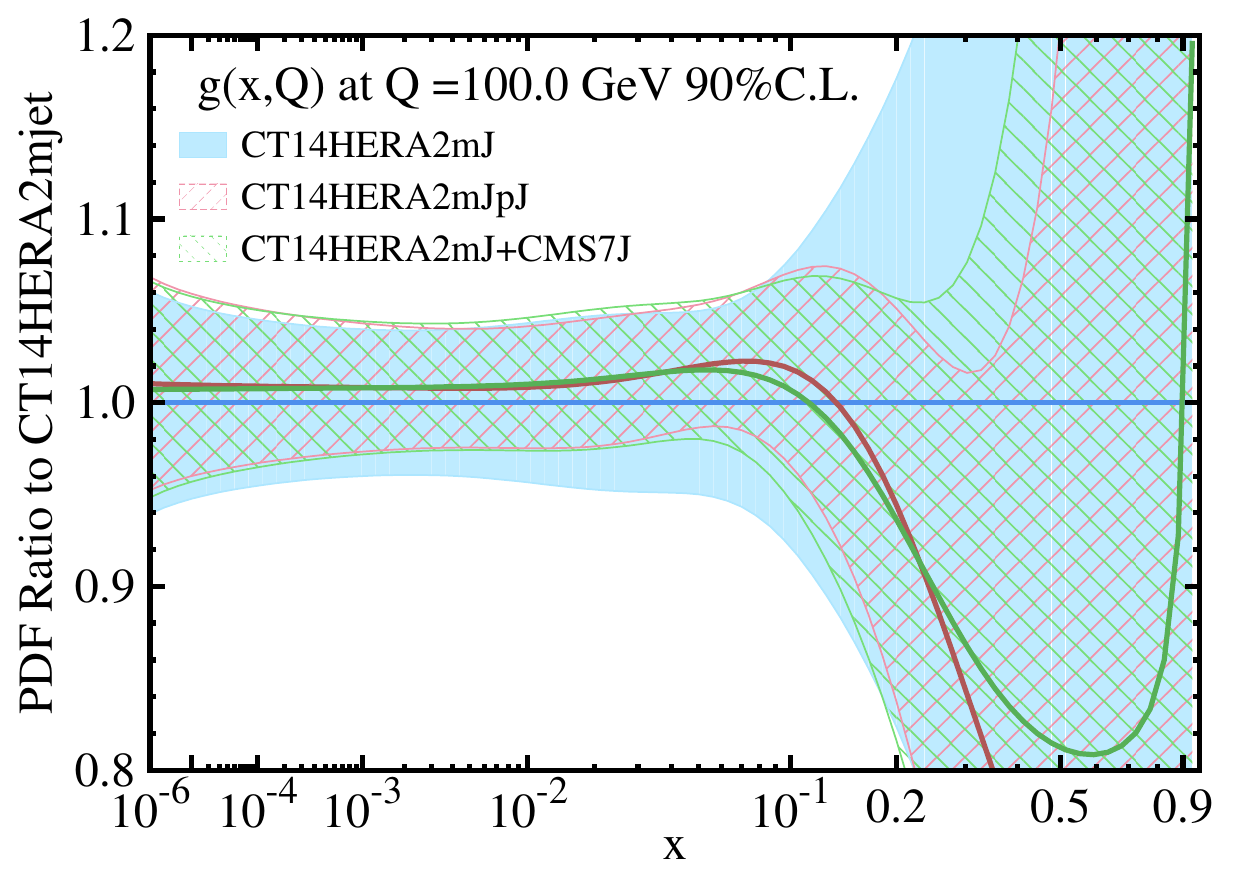}
\end{center}\caption{Left plot shows comparison of the gluon PDF error band from CT14HERA2mJ (blue band), CT14HERA2mJpJ (red shaded band),and CT14HERA2mJ+CMS7J (green shaded band), note that each error band is normalized to its own gluon central PDF.  Right plot shows, the gluon PDF ratios for 
CT14HERA2mJ (blue), CT14HERA2mJpJ (red),and CT14HERA2mJ+CMS7J (green) 
over the best-fit of the  base CT14HERA2mJ (blue band).
}\label{CT14HERA2mJetpCMS7jet}
\end{figure}

In Fig.~\ref{CT14HERA2mJetpCMS7jet}, we compare gluon PDFs from CT14HERA2mJ, CT14HERA2mJpJ and CT14HERA2mJ+CMS7J.
Here CT14HERA2mJpJ and CT14HERA2mJ+CMS7J are obtained by  {\tt \texttt{ePump}} by adding four jet data \cite{Aaltonen:2008eq,Abazov:2008ae,Aad:2011fc,Chatrchyan:2012bja} into CT14HERA2mJ fit, and including the CMS 7
TeV inclusive jet data \cite{Chatrchyan:2012bja} into CT14HERA2mJ fit.
We first observe that the CT14HERA2mJpJ
gluon PDF has smaller uncertainty band than the
CT14HERA2mJ+CMS7J gluon PDF, which tells that the four jet data have a
strong impact on the gluon PDF.
It is therefore understandable why we don't see significant
impact on the CT14HERA2 PDF from the $t\bar t$ data.
Despite  difference on uncertainty between
CT14HERA2mJpJ gluon PDF and CT14HERA2mJ + $t\bar t$ (from ATLAS and CMS)  gluon PDF,
it is worth to note that, both the $t\bar t$  (from ATLAS and CMS)  and  jet data
give similar impact on the gluon central PDF, which
shows the agreement between the impact on gluon PDF from
 $t\bar t$ and jet data.

It is obvious to see that, the CMS 7 TeV inclusive jet data dominate the contribution of constraining the gluon PDF among the four jet data. Therefore, in the following study, we consider only the CMS 7 TeV inclusive jet data.

It is worth to note that, the $t\bar t$ production data
have rather smaller number of data points than the jet data
by about a factor of 10.
After testing the impact on CT14HERA2 and CT14HERA2mJ PDFs,
it is interesting to compare the sensitivity per data point
for the jet and $t\bar{t}$ data.
In order to see this, a hypothetical weight is implemented
to the single differential $t\bar{t}$ production data with the weight to be
equal to the ratio between number of data points of the
CMS 7 TeV jet data and the $t\bar{t}$ data.
Taking the  CMS 8TeV normalized $p^t_T$ distribution as an example,
the hypothetical weight that apply to the data is equal to
$w = 133/8 = 16.6$.
In practice, a larger weight can arise from increasing the
event statistics or reducing the experimental errors.

\begin{figure}[H]
    \begin{center}
\includegraphics[width=0.49\textwidth]{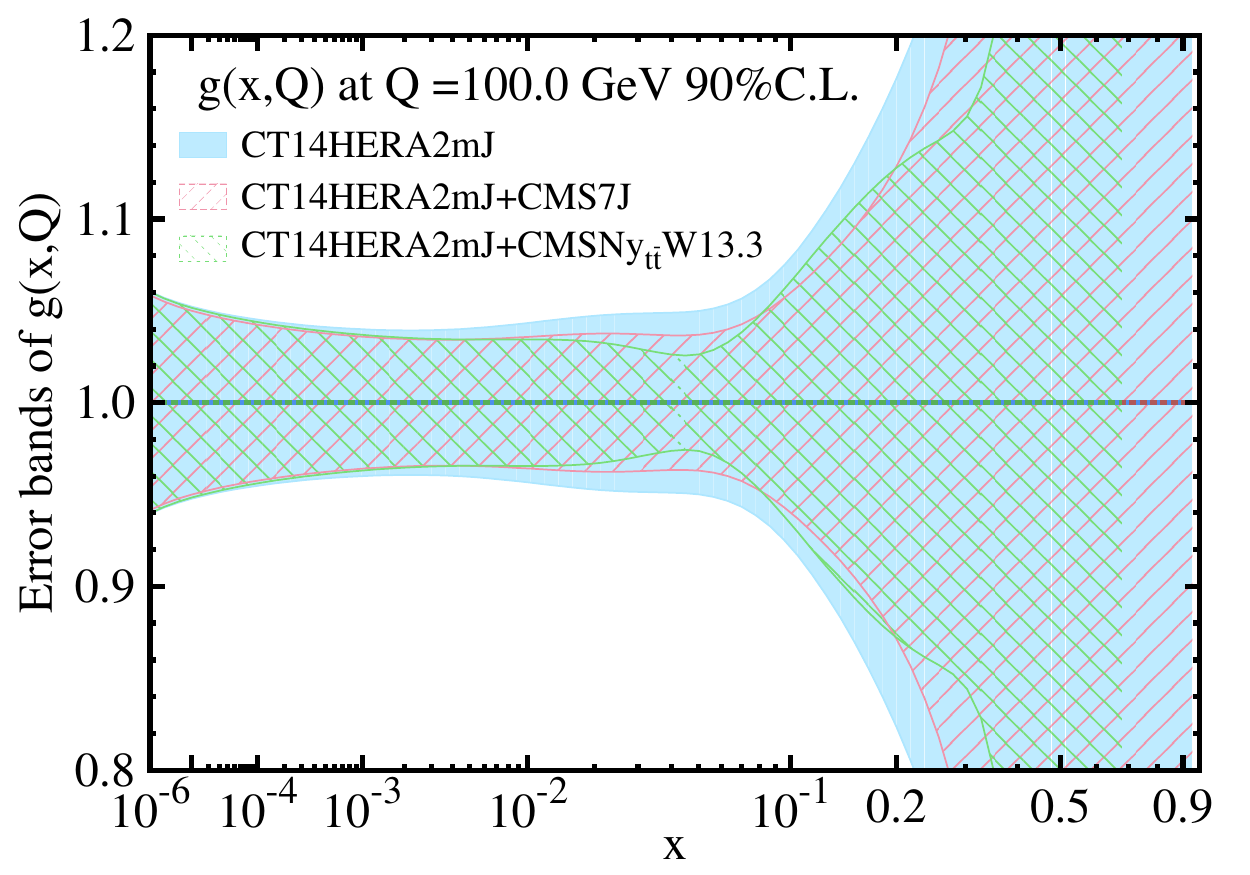}
\includegraphics[width=0.49\textwidth]{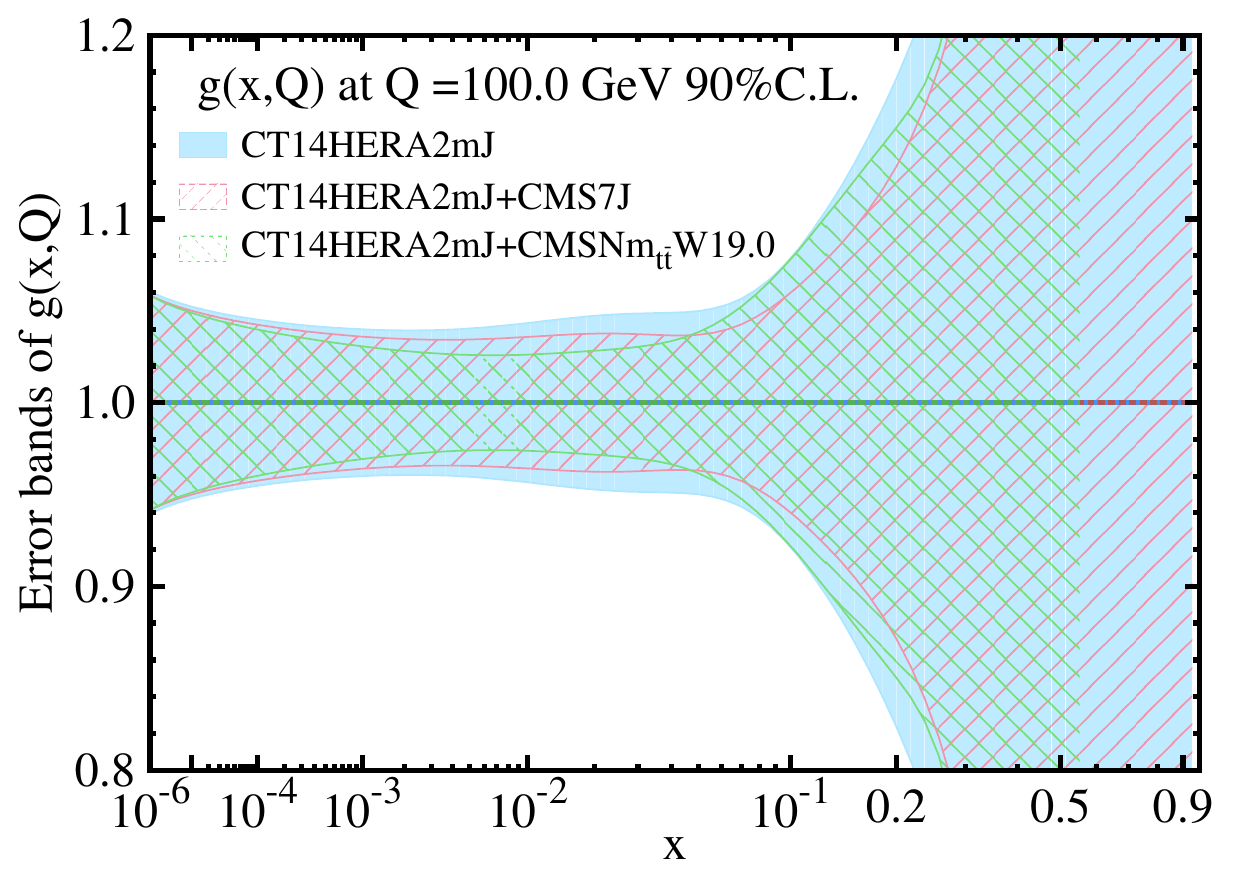}
\includegraphics[width=0.49\textwidth]{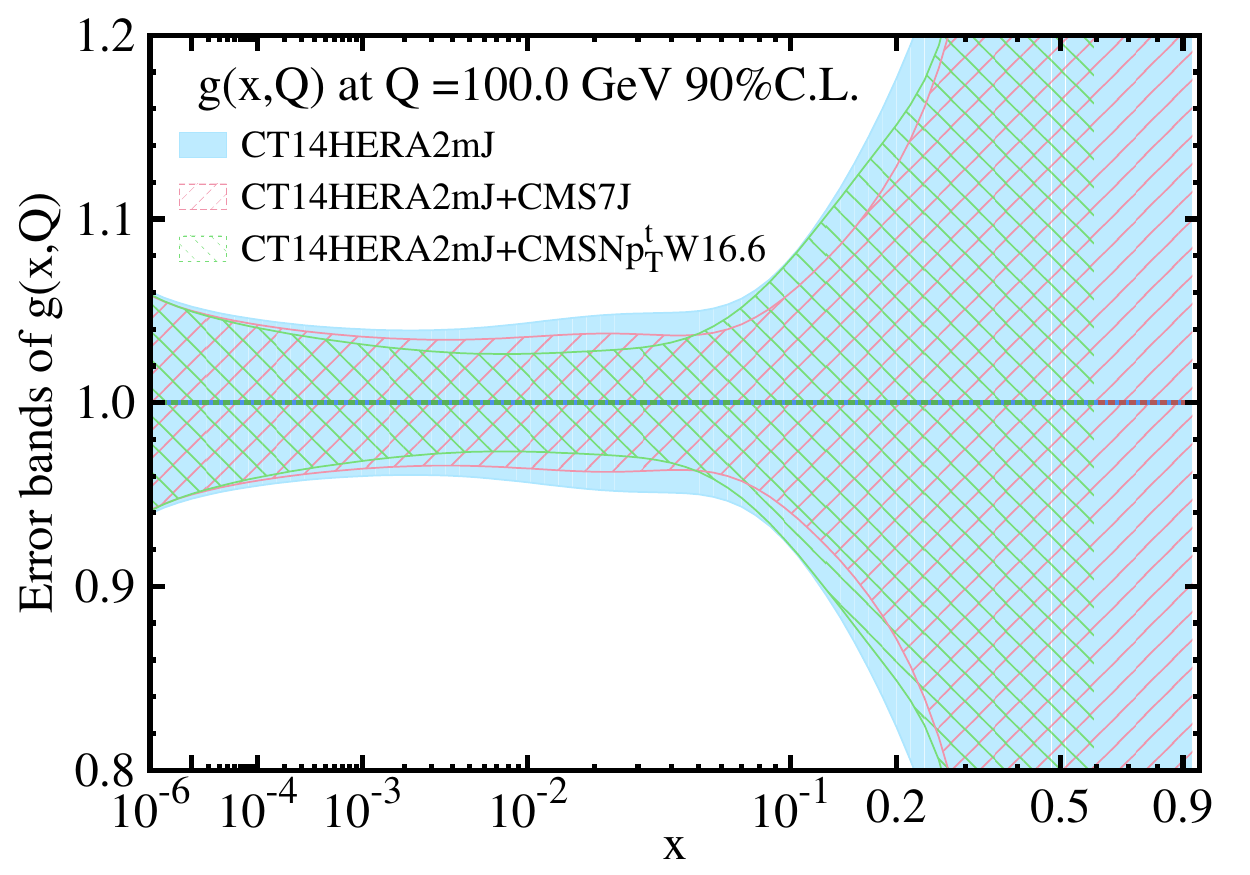}
\includegraphics[width=0.49\textwidth]{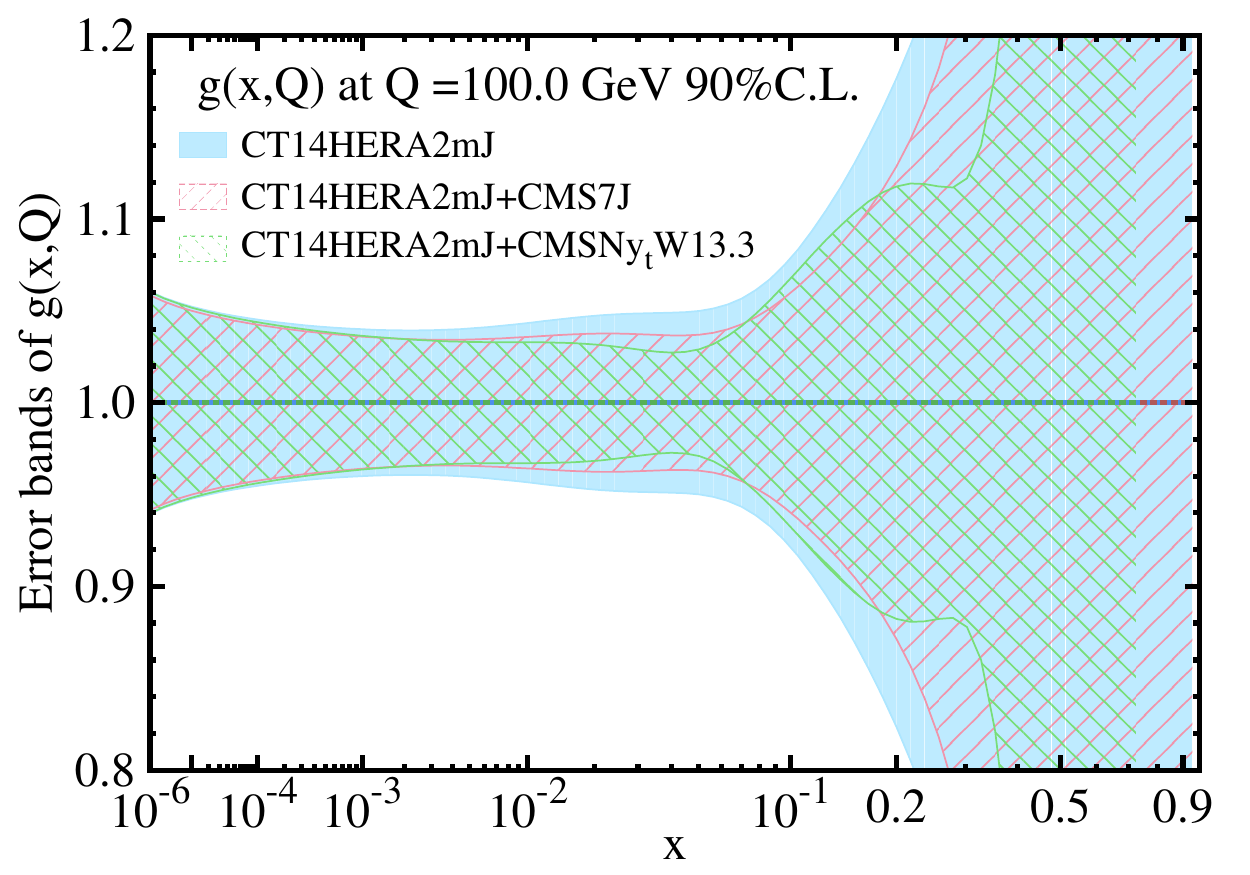}
\end{center}\caption{Comparison of the gluon PDF error band from
	CT14HERA2mJ+CMS7J (red shaded band), 
	CT14HERA2mJ+CMSN$y_{t\bar t}$W13.3 (top left, green shaded band), 
	CT14HERA2mJ+CMS8N$m_{t\bar t}$W19.0 (top right, green shaded band), 
	CT14HERA2mJ+CMS8N$p_t$W16.6 (bottom left, green shaded ban), 
	CT14HERA2mJ+CMS8N$y_t$W13.3  (bottom right, green shaded ban) PDFs, 
	that are obtained by adding CMS 7 TeV inclusive jet data,  
	CMS 8 TeV  normalized $1/\sigma \; d\sigma/dy_{t\bar t}$ data with weight 13.3,
	$1/\sigma \; d\sigma/dm_{t\bar t}$ data with weight 19.0,
	$1/\sigma \; d\sigma/dy_t$ data with weight 13.3,
	$1/\sigma \; d\sigma/dp^t_T$ data with weight 16.6 
 at Q = 100 GeV and at 90\% C.L., with the base CT14HERA2mJ  gluon PDF (blue band).
}\label{Fig:JCMSN1}
\end{figure}

In this naive estimation, we assume the central values
of the measurement do not change such that
the central prediction after updating with the
hypothetical weight is rather less meaningful.
For this reason, in the following we show the comparison of the
PDFs uncertainty.
In Fig.~\ref{Fig:JCMSN1}, we compare the impact of the CMS 7 TeV inclusive jet data and  CMS 8 TeV  normalized $t\bar{t}$
production data with the hypothetical weight on gluon PDF uncertainty.
We find that, the weighted $t\bar{t}$ production data provide stronger
constraint on gluon PDFs for $10^{-3} \lesssim x \lesssim 5 \times 10^{-2}$.
It is also true for the absolute ATLAS 8 TeV $t\bar{t}$ production data.
With the hypothetical weight equal to the ratio of number of jet
and $t\bar t$ data points, the absolute $t\bar{t}$ production data provide about the same constraint on gluon PDF as the jet data.

\begin{figure}[H]
    \begin{center}
    \includegraphics[width=0.49\textwidth]{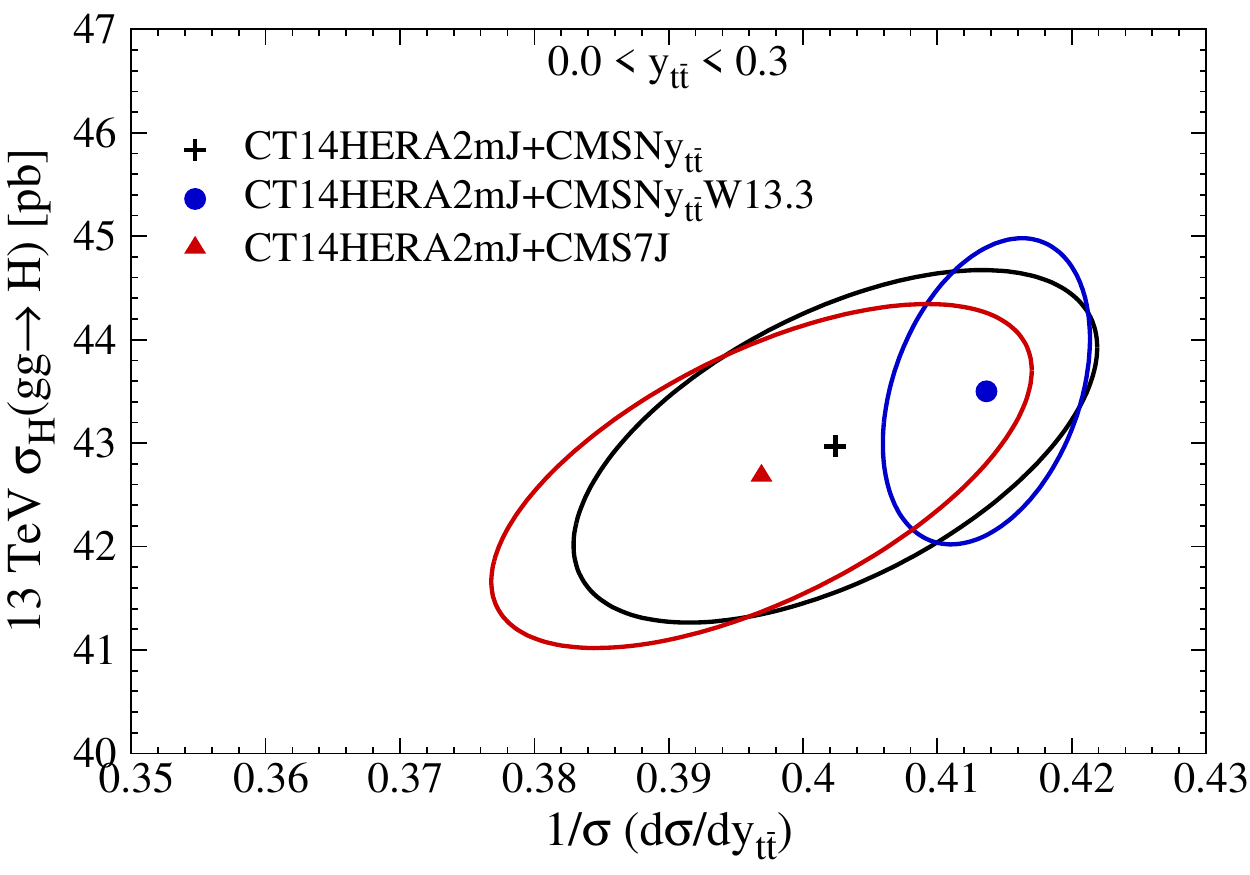}
    \includegraphics[width=0.49\textwidth]{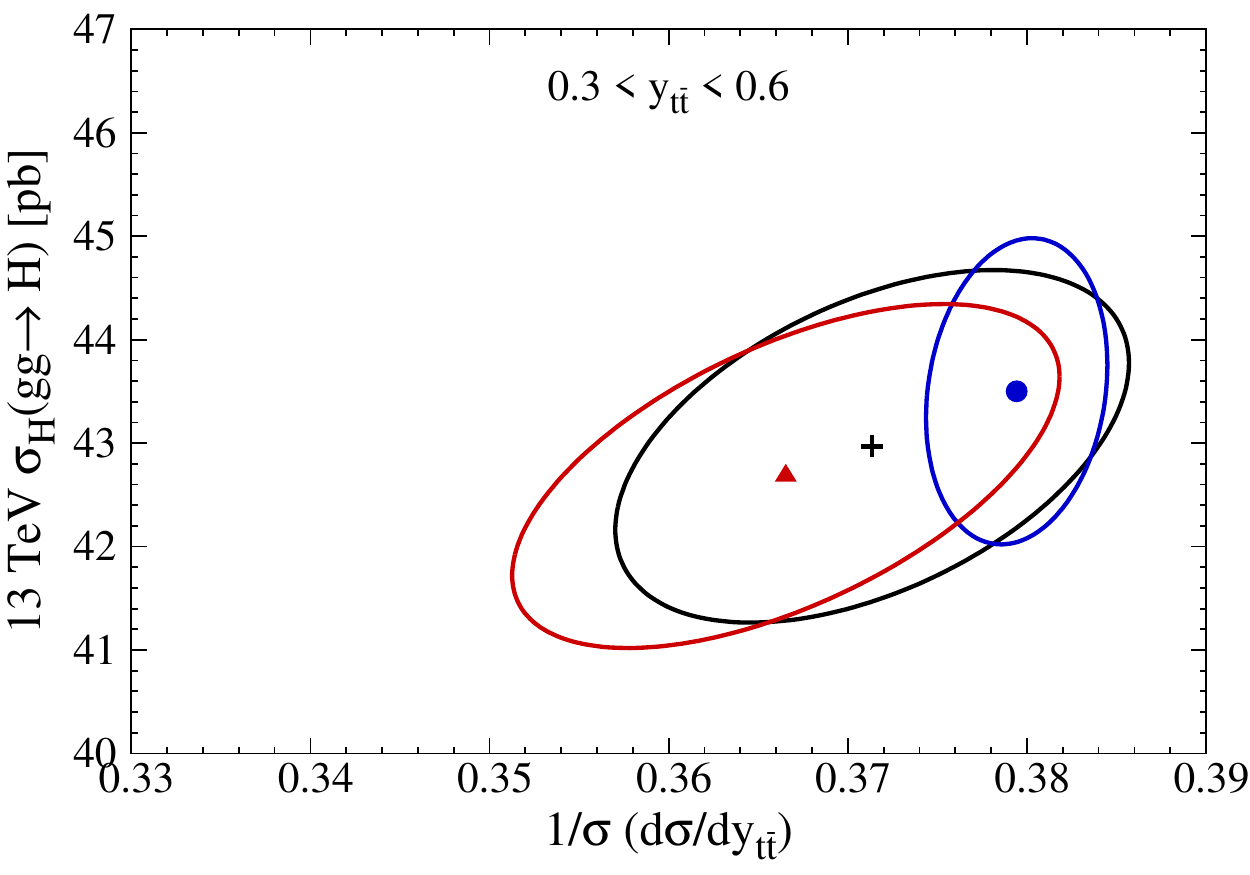}
    \includegraphics[width=0.49\textwidth]{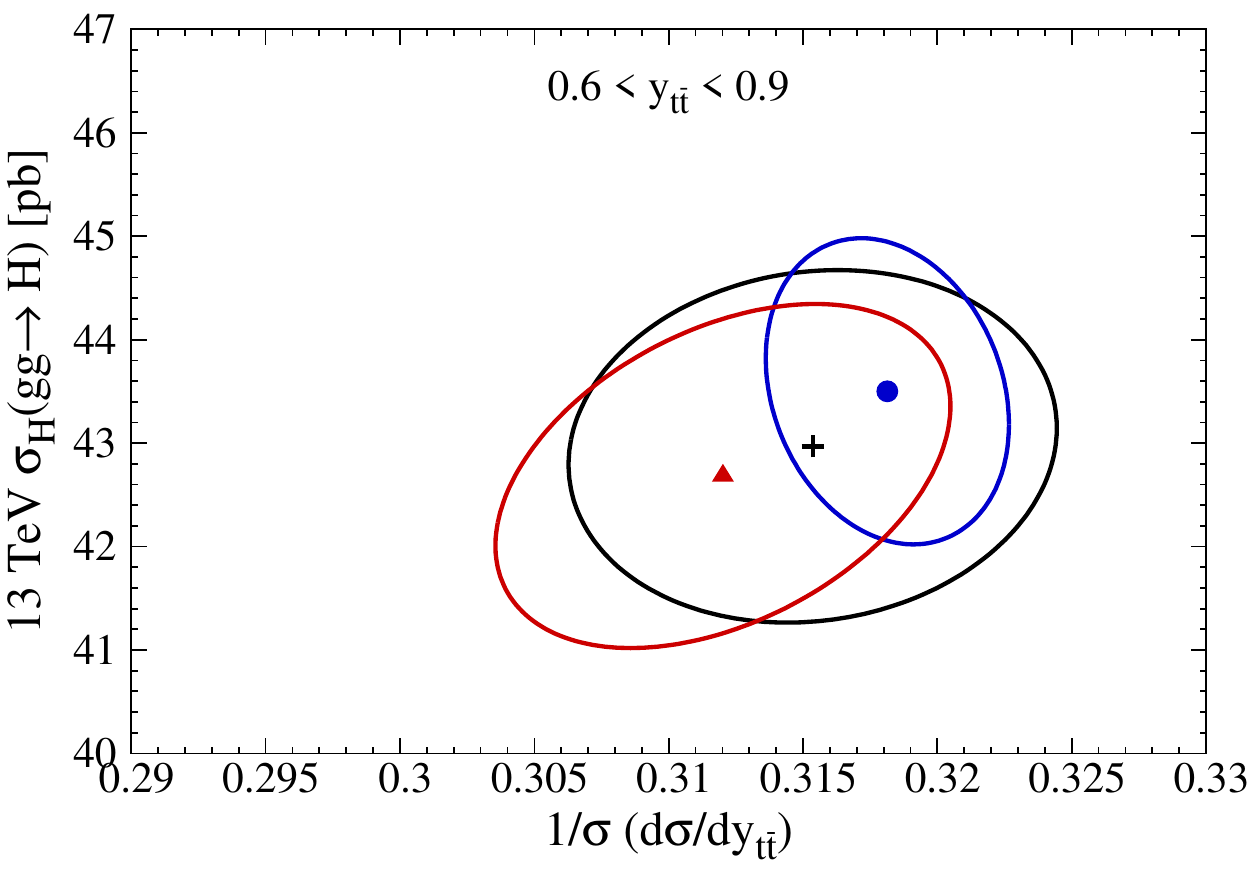}
    \includegraphics[width=0.49\textwidth]{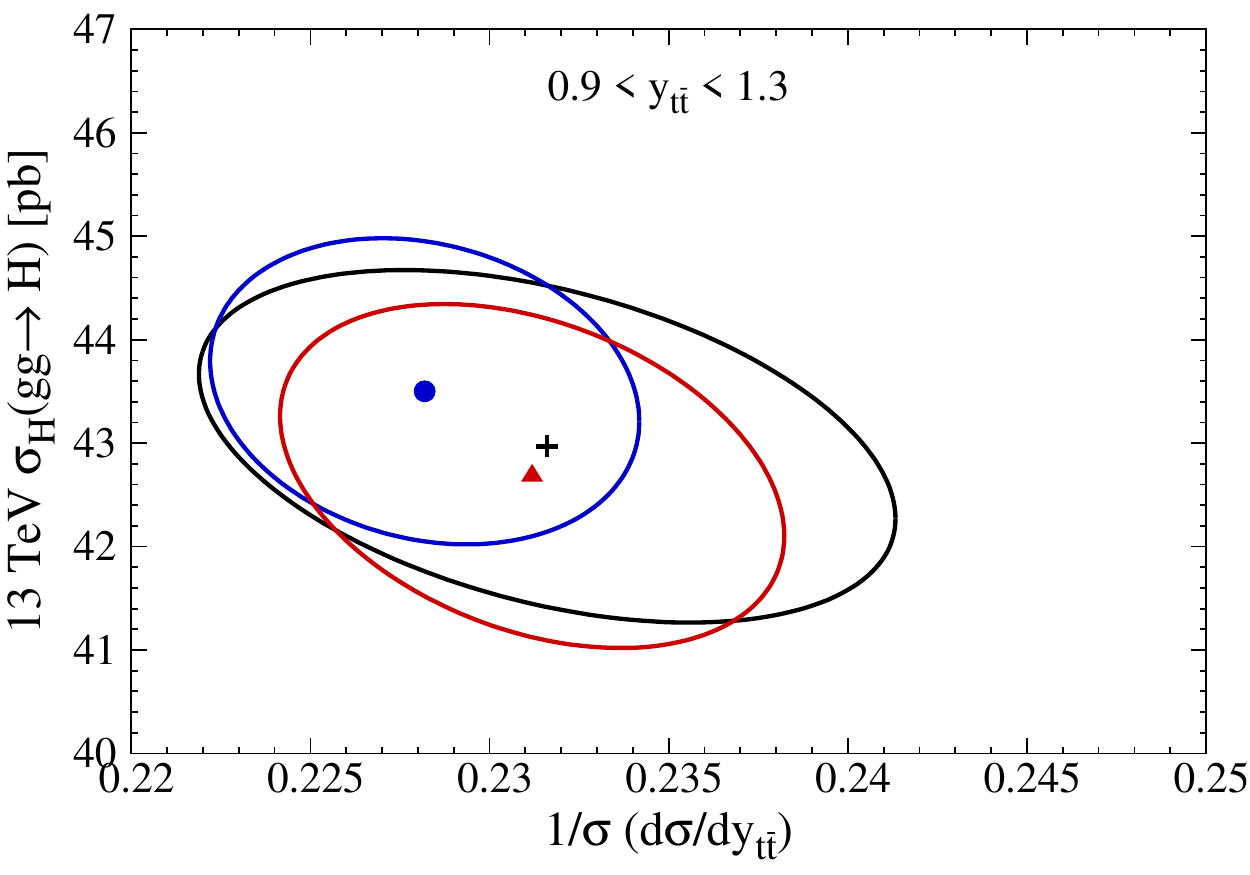}
    \includegraphics[width=0.49\textwidth]{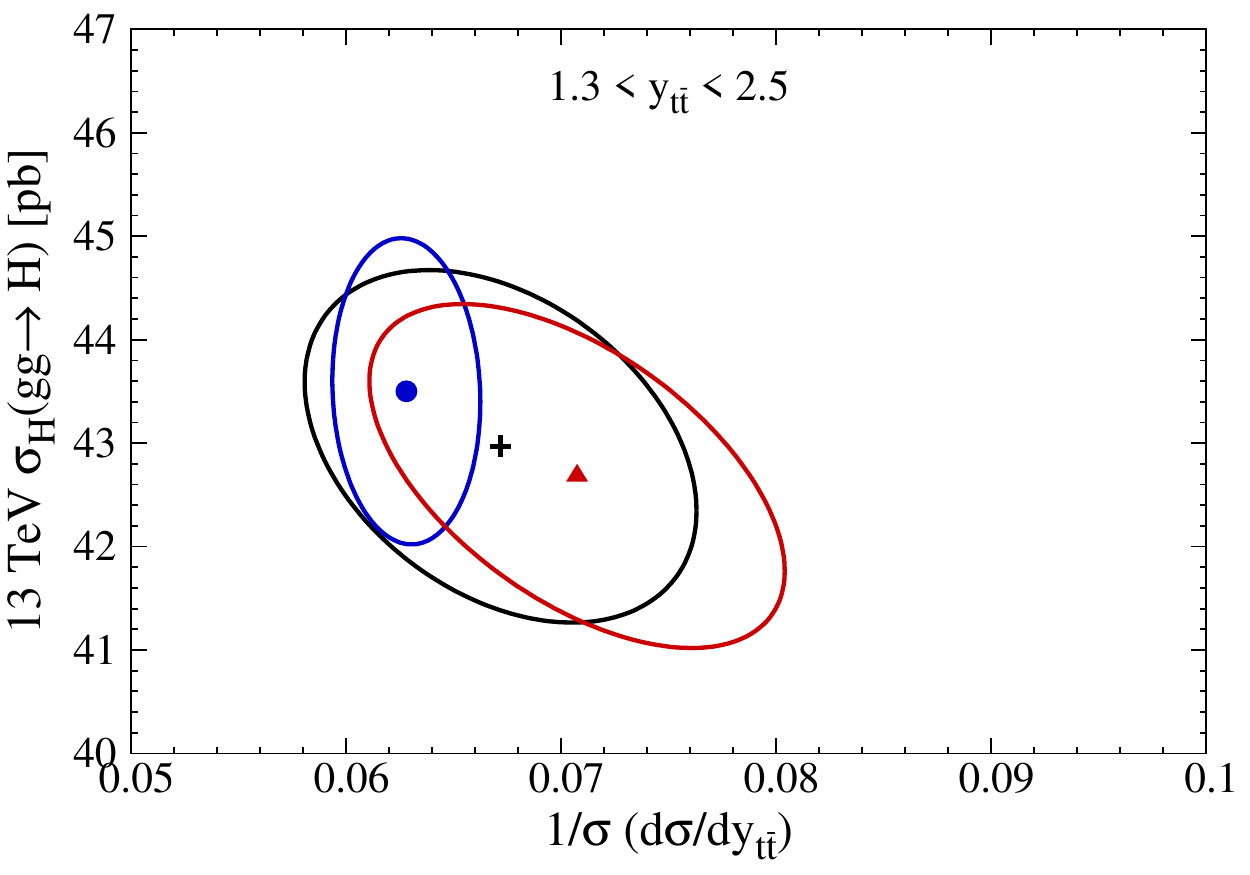}
    \end{center}
\caption{Correlation ellipses between the Higgs production rate via gluon-gluon fusion and the normalized $1/\sigma \; d\sigma/\sigma y_{t\bar t}$ differential cross section at 13 TeV for {\tt \texttt{ePump}} updated 
CT14HERA2mJ+CMSN$y_{t\bar t}$, 
CT14HERA2mJ+CMSN$y_{t\bar t}$W13.3, 
CT14HERA2mJ+CMS7J and
CT14HERA2mJ PDFs, PDF uncertainty is at the 90\% C.L..}\label{Fig:ggh}
\end{figure}

Next, we examine the impact of the  CMS 8 TeV normalized  
$1/\sigma \; d\sigma/dy_{t\bar t}$ data and CMS 7 TeV  inclusive jet via  {\tt \texttt{ePump}} on the observable. The Higgs production rate through gluon-gluon fusion at the LHC is sensitive to gluon PDF in the middle-$x$ region,
which is constrained by both CMS 7 TeV  inclusive jet and 
$1/\sigma \; d\sigma/dy_{t\bar t}$ data.
In Fig.~\ref{Fig:ggh}, we show the correlation ellipses between CMS 8 TeV normalized $d\sigma/dy_{t\bar t}$
data for various rapidity bins and Higgs production through gluon-gluon fusion
at 13 TeV for 
CT14HERA2mJ+CMSN$y_{t\bar t}$ (black), 
CT14HERA2mJ+CMSN$y_{t\bar t}$W13.3 (dark blue), and
CT14HERA2mJ+CMS7J (red). 
The central prediction of the CT14HERA2mJ+CMSN$y_{t\bar t}$W13.3 is obtained by assuming the central measurement is the same as that in CT14HERA2mJ+CMSN$y_{t\bar t}$.

\section{Top quark mass dependence }\label{mtop}

\begin{figure}[H]
	\begin{center}
		\includegraphics[width=0.49\textwidth]{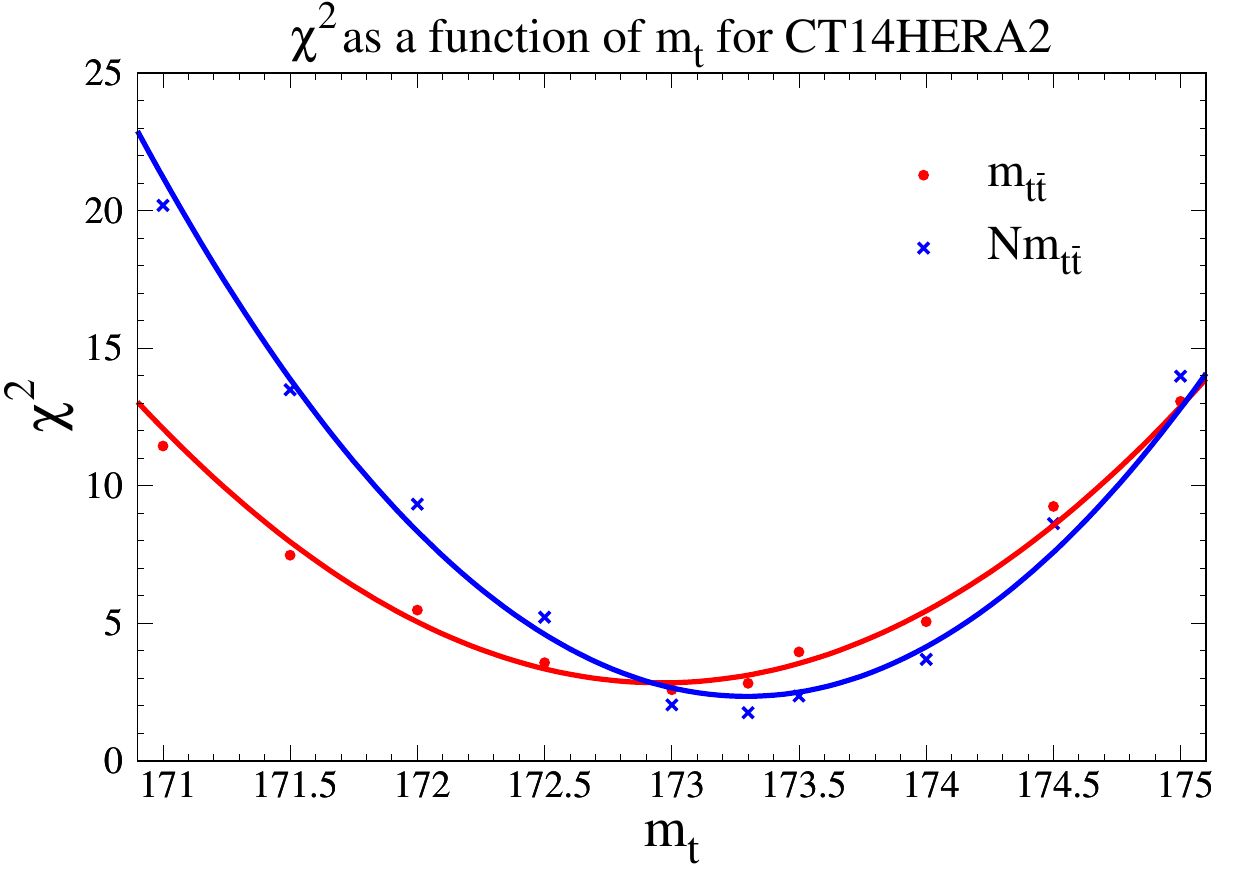}
	\end{center}
	\caption{The chi-square function ($\chi^2$)  versus the top mass for the  absolute and normalized  8 TeV single differential cross sections as a function of the invariant mass of the top pair $m_{t\bar t}$.  }\label{Fig:mt}
\end{figure}

 The top-quark mass sensitivity of the differential top pair production distributions as a function of $m_{t\bar t}$
 has been studied in Refs.~\cite{Czakon:2016vfr, Ju:2019mqc}, 
and found that the top quark mass dependence is pronounced in the  $m_{t\bar t}$ distribution.
The Ref.~\cite{Czakon:2016olj} have also studied the sensitivity upon variations of $m_t$ of the  differential distributions as a function of $m_{t\bar t}$, $y_{t\bar t}$, $y_t$ and $p^t_T$
and  found that  the invariant mass distribution of the top pair and the top transverse momentum $p_T^t$ distributions have stronger dependence on top mass than the  top rapidity  ($y_t$) and the rapidity ($y_{t\bar t}$) of the top pair.
By studying the top-quark mass dependence of the differential top pair production distributions as function of $m_{t\bar t}$ one could provide another way to determine the mass of the top quark.  
In Fig.~\ref{Fig:mt} we show the chi-square function, $\chi^2$ versus the top-quark mass, for the absolute and normalized  8 TeV single differential cross sections as a function of the invariant mass  $m_{t\bar t}$ of the top-quark pair by using the Monte-Carlo numerical calculation program MadGraph\cite{mg5} with CT14HERA2 PDFs and the same set-up as the Refs.~\cite{Czakon:2016dgf,Czakon:2017dip}.
The $d\sigma/dm_{t\bar t}$  is shown in red; the $1/\sigma \; d\sigma/dm_{t\bar t}$ is shown in green.
The parabolic curves are fitted from calculation with many values of top mass from $171.0$ GeV to $175.0$ GeV. 
As we see that the two curves have slighly different minimum, $m_t=173.0$ GeV for absolute distributions and  $m_t=173.5$ GeV for normalized distributions. It may be because the anti correlation of the the inverse of the total cross section. 
We also perform same the study for the $p^t_T$, $|y_t|$ and  $|y_{t\bar t}|$ distributions, and we find that differential cross sections as a
function of  $p^t_T$, $|y_t|$ and  $|y_{t\bar t}|$ do not depend on
top-quark mass.

\section{Conclusions}\label{sec:conclusion}

In this paper, we examined the impact of the 
$t\bar t$ data (top quark pair invariant mass $m_{t\bar t}$, top quark pair rapidity $y_{t\bar t}$, the individual top quark/antiquark transverse momentum $p^t_T$, and absolute value of the top quark rapidity $|y_t|$)  on the CT14HERA2 and CT14HERA2mJ PDFs using {\tt \texttt{ePump}} package.
From Fig.\ref{Fig:CT14HERA2pttb-sing-dis}, we observe that, when adding the $t \bar t$ data one by one to a global data set, CMS normalized distributions show larger constraining power than the ATLAS absolute and normalized data. Furthermore, we observed different impact from various distributions on the gluon PDF. The reasons are as follows. First of all, the normalized data yield stronger constraints than the absolute data, as expected, because the dominant collider luminosity error present in the absolute data cancel in the normalized data. Due to its much smaller error, the normalized data generally provide a much stronger constraint on PDFs.
Secondly, different differential cross section data could have different degree of sensitivity to gluon PDFs. For example, it is expected that the rapidity distributions, either the rapidity of $t$ or $t \bar t$ pair, are more sensitive to gluon PDFs than the other distributions. Moreover, to extract PDFs from the data, it is also important to use the state-of-art theory calculations in the global analysis. It has been known that both the transverse moment of $t$ or $t \bar t$ pair, as well as the invariant mass of the $t \bar t$ pair could suffer large higher order QCD and electroweak corrections. Yet, in the current analysis, we only include up to NNLO QCD corrections.
In Fig.\ref{Fig:mt}, we  show the top mass dependence of the differential top pair production distributions at 8 TeV.  
We found that only $ d\sigma/dm_{t\bar t}$ distribution is sensitive to the top-quark mass with the minimum at around $173.3$ GeV.
Because the invariant mass of the $t \bar t$ pair is expected to be sensitive to the value of top quark mass $m_t$, the theory predictions of both low and high mass bins are sensitive to higher order QCD and electroweak corrections. Hence, the impact of the normalized and absolute data could be different, depending on the accuracy of the integrated cross section value used in the normalized data measurement. 
Note that all the top-quark pair production data show minor impact on CT14HERA2 gluon PDF when jet data have been included in the global analysis. 
It is because the number of data points for the $t\bar t$
data is much less than the jet data.
By giving a hypothetical weight on the $t\bar t$  data as the ratio
of number of data points between jet data and $t\bar t$ data,
the $t\bar t$  data show good agreement with the impact from jet data
with similar strength.
Hence, the sensitivity per data point of $t\bar t$ data is
similar to that of jet data, while the total sensitivity of
the data set depends on the total number of data points.

\begin{acknowledgments}
We would like to thank Joey Huston and C. -P. Yuan for many helpful discussions. We are very grateful to Joey Huston for proofreading.
The work of S.~Dulat was supported by the National Natural Science Foundation of China under the Grant No. 11965020 and No. 11847160. 
\end{acknowledgments}


%

\end{document}